\documentclass[longauth]{aa}
\pdfoutput=1
\usepackage{mathtools}
\usepackage{multirow}
\usepackage{comment}
\usepackage{siunitx}
\usepackage{booktabs}
\usepackage{graphicx}
\usepackage{menukeys}
\usepackage{algorithm}
\usepackage{ulem}
\usepackage[noend]{algpseudocode}
\usepackage{ccaption}
\usepackage{caption}
\usepackage{txfonts}
\usepackage{subfigure}
\usepackage{cuted}
\usepackage{float}
\usepackage{multicol}
\usepackage{placeins}
\usepackage{lscape}
\usepackage{soul}
\setstcolor{red}
\usepackage{enumitem}

\usepackage[colorlinks=true, allcolors=blue]{hyperref}
\usepackage{upgreek} 
\usepackage[version=4]{mhchem}
\DeclareCaptionFormat{continued}{ #1~(continued)}
\usepackage[]{starfont}
\usepackage{amsmath}

\DeclareSymbolFont{starfontsym}{OT1}{sts}{m}{n}
\DeclareMathSymbol{\mathTerra}{\mathord}{starfontsym}{76}

\DeclareSIUnit \lsun {L_\odot}
\DeclareSIUnit \rsun {R_\odot}
\DeclareSIUnit \msun {M_\odot}
\DeclareSIUnit \mearth {M_\oplus}
\DeclareSIUnit \rearth {R_\oplus}

\DeclareSIUnit \year {yr}
\DeclareSIUnit \day {day}
\DeclareSIUnit \pixel {px}
\DeclareSIUnit \ppm {ppm}

\makeatletter
\renewcommand*\aa@pageof{, page \thepage{} of \pageref*{LastPage}}
\makeatother
\newcolumntype{L}[1]{>{\raggedright\arraybackslash}p{#1}}
\newcolumntype{C}[1]{>{\centering\arraybackslash}p{#1}}
\newcolumntype{R}[1]{>{\raggedleft\arraybackslash}p{#1}}

\newcommand{\teff}{T_{\rm eff}}
\newcommand{\logg}{\log g}
\newcommand{\vmic}{\xi_{\rm t}}

\newcommand{\vsini}{V_{\rm sini}}

\DeclareSIUnit \parsec {pc}
\DeclareSIUnit \astrounit {au}
\DeclareSIUnit \mjup {M_J}
\DeclareSIUnit \rjup {R_J}

\providecommand{\orcit}[1]{\protect\href{https://orcid.org/#1}{\protect\includegraphics[width=8pt]{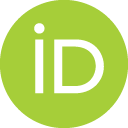}}}

\begin{document}
   \title{The panchromatic JWST dayside spectrum of WASP-121\,b \\ reveals
a refractory-rich formation}
   \author{
K. Angelique Kahle~\orcit{0000-0001-7714-7551}\inst{1,2}\thanks{E-mail: kahle@mpia.de}
\and Paul Molli\`ere~\orcit{0000-0003-4096-7067}\inst{1}
\and Laura Kreidberg~\orcit{0000-0003-0514-1147}\inst{1}
\and Bertram Bitsch~\orcit{0000-0002-8868-7649}\inst{3}
\and Mara Attia~\orcit{0000-0002-7971-7439}\inst{4}
\and Silke S. Dainese~\orcit{0000-0001-7568-6674}\inst{5}
\and Nicholas Storm~\orcit{0000-0002-5259-3974}\inst{1,2}
\and Daniel Valentine~\orcit{0000-0002-2643-6836}\inst{6}
\and Thomas M. Evans-Soma~\orcit{0000-0001-5442-1300}\inst{1,7}
\and H.J. Hoeijmakers~\orcit{0000-0001-8981-6759}\inst{8}
\and Stefan Pelletier~\orcit{0000-0002-8573-805X}\inst{9}
\and Ludmila Carone~\orcit{0000-0001-9355-3752}\inst{10}
\and Cyril Gapp~\orcit{0009-0007-9356-8576}\inst{1,2}
\and Yoav Rotman~\orcit{0000-0003-4459-9054}\inst{11}
\and Sophia R. Vaughan~\orcit{0000-0002-8199-9818}\inst{1}
\and D. A. Christie~\orcit{0000-0002-4997-0847}\inst{1}
\and Louis-Philippe Coulombe~\orcit{0000-0002-2195-735X}\inst{12,13}
\and Christiane Helling~\orcit{0000-0002-8275-1371}\inst{9,14}
\and Thomas Henning~\orcit{0000-0002-1493-300X}\inst{1}
}
\institute{
\inst{1} Max Planck Institute for Astronomy, K\"{o}nigstuhl 17, 69117 Heidelberg, Germany \\
\inst{2} Department of Physics and Astronomy, Heidelberg University, Im Neuenheimer Feld 226, 69120 Heidelberg, Germany \\
\inst{3} Department of Physics, University College Cork, Cork, T12 R229, Ireland \\
\inst{4} Kapteyn Astronomical Institute. University of Groningen, Groningen, the Netherlands \\
\inst{5} Department of Physics and Astronomy, Aarhus University, Aarhus, Denmark \\
\inst{6} School of Physics, University of Bristol, Bristol, UK \\
\inst{7} School of Science, University of Newcastle, Callaghan, New South Wales,
Australia \\
\inst{8} Department of Physics, Lund University, Lund, Sweden \\
\inst{9} Observatoire astronomique de l'Universit\'e de Gen\`eve, 51 chemin Pegasi 1290 Versoix, Switzerland \\
\inst{10} Space Research Institute, Austrian Academy of Sciences, Schmiedlstrasse 6, A-8042 Graz, Austria \\
\inst{11} School of Earth and Space Exploration, Arizona State University, Tempe, AZ, USA \\
\inst{12} Plan\'etarium de Montr\'eal, Espace pour la Vie, 4801 av. Pierre-de Coubertin, Montr\'eal, Canada \\
\inst{13} Institut Trottier de recherche sur les exoplan\`etes, D\'epartement de Physique, Universit\'e de Montr\'eal, Montr\'eal, Qu\'ebec, Canada \\
\inst{14} Institute of Theoretical and Computational Physics, TU Graz, NAWI Graz, Petersgasse 16, 8010 Graz, Austria \\
}
   \titlerunning{Joint JWST dayside spectrum of WASP-121\,b}
   \authorrunning{K. A. Kahle et al.}
   \date{Received -; accepted -}
   \abstract
    {One path to understand how planets form is to link their present-day atmospheric composition to predictions from planet formation models. Traditional approaches based solely on C/O and overall metallicity are prone to degeneracies, but abundances of refractory species can provide a useful additional formation tracer for the hottest planets. Here we investigate the refractory abundance in the atmosphere of the ultra-hot Jupiter WASP-121\,b, combining a new JWST MIRI/LRS observation with archival NIRSpec/G395H and NIRISS/SOSS data to obtain a panchromatic dayside emission spectrum from $0.6$ to $\SI{12}{\micro\meter}$.
  Our retrieval analysis detects the refractory tracer SiO gas at high confidence ($\Delta\ln$(Z)$>23$), in addition to previously detected volatile species. The atmosphere is enriched in volatile and refractory species, with enhanced refractory-to-volatile abundance ratios of $\mathrm{Si}/\mathrm{O}=3.54^{+0.86}_{-0.69}\times$ stellar and $\mathrm{Si}/\mathrm{C}=3.05^{+1.12}_{-0.80}\times$ stellar, relative to new stellar abundance constraints from ESPRESSO data.
 In addition, we confirm the depletion of TiO and the need for an additional source of reflected light opacity with a geometric albedo of $0.22\pm0.03$. The retrieved dayside temperature-pressure profile has a strong inversion layer, with a more complex structure than standard parameterizations can accommodate, and an eclipse map analysis indicates a small eastward hotspot offset of $4.8^{+2.7\circ}_{-2.8}$. Comparing our results with models of planet formation, we find that the measured enrichment pattern was shaped by accretion from multiple reservoirs, either through a mixture of solid and gas accretion interior to the water ice line or through continued solid accretion during inward migration from farther out in the disk. Finally, we model the planet's dynamical history and find that it could reach its current high-obliquity orbit as a consequence of a post-formation dynamical event, such as planet-planet scattering or von Zeipel–Lidov–Kozai cycles.
    } 
   \maketitle

\section{Introduction}\label{sec:intro}
A long-standing goal of both exoplanet atmosphere research and planet formation studies is to find a link between planet formation and the observable present-day atmospheres. Elemental inventories play a key role in this, as they have the potential to determine the formation history and evolutionary pathway of the planets~\citep[e.g.,][]{oeberg2011snowlines,schneider2021IIpebblesVolRef,chachan2023formationmodel,Penzlin2024bowie}. 

Traditionally, a widely used diagnostic tool for the formation location of a planet is the abundance ratio between carbon and oxygen reservoirs in its present-day atmosphere~\citep[e.g.,][]{oeberg2011snowlines,Madhusudhan2014hotJups,Helling2014diskevol,Cridland2019formation,Schneider2021IpebblesCO}. However, when considering the uncertainties in planet formation and disk composition, structure, and evolution, metal enrichment and the carbon-to-oxygen ratio alone are insufficient to draw clear conclusions on a planet's formation~\citep[e.g.,][]{Mordasini2016imprint,Eistrup2016volatilesettling,Turrini2021tracer,molliere2022formation,Feinstein2025link}: A specific ratio of carbon-to-oxygen can be produced by a variety of different formation and evolution pathways. In addition, the exact carbon-to-oxygen ratio of a planet is often ambiguous from space-based data, as it can be highly model dependent~\citep[e.g.,][]{lueber2024informationcontent} and affected by the presence of unobserved clouds~\citep{Helling2021clouds}. 

To overcome these limitations, one path forward is to include multiple element abundances \cite[e.g.,][]{Lothringer2021VolRef,Turrini2021tracer,Pacetti2022diskchem,chachan2023formationmodel,Crossfield2023Sulfur} or isotope ratios \citep[e.g.,][]{Zhang202113CO,Stauffenberg202613CO,Gonzalez202613CO} when linking planetary atmospheres to formation models. Specifically, \citet{Lothringer2021VolRef} suggested the use of the refractory-to-volatile ratio as an additional formation tracer. This abundance ratio of refractory species, such as silicon, magnesium, and iron, to volatile species, such as carbon, oxygen, and nitrogen, can trace the ratio of rocky-to-icy material that was accreted onto the planet. While volatile species are accreted onto the planet from both gas and solids, refractory species reside in solids across most of the protoplanetary disk and hence trace the amount of solid accretion. The refractory-to-volatile abundance ratio of a planet can therefore constrain the regions in which it accreted solid disk material, which helps to disentangle formation scenarios that would be degenerate based on C/O alone~\citep{schneider2021IIpebblesVolRef,Danti2023pebbles,chachan2023formationmodel}.

For most gas giants, precisely measuring their refractory inventory is a hurdle we have yet to overcome, even in our own solar system: Refractory species are condensed into the deepest layers of the giant planet atmospheres, and therefore inaccessible via remote sensing. On hot Jupiter exoplanets, refractory elements can be observed as cloud species~\citep[e.g.,][]{Wakeford2015clouds,Powell2018clouds,Grant2023quartz,inglis2024clouds}, which inhibits an exact measurement of their planet-wide abundance. In contrast, ultra-hot Jupiters provide us with the opportunity to directly measure refractory elements in the planet's observable gas phase: with dayside temperatures surpassing 3000\,K~\citep{parmentier2018dissociation,Evans-Soma2025SiO, Sanchez2026}, silicate cloud species are evaporated on the dayside, allowing us a glimpse into the planet's silicon reservoir using eclipse spectroscopy. 

Indeed, the major silicon-bearing species SiO was successfully detected on the dayside of WASP-121\,b~\citep{Evans-Soma2025SiO}, using phase curve data taken with the Near-Infrared Spectrograph \citep[NIRSpec,][]{Jakobsen2022NIRSpec} on the James Webb Space Telescope~\citep[JWST,][]{Gardner2006JWST}. This makes WASP-121\,b a prime target for a more detailed characterization of its elemental inventories, by combining available dayside data from the JWST Near Infrared Imager and Slitless Spectrograph~\citep[NIRISS,][]{Doyon2023NIRISS}, NIRSpec, and the Mid-Infrared Instrument~\citep[MIRI,][]{Rieke2015MIRI}. The mid-infrared specifically contains strong Si-O stretching vibrations, opening up the possibility to more precisely measure the silicon content.

The ultra-hot Jupiter WASP-121\,b \citep[$\mathrm{mass}=\SI{1.2}{\mjup}$, $\mathrm{radius}=\SI{1.75}{\rjup}$, $\mathrm{equilibrium\,\, temperature}=\SI{2410}{\kelvin}$,][]{Bourrier2020parameters, Sing2024wasp121bagemass} is a low density planet orbiting its F-type host star on a $1.3\,$day orbit~\citep{Kokori2022exoclockII}, making it a prime target for atmospheric characterization. This has led to many atmospheric studies in the past, using JWST NIRISS~\citep[][]{Pelletier2026NIRISSWASP121b,Splinter2025WASP121bAlbedo,Frazier2026WASP121bNIRISSphaseCModel}, NIRSpec~\citep{Mikal-Evans2023WASP121bNIRSpecPhaseC,Gapp2025SiO,Evans-Soma2025SiO}, the Hubble Space Telescope~\citep[HST,][]{Evans2018WASP-121btransm,mikal-Evans2020wasp121bEclipse,Changeat2024wasp121bvariable}, Spitzer~\citep{Dravenport2025WASP121bspitzerPhaseC}, and various ground based instruments~\citep[e.g.,][]{Hoeijmakers2020HEARTS,Seidel2023sodium,Maguire2023espressoWasp121b,Smith2024IGRINSWASP121b,Pelletier2025CRIRESWASP121b,Prinoth2025TiO,Seidel2025WASP121bJet}. The dayside of WASP-121\,b shows an inverted temperature structure~\citep{mikal-Evans2020wasp121bEclipse, Evans-Soma2025SiO}, leading to a dissociation of major molecular species in the upper atmosphere~\citep{parmentier2018dissociation,Bazinet2025dissociation}. In addition, Ti-bearing species are depleted on WASP-121\,b's dayside and terminator~\citep{Hoeijmakers2020HEARTS,Smith2024IGRINSWASP121b,Prinoth2025TiO,Pelletier2026NIRISSWASP121b}. Current theories suggest that Ti is cold-trapped on the planet's night side, which is observed to be much colder than suggested by previous general circulation models~\citep{Evans-Soma2025SiO,Frazier2026WASP121bNIRISSphaseCModel}. Whether VO alone drives the planet's temperature inversion, or if other species and mechanisms are needed, is still an open question. Finally, the NIRISS phase curve revealed that the planet is enshrouded by a hydrogen-helium outflow, which spans roughly 60\% of its orbit~\citep{Allart2025Heoutflow}.

Both ground and space-based instruments constrained the planet's refractory-to-volatile ratio, which is inconsistent across the literature: On the ground,~\citet{Pelletier2025CRIRESWASP121b} find WASP-121\,b to be volatile-rich with a volatile-to-refractory abundance ratio of $1.75^{+0.57}_{-0.41}\times$ stellar, based on the measurement of H$_2$O, CO, and atomic Fe and Ni with ESPRESSO~\citep{Pepe2021ESPRESSO} and CRIRES$^+$~\citep{Follert2014crires,Dorn2023crires} at the Very Large Telescope (VLT). \citet{Smith2024IGRINSWASP121b} observed with IGRINS~\citep{Park2014IGRINS,Mace2018IGRINS} on Gemini South II and derive an enhancement in refractories, with a refractory-to-volatile abundance ratio of $3.83^{+3.62}_{-1.67}\times$ stellar. In space, \citet{Lothringer2021VolRef} analyzed the ultraviolet to infrared (IR) transmission spectrum obtained with the HST, and found a refractory-to-volatile enrichment of $5.0^{+6.0}_{-2.7}\times$ solar. Single-instrument JWST studies with NIRISS and NIRSpec find an enhancement in both volatiles and refractories, without being able to distinguish between volatile or refractory enrichment~\citet{Pelletier2026NIRISSWASP121b, Evans-Soma2025SiO}. \citet{Saha2025Titanate} combine these JWST spectra, and derive a sub-solar Si/O abundance ratio of $0.034\pm0.024$, with their free chemistry model that includes CaTiO$_3$ clouds.

To resolve these discrepancies,  we analyze the combined JWST dayside spectrum from $0.6$ to $\SI{12}{\micro\meter}$. Section~\ref{sec:data} describes the datasets used in this analysis and details the data reduction of the MIRI dataset. The joint retrieval analysis is presented in Sect.~\ref{sec:retrieval}. Section~\ref{sec:mapping} presents an eclipse mapping of WASP-121\,b\,s dayside, based on the MIRI/LRS data. The results of the analysis and their implications for WASP-121\,b's formation and evolution are discussed in Sect.~\ref{sec:discussion}. Our findings are summarized in Sect.~\ref{sec:summary}.

\section{Data}\label{sec:data}
We obtained a secondary eclipse observation with JWST MIRI Low-Resolution Spectroscopy (LRS) in \texttt{SLITLESS} mode, covering wavelengths between $5-12\,\si{\micro\meter}$. The time series observations were taken from April 10 to 11, 2024 (GO 2961, P.I.:~P.~Mollière) and span 6.8\,h. This is equal to twice the eclipse time and an additional 1\,h to allow for scheduling and instrument settling. The observations started 2\,h before eclipse ingress. We observed one exposure with 1849 integrations, using 83 groups per integration to reach approximately 80\% of the detector saturation limit. The reduction of this dataset is detailed in Sect.~\ref{subsec:EurekaReduction}.

In addition to the MIRI/LRS data, our analysis on WASP-121\,b's dayside makes use of all its JWST eclipse spectra taken to date: The $0.6-2.8\,\si{\micro\meter}$ spectrum was observed as part of the NIRISS Single Object Slitless Spectroscopy~\citep[SOSS,][]{Albert2023NIRISSSOSS} phase curve observations obtained from October 26 to 28, 2023 (GTO 1201, P.I.:~D.~Lafrenière). We utilize both orders of the eclipse spectrum from~\citet{Pelletier2026NIRISSWASP121b}, which was reduced with the \texttt{NAMELESS} pipeline~\citep{Coulombe2023NamelessWASP18b,coulombe2025LTTNAMELESS}. For wavelengths between $2.7-5.2\,\si{\micro\meter}$, a phase curve of the planet was observed using the NIRSpec/G395H grating from October 14 to 15, 2022 (GO 1729, P.I.:~T.~Evans-Soma, co-P.I.: T. Kataria). The dayside spectrum used in this analysis was derived from the data reduction of~\citet{Evans-Soma2025SiO}, who published phase-binned emission spectra\footnote{Zenodo: \url{https://doi.org/10.5281/zenodo.14728976}}. Since~\citet{Evans-Soma2025SiO} did not explicitly derive an eclipse spectrum, we instead use their emission spectra from phase bins before and after the two secondary eclipses together with their phase-curve models, to convert the spectra to a phase comparable to the other eclipse spectra. This conversion is detailed in Appendix~\ref{app:nirspec}. 

\subsection{MIRI/LRS data reduction}\label{subsec:EurekaReduction}
\begin{table*}[t]
    \centering
    \caption{Parameters used for the MIRI/LRS broadband light curve fit.}
    \label{tab:LC:params}
    \begin{tabular}{cccc}
    \hline\hline
        Parameter & Description & Prior & Posterior \\\hline
        F$_\mathrm{planet}$/F$_\mathrm{star}$ & Planet-to-star flux ratio [ppm] & $\mathcal{U}(1000, 10000)$ & $6188_{-36}^{+36}$ \\
        $p$ & Orbital period$^1$ [day] & $\mathcal{N}(1.274924762,0.000000046)$ & $1.274924790_{-0.000000037}^{+0.000000036}$ \\
        $t_0$ & Transit mid-time$^1$ [BJD] & $\mathcal{N}(58661.063783,0.000030)$& $58661.063794_{-0.000028}^{+0.000029}$ \\
        $a$ & Planet semi-major axis$^2$ [R$_\mathrm{star}$] & $\mathcal{N}(3.8131 ,0.0075)$ & $3.7927_{-0.0043}^{+0.0046}$ \\
        R$_\mathrm{star}$ & Stellar radius$^3$ [R$_\odot$] & $\mathcal{N}(1.437, 0.022)$ & $1.437_{-0.020}^{+0.020}$ \\
        $i$ & Planet inclination$^3$ [$^\circ$] & $\mathcal{N}(88.09, 0.15)$& $88.33_{-0.12}^{+0.13}$ \\
        $A_\mathrm{cos}$ & Sinusoid Amplitude & $\mathcal{N}(0.5, 0.1)$ & $0.463_{-0.074}^{+0.093}$ \\
        $c_0$ & Baseline coefficient & $\mathcal{N}( 1, 0.001 )$& $0.996315_{-0.000073}^{+0.000037}$ \\
        $c_1$ & Baseline coefficient & $\mathcal{N}(0, 0.01 )$& $0.0004_{-0.0008}^{+0.0013}$ \\
        $r_0$ & Ramp coefficient & $\mathcal{N}(0, 0.001)$& $0.00136_{-0.00021}^{+0.00038}$ \\
        $r_1$ & Ramp coefficient &$\mathcal{U}(6, 200 )$ & $27.3_{-5.6}^{+7.4}$ \\
        $\sigma_\mathrm{mult}$ & Uncertainty inflation factor & $\mathcal{U}(0.5, 4)$ & $1.452_{-0.024}^{+0.025}$ \\ \hline
        R$_\mathrm{planet}$/R$_\mathrm{star}$ & Ratio of planet-to-stellar radius$^4$ & fixed & $0.123180$ \\
        $e$ &Planet orbit eccentricity$^3$ & fixed & 0 \\
        $\omega$& argument of periastron$^3$ [$^\circ$] & fixed & 10 \\ \hline
    \end{tabular}
    \tablefoot{The fixed parameters and narrow prior distributions are based on the values found by: $^1$\citet{Kokori2022exoclockII}, $^2$\citet{Bourrier2020parameters}, $^3$\citet{Mikal-Evans2023WASP121bNIRSpecPhaseC}, $^4$\citet{Gapp2025SiO}. Normal prior distributions are reported as $\mathcal{N}$(center, width), and $\mathcal{U}$ marks uniform priors within the given bounds. The posteriors are reported with their $1\,\sigma$ uncertainty intervals.}
\end{table*}
\begin{figure}[t]
	\centering
\includegraphics[width=0.5\textwidth]{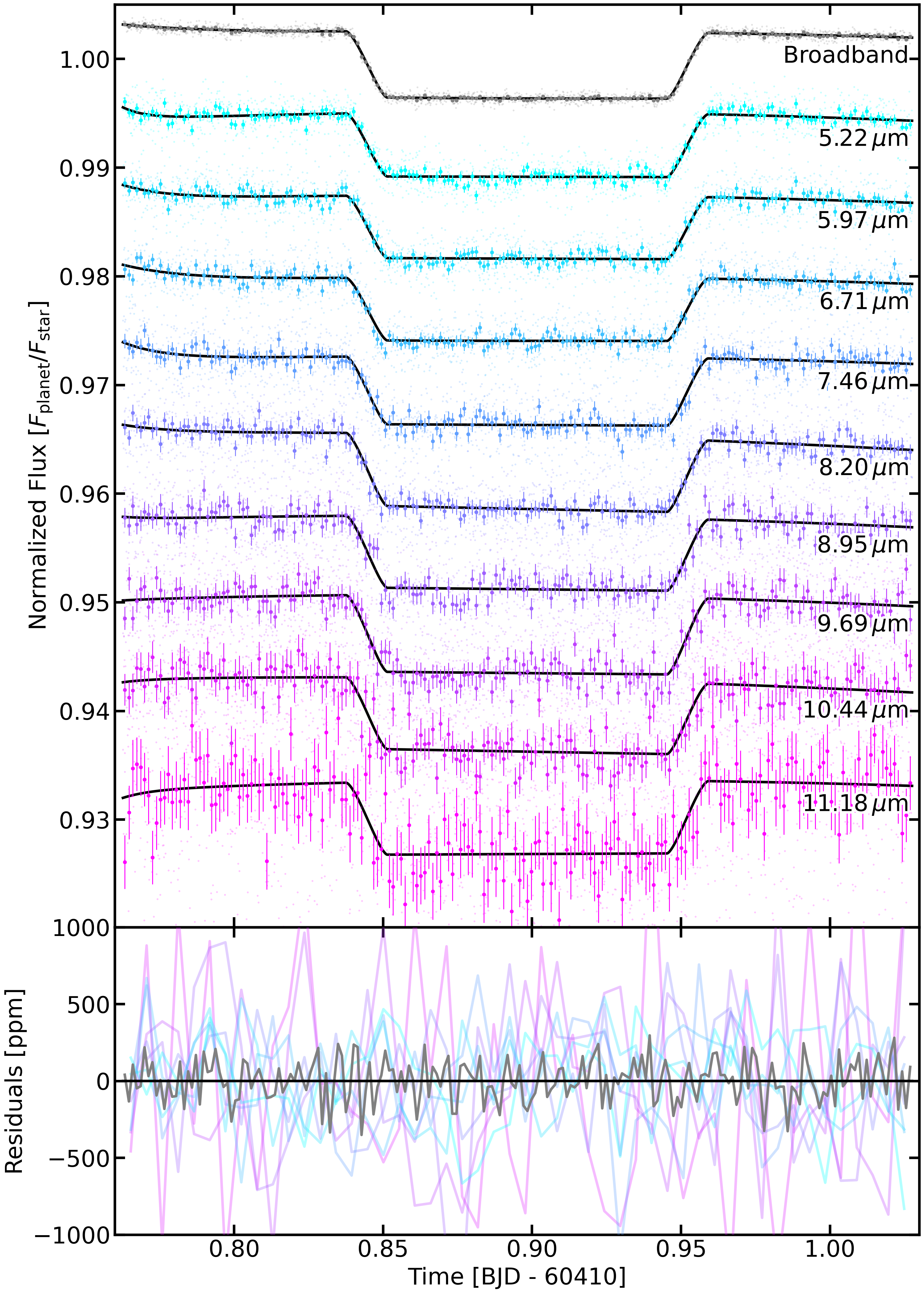}
	\caption{Light curves and residuals of the MIRI/LRS secondary eclipse of WASP-121\,b. Top: Broadband and spectroscopic light curves with the center wavelength indicated on the right. We show every fifth spectroscopic bin, starting with the second bin. Binned data and uncertainties are shown in color, and the best-fit models are shown in black. Transparent dots indicate the unbinned data. Bottom: Residuals for light curves at wavelengths shorter than $\SI{10.5}{\micro\meter}$. The data are binned to 50 bins in the residual panel and to 200 bins in the light curve panel.}
	\label{fig:LC}
\end{figure}
We reduced the MIRI/LRS data with the widely used open source pipeline \texttt{Eureka!}~\citep[v1.0,][]{Bell2022Eureka}. \texttt{Eureka!} starts from the uncalibrated \texttt{.fits} files and is divided into six stages: Stage 1 and 2 wrap the \texttt{JWST} pipeline to apply basic calibrations to the observations and conduct the up-the-ramp fitting. Stage 3 extracts the spectrum, and Stage 4 is used to bin the extracted spectra to a defined spectral resolution and for clipping outliers. Stage 5 is a light curve fitting module, and Stage 6 is used to create plots. We used all stages of the pipeline. 

To calibrate the data, we acquired calibration reference files from the Calibration References Data System (CRDS) of the JWST pipeline~\citep{bushouse_2026jwstpipeline}. The CRDS can associate a JWST dataset with the best reference files, based on specified context maps. The calibration files used in this analysis are based on the context map \texttt{pmap 1293}. We exclude the first and last group for up-the-ramp fitting, as these are typically affected by readout effects. We apply a jump rejection threshold of 5$\,\sigma$ during the ramp fitting and use the default non-linearity correction of the JWST pipeline. We skip the photom step of the pipeline to avoid introducing extra noise to the extracted data. The data are extracted in a detector region covering pixel rows 22-393 in the spatial direction, and pixel columns 15-57 in the spectral direction, using the optimal extraction algorithm~\citep{Horne1986optimal}. The extraction window spanned a half-width of 3 pixels from the trace center, including the core of the point spread function (PSF) of the stellar emission. The background was calculated per spectral column, using the median flux outside a background half-width of 12 pixels to avoid including the side lobes of the stellar trace. We tested different combinations of extraction aperture half-widths (2 to 8 pixels in steps of 1) and background half-widths (9 to 18 pixels in steps of 1), and chose the combination that resulted in the smallest root mean squared (RMS) residual noise in the broadband light curve fit. On the detector images, we rejected 10$\,\sigma$ outlier pixels during the optimal extraction of the spectral trace, and 6$\,\sigma$ outlier pixels during the background subtraction. Furthermore, we used \texttt{Eureka}'s full frame outlier detection to iteratively reject 5$\,\sigma$ outlier pixels along the time axis, using 5 iteration steps.

We extracted broadband and spectroscopic light curves between $\SI{5}{\micro\meter}$ and $\SI{12}{\micro\meter}$. The spectroscopic light curves were extracted in 47 bins of $\SI{0.15}{\micro\meter}$ width, which is wider than the native resolution of MIRI/LRS at $\SI{5}{\micro\meter}$. For all spectroscopic light curves, we masked $5\,\sigma$ outlier integrations in 5 iterations of a 20 data point box-car filter in time. 

\subsection{MIRI/LRS light curve fitting}\label{subsec:EurekaFitting}
We modeled the eclipse light curve using Stage 5 of \texttt{Eureka!} (version 1.2.1). Our model consists of an astrophysical model and a systematics model, which are applied multiplicatively. The astrophysical model consists of a \texttt{batman}~\citep{Kreidberg2015batman} eclipse model and a single sinusoid that describes the planet's phase variations over the course of the observations. We fitted for the amplitude of the sinusoid $A_\mathrm{cos}$, but fixed the phase offset such that the sinusoid has its maximum at phase 0.5, during the eclipse mid-time. We also tested the need for a phase offset of the sinusoid, by additionally fitting for the amplitude of a second sinusoid peaking at quadrature phase, but find that the inclusion of the second sinusoid is disfavored (difference in broadband evidence $\Delta\ln{(Z)}=-2.0$, see Sect.~\ref{subsec:retrievalresults} for an explanation). This finding is consistent with our eclipse mapping, in Section~\ref{sec:mapping}, which shows that the hotspot offset measured with MIRI/LRS is consistent with zero within $2\,\sigma$. In contrast, the use of a single sinusoid in the astrophysical model is preferred by $\Delta\ln{(Z)}=8.7$ over a model without a sinusoid.

\begin{figure}[t]
	\centering
\includegraphics[width=0.5\textwidth]{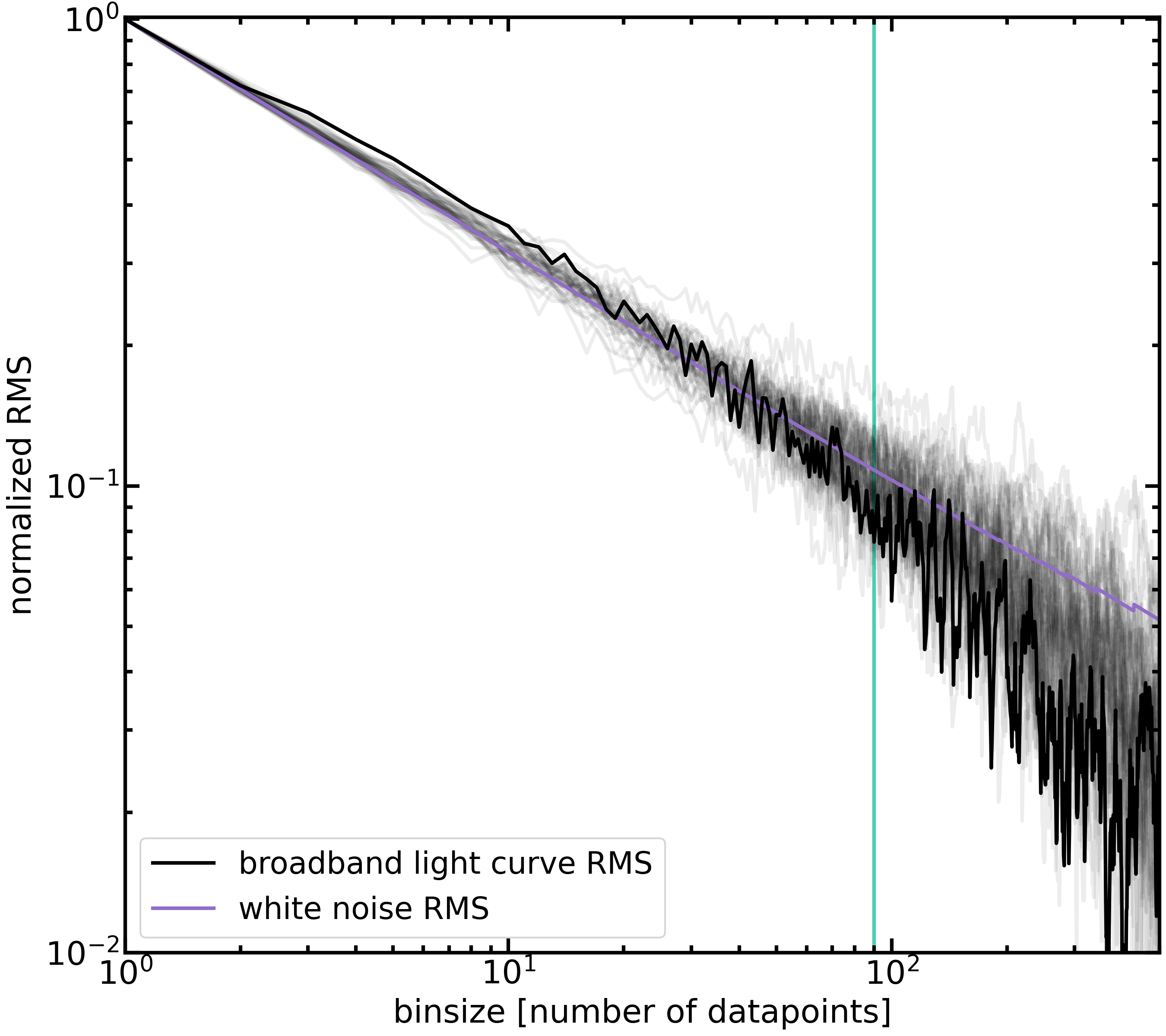}
	\caption{Normalized root-mean square (RMS) noise as a function of the number of binned integrations. The opaque solid line corresponds to the broadband light curve, the transparent lines correspond to individual spectroscopic light curves. The green vertical line marks a binsize of 90, which approximately corresponds to the length of eclipse ingress and egress.}
	\label{fig:notAllan}
\end{figure}

The systematics model consists of a linear trend of the flux, $F$, with time, $t$, and a single exponential ramp of the flux with time:
\begin{equation}\label{eq:systematicsFit}
F_\mathrm{sys}(t) = (c_0 + c_1 t )  (1 +  r_0 \exp(- r_1  t)) 
\end{equation}
Where we freely fitted for the coefficients $c_1$, $c_0$, $r_0$, and $r_1$ with the prior distributions described in Table~\ref{tab:LC:params}. To account for a possible underestimation of the light curve uncertainties by the \texttt{Eureka!} extraction, we additionally fitted for an uncertainty inflation parameter, $\sigma_\mathrm{mult}$. We further tested decorrelating the light curve flux against the spectral trace width and spatial position on the detector. The additional parameters for this decorrelation are disfavored by $\Delta\ln({Z})=-8.5$ and $\Delta\ln({Z})=-2.5$, respectively.

We excludeed the first 135 integrations from the light curve fitting, as these show the typical flux-settling behavior of MIRI/LRS~\citep[e.g.,][]{Bouwman2023miri,Bell2023wasp43_firstlook}. The orbital parameters $p$, $t_0$, $a$, and $i$ of WASP-121\,b (see Table~\ref{tab:LC:params}) were free parameters with normal priors around literature values for the broadband light curve fit. We further fixed the orbit eccentricity to 0 and the argument of periastron to 10. The orbital parameters of the spectroscopic light curve fits were then fixed to the results of the broadband light curve. We used the \texttt{dynesty} sampler~\citep{Speagle2020dynesty} with 1024 live points and a convergence tolerance of $0.01$ for the broadband light curve fits. The \texttt{dynesty} sampler was configured to decompose the prior cube with multiple ellipsoids and to use a random walk for sampling the next live point. For the spectroscopic fits, we found it faster to sample the posteriors using the Markov Chain Monte Carlo (MCMC) algorithm implemented by the \texttt{emcee} python package~\citep{Foreman-Mackey2013emcee}. For this, we initiated the sampling based on an initial least-squares fit. We used 200 walkers with chains of 30000 steps and removed 5000 samples as burn-in. The number of steps is longer than 130 times the mean autocorrelation time for each spectroscopic fit, ensuring convergence of the sampling.

Broadband light curve and characteristic spectroscopic light curves are shown in Fig.~\ref{fig:LC}. The broadband light curve is well-fit and shows only minor contributions of correlated noise. We achieve an accuracy of $355\,$ppm per $\SI{13.2}{\second}$ integration, approximately $1.4$ times the photon noise. For spectroscopic light curve fits, the RMS deviations of the residuals range from $1300\,$ppm at $\SI{5}{\micro\meter}$ to $2600\,$ppm at $\SI{10.5}{\micro\meter}$ and $11000\,$ppm at $\SI{12}{\micro\meter}$, corresponding to $1.2$, $1.3$, and $1.5$ times the photon noise. Similar to~\citet{Valentine2024WASP17bRamp}, we observe a dampening of the MIRI/LRS ramp behavior between $\SI{7.5}{\micro\meter}$ and $\SI{9.5}{\micro\meter}$. As shown in Figs.~\ref{fig:LC} and~\ref{fig:notAllan}, minor correlated noise is seen for some spectroscopic bins. In Fig.~\ref{fig:notAllan}, this is indicated by the RMS being above the white noise prediction for large integration bin sizes along the time axis. At short wavelengths, this might be caused by our simplistic non-linearity treatment~\citep[see e.g.,][]{Grant2023quartz}. Stronger correlated noise is only seen in the $\SI{5.22}{\micro\meter}$ bin, for which we account using the method described by~\citet{Pont2006beta} and~\citet{Cubillos2017beta}: 
We apply an additional uncertainty inflation by the fraction between fit RMS and expected white noise RMS at a bin size of 90 integrations~\citep[see Fig.~\ref{fig:notAllan} and e.g.,][for application examples]{Cubillos2017beta, Kahle2025hd86226c}. This bin size approximately corresponds to the duration of ingress and egress, and leads to an uncertainty inflation by a factor of $1.5$.

\subsection{Stellar abundances}\label{subsec:star}
\begin{table}[]
    \centering
    \caption{Adapted and derived stellar parameters}
    \label{tab:star}
    \begin{tabular}{lcc}
    \hline\hline
        Parameter & Description & Value \\ \hline
        $A(\mathrm{O})_\mathrm{NLTE}$ & Oxygen abundance [dex] & $9.01 \pm 0.01$ \\
        $A(\mathrm{C})_\mathrm{LTE}$ & Carbon abundance [dex] & $8.69\pm0.04$ \\
        $A(\mathrm{Si})_\mathrm{NLTE}$ & Silicon abundance [dex] & $7.71 \pm0.07$ \\
        $A(\mathrm{Fe})_\mathrm{NLTE}$ & Iron abundance [dex] & $7.68\pm0.10$ \\        $A(\mathrm{Mg})_\mathrm{NLTE}$ & Magnesium abundance [dex] & $7.58 \pm 0.04$ \\   $A(\mathrm{S})_\mathrm{LTE}$ & Sulfur abundance [dex] & $7.46 \pm 0.11$ \\ \hline
        $T_\mathrm{eff}$ & Effective temperature$^1$ [K] & $6459\pm 140$ \\ 
        $\log g$& Surface gravity$^2$ [log$_{10}$(cm/s$^2$)] &$4.251\pm0.003$ \\
        $[\mathrm{M}/\mathrm{H}]$ & Metallicity$^2$ [dex] & $0.17\pm0.05$ \\
        R$_\mathrm{star}$ & Radius$^2$ [R$_\odot$]& $1.461\pm0.005$\\ \hline 
    \end{tabular}
    \tablefoot{Posterior distributions for stellar abundances derived in Sect.~\ref{subsec:star} are reported with their $1\,\sigma$ uncertainties. The lower section provides stellar parameters derived by $^1$\citet{delrezWASP121HotJupiter2016} and $^2$\citet{Sing2024wasp121bagemass}. The stellar metallicity was only used during eclipse mapping in Sect.~\ref{sec:mapping}.}
\end{table}
The stellar abundances of WASP-121 were derived using data from the Echelle SPectrograph for Rocky Exoplanets and Stable Spectroscopic Observations~\citep[ESPRESSO,][]{Pepe2021ESPRESSO} at the VLT in Chile. To obtain a single spectrum with a high signal-to-noise ratio, we averaged a total of 98 spectra from three observing nights on April 4, 7, and 8, 2021 (program 106.21QM, P.I.:~Hoeijmakers). These observations were selected to be within four nights of one another, ensuring that telluric contamination affected nearly the same wavelength channels in all spectra.

We adopted stellar parameters from \citet{delrezWASP121HotJupiter2016}, specifically, $\teff = 6459 \pm 140$ K, $\logg = 4.2 \pm 0.2$ dex, $\vmic = 1.5 \pm 0.1$ kms$^{-1}$ and $\vsini = 13.5 \pm 0.7$ kms$^{-1}$, which are completely compatible with the recently derived ones by \citet{Evans-Soma2025SiO}. We use the \texttt{Turbospectrum NLTE} code, and its wrapper \texttt{TSFitPy} \citep{Gerber2023, Storm2023} to derive 1D LTE and NLTE (wherever possible) abundances of Fe, C, O, Mg, S, and Si. We note that, unlike \citet{Evans-Soma2025SiO}, we used direct NLTE synthesis during the fitting. We adopted the standard composition MARCS models \citep{Gustafsson2008}, the Gaia-ESO line list \citep{Heiter2021} with new atomic data for C, N, O, Si, Mg, as described in \citet{Magg2022}, and used VALD for its gaps \citep{Ryabchikova2015}. We also adopt the following NLTE model atoms: Fe \citep{Bergemann2012a, Semenova2020}, O \citep{Bergemann2021}, Mg \citep{Bergemann2017} and Si \citep{Bergemann2013, Magg2022}. Based on the fitting, our best-fit abundances are reported in Table~\ref{tab:star}.

\section{Retrieval analysis}\label{sec:retrieval}
\begin{figure*}[t]
	\centering
\includegraphics[width=0.999\textwidth]{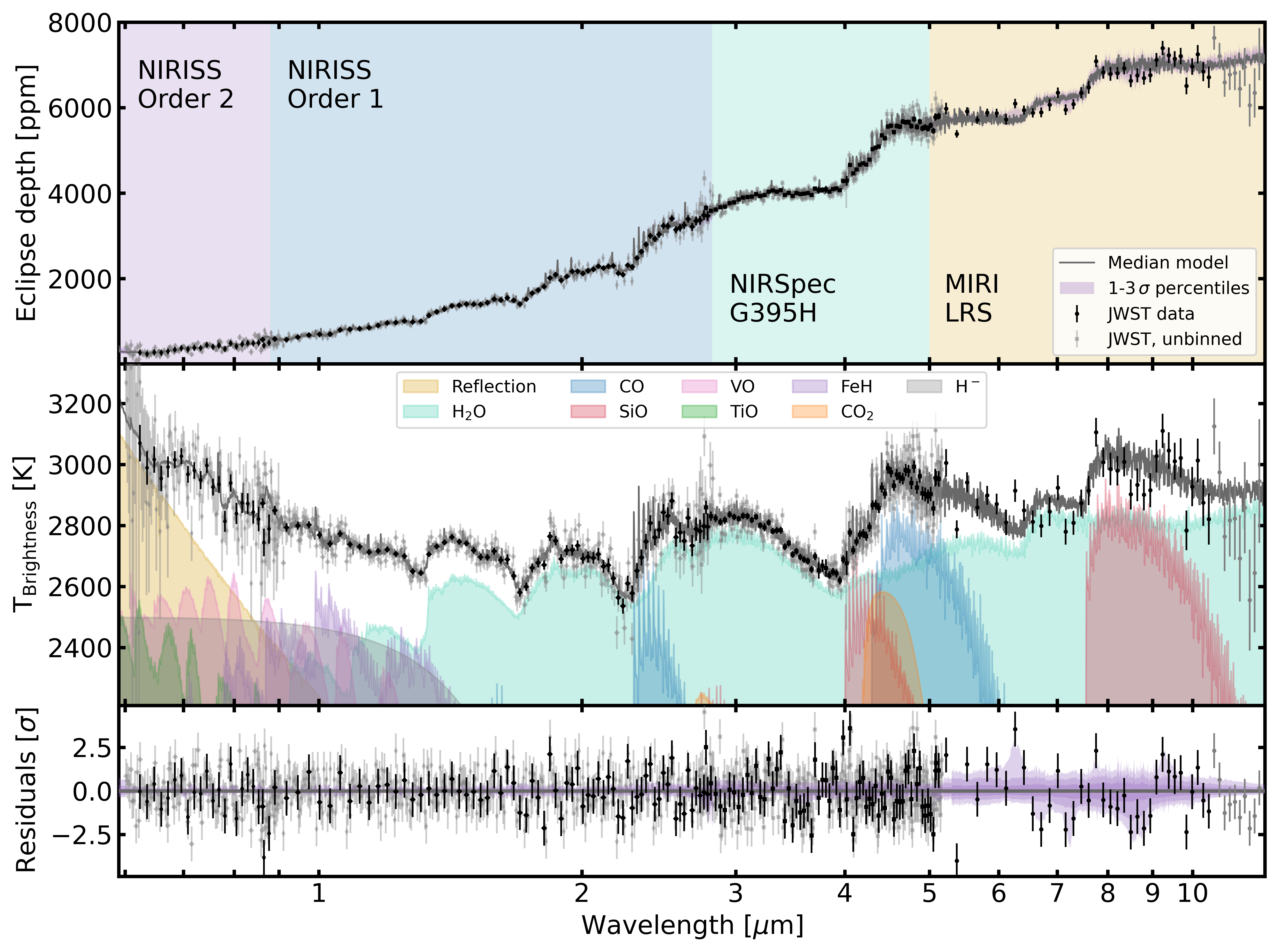}
	\caption{Panchromatic spectrum of WASP-121\,b and opacity profiles from the most relevant species. NIRISS and NIRSpec data are shown as transparent, grey datapoints, with black datapoints binning these by a factor of five for visualization. The unbinned MIRI/LRS spectrum is shown in black, with grey datapoints beyond $\SI{10.5}{\micro\meter}$ being excluded from the retrievals. The median model from our 1D cloud-free dayside retrievals is shown in a darker grey and purple shaded regions in the top and bottom panel mark 1, 2 and 3$\,\sigma$ percentiles of retrieved spectra.
    \textit{Top}: Measured eclipse depths with background colors indicating the wavelength coverage of the instruments. \textit{Middle}: Brightness temperatures and opacity profiles of species as indicated in the panel. \textit{Bottom}: Residuals to the median model.}
	\label{fig:jointspectrum}
\end{figure*}

To constrain WASP-121\,b's formation based on its elemental inventories, we performed atmospheric retrievals on the joint JWST dayside spectrum. For this, we used the radiative transfer code petitRADTRANS~\citep[version 3.3.3,][]{Molliere2019pRT,Blain2024pRT3} alongside its retrieval package~\citep{Nasedkin2024_retrieval} to simulate the dayside atmosphere of WASP-121\,b and derive corresponding atmospheric parameters. We included the full dayside spectra from NIRISS and NIRSpec, and masked MIRI/LRS datapoints with wavelengths longer than $\SI{10.5}{\micro\meter}$ (see Fig.~\ref{fig:jointspectrum}). At these long wavelengths, the eclipse depths show a downward trend, which could be caused by the ``shadowed region’’ effect, commonly seen in MIRI data~\citep{Bell2023wasp43_firstlook}. Our models generally do not reproduce this flux decline at long wavelengths, so we masked the data to avoid biasing our retrieval results.

\subsection{Nominal retrieval setup}\label{subsec:ret:nominal}
In this Section, we describe the nominal retrieval setup for our analysis. We tested different variations of this setup, which are described in Sect.~\ref{subsec:retrieval:variations}. An overview of all parameters and their prior distributions is shown in Table~\ref{tab:retrievalprior}. To simulate the dayside atmosphere, we used 100 pressure layers, uniformly spaced on a logarithmic scale between $10^2$ and $10^{-8}$\,bar. We used the self-scattering implementation of~\citet{Molliere2020pRTscatter} to account for scattering processes inside WASP-121\,b's atmosphere. Since the joint dayside spectrum extends to optical wavelengths, it is further needed to account for the reflection of WASP-121's star light~\citep{Pelletier2026NIRISSWASP121b}. 
For this, we adopted the analytical scattering prescription used by~\citet{Pelletier2026NIRISSWASP121b}, and expanded it by an additional wavelength scaling:
\begin{equation}\label{eq:scatter}
\mathrm{eclipse \, depth} = \frac{F_\mathrm{planet}}{F_\mathrm{star}} \left( \frac{R_\mathrm{planet}}{R_\mathrm{star}}\right)^2+ A_g  \left( \frac{\lambda}{\SI{0.6}{\micro\meter}} \right)^\delta \left(\frac{R_\mathrm{planet}}{a}\right)^2
\end{equation}
In this equation, $a$ is the star-planet distance, and $F$ and $R$ are the flux and radius, respectively. We assumed the planet radius is constant across wavelengths, as we do not have access to JWST transmission spectra at NIRISS and MIRI wavelengths. The geometric albedo, $A_g$, is modified by a power law with coefficient $\delta$. This allows for a decline of the albedo with wavelength $\lambda$, such that the scattered-light contribution is allowed to be negligible at mid-IR wavelengths. Our base model does not include self-consistent reflection~\citep[see e.g.,][]{Alei2022LIFEpRTreflection} of WASP-121's star light, as we found that it is insufficient to explain the flux increase seen at the shortest NIRISS wavelengths. 

The panchromatic eclipse spectrum is especially sensitive to the temperature structure of the planet. Due to this, we adopted a similar approach as~\citet{Pelletier2026NIRISSWASP121b} and freely retrieved the temperature-pressure structure: We retrieved the temperatures at 11 fixed pressures between $10^0$ and $10^{-6}$, which are equally spaced on the logarithmic scale by $0.6\,$dex. Preliminary retrievals did not constrain temperatures below or above this pressure region, so we set them to be isothermal and equal to the bordering temperatures at $10^0$ or $10^{-6}$\,bar, respectively. Based on the retrieved temperatures, we calculated the full temperature-pressure structure with a cubic interpolation, using the \texttt{CubicSpline} function of \texttt{scipy.interpolate}~\citep{Virtanen2020SciPy-NMeth}. As this interpolation can introduce an unphysically alternating temperature-pressure profile, we further penalized high second derivatives of the temperature structure according to the formalism introduced by~\citet{Line2015dwarfs}. We fixed the smoothing weight $\gamma$ to $10^{-2} = \partial^2(\Delta(\log T)) / \partial^2(0.6\,\mathrm{K})$, as freely fitting this parameter led to overly smooth temperature structures. We tested smoothing weights between $10^{1}$ and $10^{-5}$ and found that $10^{-2}$ simultaneously inhibits interpolation overshoots while still allowing sufficient flexibility to properly fit the temperature structure.

The temperatures in the upper atmosphere of WASP-121\,b are high enough to dissociate many major molecular species~\citep{parmentier2018dissociation}. Therefore, models assuming constant molecular abundances will bias the results, even when the dissociation of H$_2$ and H$_2$O is considered (see Appendix~\ref{app:freeAbund}). This makes a coupling between the chemistry and the temperature-pressure profile crucial for a correct inference of the elemental inventories. For this, we used the \texttt{easyCHEM} equilibrium chemistry code~\citep{Lei2025easychem}, which computes abundances of all required species according to Gibbs free energy minimization at each layer of the atmosphere given its pressure and temperature. We used the default \texttt{easyCHEM} species selection and thermodynamic data, but only included the condensed and liquid forms of MgSiO$_3$, SiO$_2$, MgO, VO, TiO$_2$, and Fe for the thermal stability calculations. This selection includes at least one condensate species for each relevant refractory tracer. This is needed to mimic the effect of cloud condensation on molecular inventories if the retrieval probes temperatures below approximately $2000\,$K. We excluded the other condensate species to ensure computational stability. Therefore, we included 92 species in total for computing the Gibbs free energy minimization. 

\begin{table}[]
    \centering
    \caption{Parameters and prior distributions for the retrievals with \texttt{petitRADTRANS}.}
    \label{tab:retrievalprior}
    \begin{tabular}{lr}
    \hline\hline
        Parameter, Description & Prior \\ \hline
         Nominal model & \\ \hline
        $m_\mathrm{planet}$, Planet mass [M$_\mathrm{Jupiter}$]& $\mathcal{N}(1.157, 0.070)$ \\
        $R_\mathrm{planet}$, Planet radius [R$_\mathrm{Jupiter}$]& $\mathcal{N}(1.753, 0.036)$ \\
        $a$, Semi-major axis [au]& $\mathcal{N}(0.026, 0.002)$ \\
        $T_{i=1,..,11}$, Temperature $i$ [K] & $\mathcal{U}(300, 10000)$\\
        C/O, Carbon-to-oxygen ratio & $\mathcal{U}(0,2)$\\
        $[\mathrm{Vol}/\mathrm{H}]$, Volatile enrichment & $\mathcal{U}(-3,2)$\\
        $[\mathrm{Ref}/\mathrm{H}]$, Refractory enrichment & $\mathcal{U}(-3,2)$\\
        $[\mathrm{Ti}/\mathrm{H}]$, Titanium enrichment & $\mathcal{U}(-3,2)$\\
        $A_g$, Geometric albedo & $\mathcal{U}(0,1)$\\
        $\delta$, $A_g$ power-law coefficient& $\mathcal{U}(-6,1)$\\
        Uncertainties NIRISS:& \\
        $A_\mathrm{GP}$, Global kernel amplitude [ppm]&$\mathcal{U}(0,1000)$\\
        $L_\mathrm{GP}$, Global kernel length [$\mu$m]& $\mathcal{U}(0,0.2)$ \\
        $A_\mathrm{locGP}$, Local kernel amplitude [ppm]&$\mathcal{U}(0,7000)$ \\
        $L_\mathrm{locGP}$, Local kernel length [$\mu$m]&$\mathcal{U}(0,1.12)$ \\
        $\mu_\mathrm{locGP}$, Local kernel position [$\mu$m]&$\mathcal{U}(0.60,2.35)$ \\
        Uncertainties NIRSpec:&\\
         10$^b$, Uncertainty inflation  & (a) \\
        $A_\mathrm{locGP}$, Local kernel amplitude [ppm]&$\mathcal{U}(0,7000)$ \\
        $L_\mathrm{locGP}$, Local kernel length [$\mu$m]&$\mathcal{U}(0,1.22)$ \\
        $\mu_\mathrm{locGP}$, Local kernel position [$\mu$m]&$\mathcal{U}(2.73, 5.17)$ \\  
        Uncertainties MIRI/LRS: & \\
        10$^b$, Uncertainty inflation  & (a) \\
        $A_\mathrm{locGP}$, Local kernel amplitude [ppm]&$\mathcal{U}(0,7000)$ \\
        $L_\mathrm{locGP}$, Local kernel length [$\mu$m]&$\mathcal{U}(0,2.75)$ \\
        $\mu_\mathrm{locGP}$, Local kernel position [$\mu$m]&$\mathcal{U}(5.0, 10.5)$ \\
        \hline
         Model Variations&\\\hline
        $A_\mathrm{HS}$, hotspot area factor & $\mathcal{U}(0.75, 1)$\\
        $f_\mathrm{NIRSpec}$, detector offset [ppm] & $\mathcal{U}(-2000, 2000)$\\
        $f_\mathrm{MIRI}$, detector offset [ppm] & $\mathcal{U}(-2000, 2000)$\\
        $[\mathrm{V}/\mathrm{H}]$, Vanadium enrichment & $\mathcal{U}(-3,2)$\\
        
       Guillot temperature profile: & \\
        $T_\mathrm{eq}$, Equilibrium temperature [K] & $\mathcal{U}(100,10000)$\\
        $\log_{10}(\kappa_\mathrm{IR}$), IR mean opacity [log$_{10}$(cm$^2$/s)]& $\mathcal{U}(-5,7)$ \\
        $\log_{10}(\gamma)$, visual-to-infrared opacity ratio& $\mathcal{U}(-2,5)$ \\ \hline
    \end{tabular}
    \tablefoot{All atmospheric enrichments were retrieved relative to the solar abundances of~\citet{Asplund2009}. (a) Uncertainty inflations were retrieved with a prior of $\mathcal{U}(0.01\cdot\min(\sigma^2), 10\cdot\max(\sigma^2))$, where $\sigma$ are the uncertainties of the corresponding dataset. Inflated uncertainties are computed according to $\sigma_\mathrm{inf} = (\sigma^2 + 10^b)^{1/2}$, following~\citet{Line2015dwarfs}.}
\end{table}

Apart from the species selection and thermochemical data, \texttt{easychem} requires an input temperature profile and abundances for each involved element. We computed the input element reservoirs by starting from a solar composition~\citep{Asplund2009} and retrieved the atmospheric element enrichments. Instead of retrieving each element enrichment individually, our free parameters are the volatile enrichment [Ref/H], the C/O abundance ratio, the refractory enrichment [Ref/H], and, separated from this, the titanium enrichment [Ti/H]. While titanium is a refractory element, past studies showed that it is strongly depleted in WASP-121\,b's atmosphere, possibly due to nightside cold-trapping~\citep{Hoeijmakers2024tio,Prinoth2025TiO,Pelletier2026NIRISSWASP121b}. Therefore, we decouple the titanium enrichment from the other refractory elements to avoid biasing our refractory reservoir. Based on the retrieved parameters, the solar oxygen composition is first scaled to match the C/O ratio using the corresponding easychem function. Oxygen, carbon, and nitrogen are then scaled according to the retrieved [Vol/H], while titanium is scaled according to [Ti/H]. Finally, the remaining elemental abundances, including Si, V, and Fe but excluding H and He, are scaled according to [Ref/H]. The whole composition is then normalized before computing the Gibbs free energy minimization.

For computing the spectra, we included the line opacities of H$_2$O~\citep{Polyansky2018H2O}, CO~\citep{Rothman2010CO}, SiO~\citep{Yurchenko2022SiO}, VO~\citep{McKemmish2016VO}, TiO~\citep{McKemmish2019TiO}, Na~\citep{Piskunov1995database,Allard2019Na}, K~\citep{Piskunov1995database,Schweitzer1996broadening}, FeH~\citep{Wende2010FeH}, CO$_2$~\citep{Yurchenko2020CO2}, CH$_4$~\citep{Hargreaves2020CH4}, HCN~\citep{Barber2014HCN}, SiO$_2$~\citep{Owens2020SiO2}, and C$_2$H$_2$~\citep{Chubb2020C2H2} based on correlated-k tables with a resolution of $R=1000$. For MIRI/LRS, we computed the spectra with a lower resolution of $R=200$, to reduce the retrieval runtime. We also included the continuum emission from H$^{-}$~\citep[free-free and bound-free,][]{Gray2008stars}, and for the collision-induced absorptions between H$_2$ atoms~\citep{Borysow2001H2H2} and between H$_2$ and Helium~\citep{Borysow1988H2HeI,Borysow1989H2HeII,Borysow1989H2HeIII}. For self-consistently computing scattering within the planet's atmosphere, we included Rayleigh scattering cross-sections for H$_2$~\citep{Dalgarno1962H2scatter} and He~\citep{Chan1965HeScatter}. 

For computing the eclipse depths with Eq.~\ref{eq:scatter}, we selected the stellar model from the~\texttt{PHOENIX} grids~\citep{Husser2013phoenix} that best matches the stellar parameters in Table~\ref{tab:star}. We further assume a fixed stellar radius of $R_\mathrm{star}=\SI{1.46}{\rsun}$. We do not apply additional wavelength shifts or rotational broadening to the stellar spectrum, as the analyzed low-resolution JWST spectra are not sensitive to the effects of such modifications.

We use the retrieval package of petitRADTRANS~\citep{Nasedkin2024_retrieval} to estimate best-fit parameters, uncertainties, and model evidences using the nested sampling algorithm implemented in \texttt{PyMultiNest}~\citep{Buchner2014PyMultiNest,Feroz2009nested}. For most retrievals, we used 1500 live points and a constant sampling efficiency of $5\%$. The constant-efficiency mode of \texttt{PyMultiNest} adjusts the sampling volume of each iteration such that it matches the user-defined sampling efficiency. While this only slightly affects posterior distributions of the parameters \citep[possibly leading to uncertainty underestimations,][]{Chubb2022consteff}, the decoupling of the sampling from the prior volume biases the evidence estimates~\citep{Feroz2019INS}. As a consequence, we recomputed retrievals used for species detection with a flexible target sampling efficiency of $0.3$. Each of these retrievals costs approximately 200000\,CPUh on the \texttt{viper} and \texttt{vera} high-performance computation clusters, which is over ten times more than retrievals with a constant sampling efficiency. This higher cost is the reason why we do not compute all retrievals with a flexible sampling efficiency.

\subsection{Uncertainty treatment}\label{subsec:retireval:errors}
We accounted for the possibility of underestimated uncertainties in the dayside spectra~\citep[e.g.,][]{Line2015dwarfs,Molliere2025PSO}, and wavelength-correlated noise due to detector systematics or unknown spectral features~\citep{Rotman2025GPs}. For the latter, we implemented the Gaussian process (GP) prescription of~\citet{Rotman2025GPs} in \texttt{petitRADTRANS}. For each dataset, we allowed the inclusion of square-exponential kernels for global (instrument-wide) and local Gaussian correlations. These kernels alter the covariance matrix used for likelihood computation during the retrievals and can induce off-diagonal elements that correlate different datapoints with each other. While global kernels induce a correlation of neighboring datapoints with a given amplitude, $A_\mathrm{GP}$, and length scale, $L_\mathrm{GP}$, local kernels induce a correlation with a given amplitude, $A_\mathrm{locGP}$, and length scale $L_\mathrm{locGP}$ around a specific wavelength $\mu_\mathrm{locGP}$. Both global and local kernels were constrained to affect only the covariance matrices of the corresponding instruments, as we assume model-data misfits originate from instrument systematics. This implementation could potentially miss broad-wavelength spectral features that extend over multiple datasets. However, we do not see evidence for this type of feature in the residuals of WASP-121\,b's dayside spectrum. 

Initial tests showed that the length scales of global kernels converged to zero for the NIRSpec and MIRI data, suggesting that these do not exhibit instrument-wide correlated noise. For these datasets, we fitted for an uncertainty-inflation factor~\citep{Line2015dwarfs,Molliere2025PSO}, which effectively acts as uncorrelated white noise. For NIRISS, the global kernel converged to length scales of $\SI{0.03}{\micro\meter}$, characteristic of the size of the PSF~\citep{Rotman2025GPs}. For this instrument, we did not fit for the white noise uncertainty inflation, but instead for amplitude, $A_\mathrm{GP, NIRISS}$, and length scale, $L_\mathrm{GP, NIRISS}$, of a global GP kernel. Inspecting the best fits of early retrieval tests, we observed that each dataset possibly shows one model-data mismatch at specific wavelengths. Due to this, we allowed for one local GP kernel per instrument, described by amplitude $A_\mathrm{locGP}$, length scale $L_\mathrm{locGP}$, and wavelength position $\lambda_\mathrm{locGP}$.

\subsection{Variations of the retrieval setup}\label{subsec:retrieval:variations}
\begin{figure}
	\centering
\includegraphics[width=0.499\textwidth]{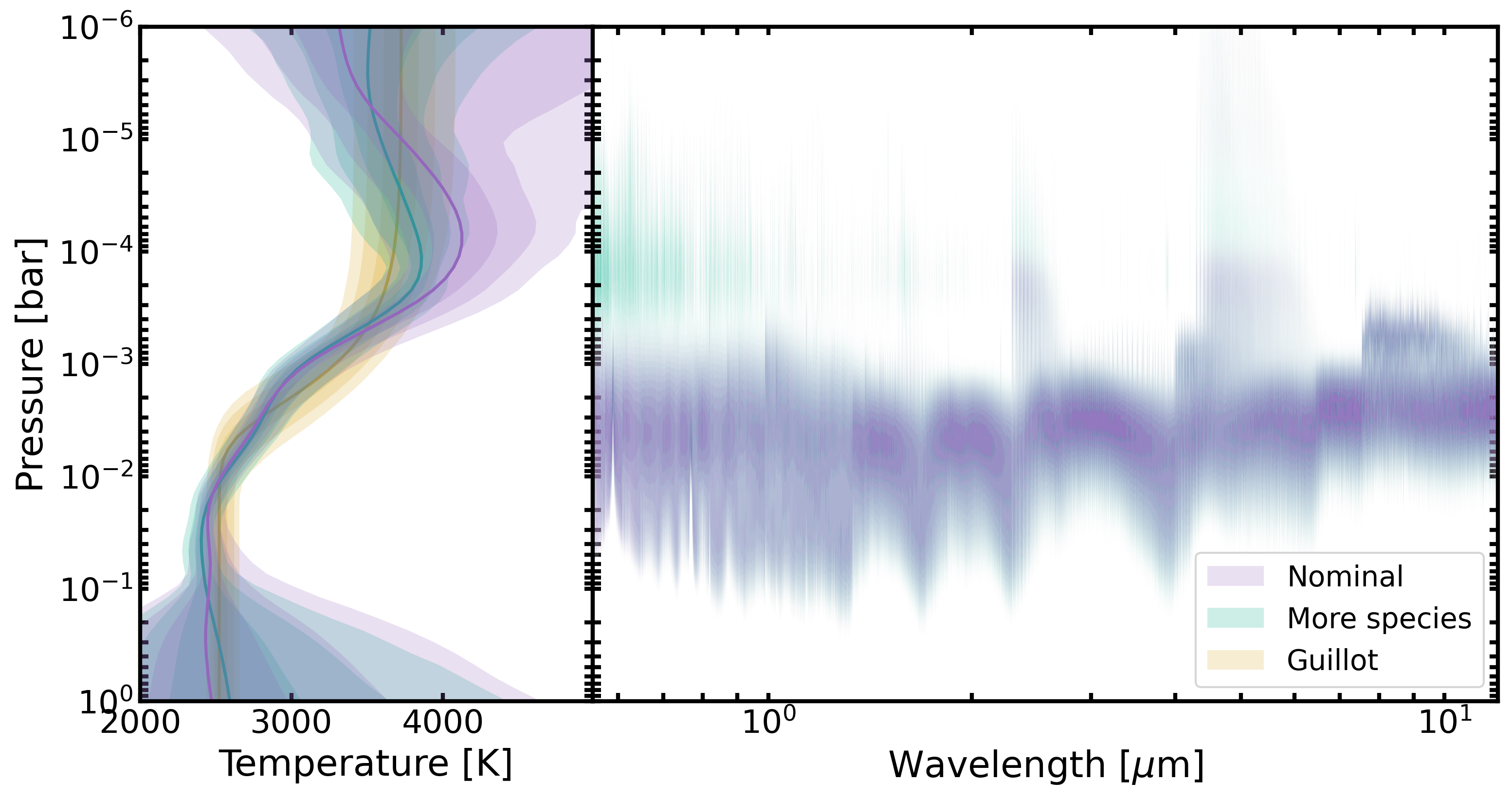}
\caption{Retrieved temperature structures (left panel) and logarithmic emission contribution (right panel) for WASP-121\,b. Colored solid lines show the median retrieved temperature structure, while 1, 2, and $3\,\sigma$ percentiles are shown with shaded regions. 
Results are shown in purple for the nominal model and in green for the ``More species’’ model, which includes more opacities from atomic and refractory absorbers (see Sect.~\ref{subsec:retrieval:variations}). Results from retrievals with a~\citet{Guillot2010guillot} temperature structure are only shown in the left panel in yellow.}
\label{fig:TPcontribution}
\end{figure}
\begin{table*}[]
    \centering
    \caption{Overview of all model variations with best fit chemical parameters and various criteria for model preference estimation. }
    \label{tab:retrievalresults}
    \begin{tabular}{lcccccrrrrr}
\hline\hline
Model Name & [Vol/H] & [Ref/H] & [Ti/H] & C/O & $\chi^2_\mathrm{red}$ & $\Delta \ln$(Z) & $\Delta$BIC & $\Delta$AIC & $\Delta$BPICS \\ \hline
Nominal & $0.79^{+0.09}_{-0.10}$ & $1.22^{+0.07}_{-0.07}$ & $-2.17^{+0.52}_{-0.46}$ & $0.57^{+0.07}_{-0.08}$ & 0.96 & 0 & 0 & 0 & 0\\
More species & $0.78^{+0.08}_{-0.08}$ & $1.16^{+0.07}_{-0.07}$ & $-2.23^{+0.52}_{-0.44}$ & $0.64^{+0.05}_{-0.06}$ & 0.91 & 6.9 & -16.0 & -16.0 & -15.0\\
No extra reflection & $0.75^{+0.09}_{-0.09}$ & $1.19^{+0.07}_{-0.07}$ & $-1.95^{+0.59}_{-0.58}$ & $0.42^{+0.07}_{-0.06}$ & 0.91 & 3.0 & -0.1 & 9.5 & 5.0\\
Fit VO* & $0.42^{+0.12}_{-0.11}$ & $0.90^{+0.09}_{-0.08}$ & $-1.93^{+0.62}_{-0.60}$ & $0.64^{+0.05}_{-0.06}$ & 0.89 & 1.3 & 1.2 & -3.6 & -0.0\\
Two MIRI kernel & $0.77^{+0.09}_{-0.10}$ & $1.25^{+0.07}_{-0.07}$ & $-2.16^{+0.52}_{-0.47}$ & $0.51^{+0.08}_{-0.09}$ & 0.94 & -1.4 & 21.5 & 7.2 & 10.5\\
Hotspot area factor & $0.75^{+0.09}_{-0.09}$ & $1.22^{+0.07}_{-0.06}$ & $-2.16^{+0.52}_{-0.46}$ & $0.52^{+0.08}_{-0.09}$ & 0.89 & -2.5 & 5.8 & 1.0 & 3.0\\
Detector offsets & $0.73^{+0.09}_{-0.09}$ & $1.24^{+0.07}_{-0.07}$ & $-2.13^{+0.51}_{-0.47}$ & $0.48^{+0.07}_{-0.08}$ & 0.93 & -5.8 & 22.5 & 13.0 & 11.2\\
No GP & $0.68^{+0.12}_{-0.11}$ & $1.28^{+0.07}_{-0.07}$ & $-2.12^{+0.54}_{-0.52}$ & $0.54^{+0.07}_{-0.07}$ & 0.95 & -12.6 & -14.9 & 32.7 & 27.4\\
No reflection no GP & $0.96^{+0.14}_{-0.22}$ & $1.41^{+0.08}_{-0.08}$ & $-0.36^{+0.10}_{-0.10}$ & $0.50^{+0.08}_{-0.12}$ & 1.00 & -37.4 & 26.5 & 88.5 & 82.1\\
Guillot & $0.80^{+0.09}_{-0.09}$ & $1.09^{+0.10}_{-0.11}$ & $-2.26^{+0.51}_{-0.44}$ & $0.47^{+0.10}_{-0.09}$ & 1.24 & -117.7 & 238.4 & 276.5 & 274.8\\
\hline
Species detection&&&&&&&&&\\
\hline
Nominal & $0.79^{+0.09}_{-0.10}$ & $1.22^{+0.07}_{-0.07}$ & $-2.17^{+0.52}_{-0.46}$ & $0.57^{+0.07}_{-0.08}$ & 0.96 & 0 & 0 & 0 & 0\\
No SiO & $0.72^{+0.14}_{-0.12}$ & $0.94^{+0.18}_{-0.21}$ & $-2.33^{+0.49}_{-0.42}$ & $0.40^{+0.08}_{-0.07}$ & 0.92 & -23.2 & 49.8 & 49.8 & 49.5\\
No VO & $0.79^{+0.09}_{-0.10}$ & $1.26^{+0.07}_{-0.07}$ & $-2.51^{+0.34}_{-0.29}$ & $0.50^{+0.07}_{-0.07}$ & 0.93 & 1.0 & -1.8 & -1.8 & 5.4\\
\hline
No GP & $0.68^{+0.12}_{-0.11}$ & $1.28^{+0.07}_{-0.07}$ & $-2.12^{+0.54}_{-0.52}$ & $0.54^{+0.07}_{-0.07}$ & 0.95 & 0 & 0 & 0 & 0\\
No SiO no GP & $1.33^{+0.06}_{-0.08}$ & $1.77^{+0.08}_{-0.08}$ & $-1.76^{+0.52}_{-0.58}$ & $0.69^{+0.04}_{-0.05}$ & 0.94 & -50.7 & 84.1 & 84.1 & 82.3\\
no VO no GP & $0.58^{+0.07}_{-0.07}$ & $1.25^{+0.08}_{-0.08}$ & $-2.66^{+0.30}_{-0.22}$ & $0.40^{+0.05}_{-0.05}$ & 1.04 & -22.3 & 49.7 & 54.5 & 54.4\\
\hline
Better sampling&&&&&&&&&\\
\hline
No GP & $0.75^{+0.14}_{-0.13}$ & $1.28^{+0.09}_{-0.09}$ & $-2.08^{+0.66}_{-0.60}$ & $0.59^{+0.08}_{-0.09}$ & 0.97 & 0 & 0 & 0 & 0\\
No SiO no GP & $1.38^{+0.10}_{-0.11}$ & $1.72^{+0.12}_{-0.12}$ & $-1.75^{+0.72}_{-0.82}$ & $0.73^{+0.06}_{-0.07}$ & 0.92 & -44.7 & 84.7 & 84.7 & 83.4\\
No VO no GP & $0.67^{+0.15}_{-0.12}$ & $1.25^{+0.09}_{-0.11}$ & $-2.65^{+0.35}_{-0.24}$ & $0.47^{+0.12}_{-0.10}$ & 1.00 & -23.1 & 47.4 & 52.2 & 52.1\\
\hline
Updated VO opacity&&&&&&&&&\\\hline
Nominal & $0.79^{+0.08}_{-0.10}$ & $1.09^{+0.08}_{-0.07}$ & $-1.15^{+0.47}_{-0.82}$ & $0.66^{+0.05}_{-0.06}$ & 0.90 & 0 & 0 & 0 & 0\\
No VO & $0.79^{+0.09}_{-0.10}$ & $1.26^{+0.07}_{-0.07}$ & $-2.51^{+0.34}_{-0.29}$ & $0.50^{+0.07}_{-0.07}$ & 0.93 & -1.6 & 9.2 & 9.2 & 11.5\\
\hline
No GP & $0.65^{+0.12}_{-0.11}$ & $1.11^{+0.06}_{-0.07}$ & $-1.80^{+0.58}_{-0.65}$ & $0.61^{+0.07}_{-0.08}$ & 0.98 & 0 & 0 & 0 & 0\\
No VO no GP & $0.67^{+0.15}_{-0.12}$ & $1.25^{+0.09}_{-0.11}$ & $-2.65^{+0.35}_{-0.24}$ & $0.47^{+0.12}_{-0.10}$ & 1.00 & -23.4 & 57.0 & 61.7 & 62.4\\
\hline

    \end{tabular}
    \tablefoot{The model names are descriptive for the retrieval variations discussed in Sect.~\ref{subsec:retrieval:variations}. The ``No extra reflection’’ model does not contain the analytical reflection description of Eq.~\ref{eq:scatter}, but does contain self-consistent reflection. The section ``Better sampling’’ contains results of retrievals computed without the constant efficiency mode of \texttt{PyMultiNest} and a sampling efficiency of $0.3$. Updated VO opacities in the last section are from~\citet{Bowesman2024VO} and do not affect the remaining parametrization. The differences are computed relative to the first row in each section. For calculating the reduced $\chi^2$, we account for the uncertainty inflation of NIRSpec and MIRI datapoints and the impact of the Gaussian processes on the covariance. Favored models have a high positive $\Delta$ln(Z) and a high negative $\Delta$BIC, $\Delta$AIC, and $\Delta$BPICS. (*) When retrieving vanadium separately, the ``Fit VO’’ model derived a value of $ \mathrm{[V/H]}=1.427^{+0.209}_{-0.232}$.}
\end{table*}
To verify that the results are robust to our model choices, we examined results across different variations of the nominal model. The retrieval setups are mostly identical to our nominal setup and differ only in one described aspect. 

One of the most impactful model choices is the chemistry description. In addition to the Gibbs free energy minimization method, we also attempted to retrieve the molecular abundances freely. Despite our implementation of H$_2$O and H$_2$ dissociation similar to~\citet{Gapp2025SiO}, we found that these retrievals result in unphysically high abundances of CO in the planet's atmosphere (see Appendix~\ref{app:freeAbund}). Therefore, our remaining analysis focuses on retrievals that use the \texttt{easyCHEM} chemistry. The assumption of thermochemical equilibrium on WASP-121\,b's dayside is justified, given the hot temperatures and the absence of photochemistry tracers such as SO$_2$. 

The included opacities in our nominal retrieval cover all species typically considered in low-resolution spectroscopy of WASP-121\,b. However, they do not account for the contribution of the numerous detected atomic species~\citep[e.g.,][]{Hoeijmakers2020HEARTS, Pelletier2025CRIRESWASP121b}, or for species with minor expected abundances, such as MgO or SiS~\citep{Visscher2010chemistry}. To test the impact of our incomplete chemical inventory, we set up a ``More species’’ retrieval, that included opacity contributions of MgH~\citep{GharibNezhad2013MgH}, Fe~\citep{KuruczLinelists}, Mg~\citep{KuruczLinelists}, Si~\citep{KuruczLinelists}, V~\citep{KuruczLinelists}, SiS~\citep{Upadhyay2018SiS,Chubb2021ExoMolOP}, SiH~\citep{Yurchenko2018SiH,Chubb2021ExoMolOP}, and MgO~\citep{Li2019MgO}, in addition to the species included in or nominal setup. Since gas-phase MgO is not part of the default \texttt{easyCHEM} species selection, we added its thermochemical properties manually\footnote{We obtained the thermochemical properties of MgO from the \texttt{ThermoBuild} tool~\citep{mcbride_2002} of the Glenn research center, under~\url{https://cearun.grc.nasa.gov/ThermoBuild/}}. The inclusion of the additional species does not significantly alter the retrieved elemental inventories of WASP-121\,b, which are largely identical to our nominal retrieval (see Table~\ref{tab:retrievalresults}). In contrast, the inclusion of the additional opacities increased the retrieval runtime by approximately $40\%$, which is why our nominal chemistry setup does not include them.

Since the temperature structure has a major impact on the emission spectrum, we further tested the use of the widely used parametrized temperature structure from~\citet{Guillot2010guillot}.
In this, we fitted for the parameters for the planet's equilibrium temperature $T_\mathrm{eq}$, the mean infrared opacity $\kappa_\mathrm{IR}$, and the ratio between visual and infrared opacity $\gamma$. We note that these parameters are primarily used to shape the temperature profile and do not necessarily describe the planet's actual equilibrium temperature or opacity relations. We computed the temperature structure in the limit of dayside-averaged flux (using a redistribution coefficient of $0.5$), and assumed an internal temperature of $150\,$K. We also tested to include the internal temperature as a free parameter, but it remained unconstrained. Due to this, we conclude that the exact value we assume for this parameter is not relevant to our conclusions. As shown in Fig.~\ref{fig:TPcontribution}, the~\citep{Guillot2010guillot} temperature parametrization leads to similar temperatures at the probed pressures between $10^0$ and $10^{-6}$\,bar, as compared to the nominal parametrization. However, the Guillot temperature profile does not offer the flexibility to describe the two different temperature gradients seen for $10^{-2}-10^{-3}$ and $10^{-3}-10^{-4}$\,bar in the nominal models. 

To test if our main results are driven by the Gaussian processes we implemented as an uncertainty treatment, we also computed retrievals without them. Instead, these retrievals included uncertainty inflation factors~\citep{Line2015dwarfs,Molliere2025PSO} for the uncorrelated noise of all detectors. As shown in Table~\ref{tab:retrievalresults}, the retrieved elemental enrichments and abundance ratios of these ``No GP’’ retrievals agree within one-sigma with our nominal setup. The largest deviation is seen in [Vol/H], which is found to be $0.7\,\sigma$ smaller in the ``No GP’’ retrievals.

Another crucial component for fitting the joint dayside spectrum of WASP-121\,b is the parametrization of reflected starlight. We therefore computed a retrieval without the analytical scattering description introduced in Eq.~\ref{eq:scatter}. Instead, we implemented the self-consistent description of the reflection of stellar light from \texttt{petitRADTRANS}~\citep{Alei2022LIFEpRTreflection}. Since molecular hydrogen is dissociated in the upper atmosphere of WASP-121\,b (see Appendix Fig.~\ref{app:fig:freeAbund}), we additionally tested the impact of including atomic hydrogen scattering and free-free and bound-free absorption~\citep{Gray2008stars}. In the probed atmosphere, the effect of hydrogen scattering is on the order of a few parts per million at the shortest wavelengths. While the impact of H scattering is smaller than the $>46\,$ppm uncertainties of the second order of NIRISS, we included its impact in this retrieval variation. In contrast, hydrogen bound-free and free-free absorption only become relevant when the expected photospheric temperatures exceed 6000\,K. As these temperatures are not expected at the probed pressure levels, we did not include bound-free and free-free absorption of atomic hydrogen in the final retrievals. While this retrieval variation found a good fit to the dayside spectrum, the extra emission at the shortest wavelengths is not accounted for by the astrophysical model. Instead, the NIRISS local GP kernel produces a power-law-like slope that could have an astrophysical origin, such as enhanced scattering. We therefore also recomputed a retrieval with the alternative scattering description and without Gaussian processes, which will be discussed further in Sect.~\ref{subsec:GPresult}.

A few of our tested retrieval variations have only a minor impact on retrieval results and are not preferred by our model selection criteria. These retrievals include a wavelength-independent hotspot area scaling factor~\citep[$A_\mathrm{HS}$, see e.g.,][]{Taylor2020biases,Coulombe2023NamelessWASP18b,Pelletier2026NIRISSWASP121b}, according to
\begin{equation}\label{eq:HS}
\mathrm{eclipse \, depth} = A_\mathrm{HS} \frac{F_\mathrm{planet}}{F_\mathrm{star}} \left( \frac{R_\mathrm{planet}}{R_\mathrm{star}}\right)^2+ A_g \left( \frac{\lambda}{\SI{0.6}{\micro\meter}} \right)^\delta \left(\frac{R_\mathrm{planet}}{a}\right)^2.
\end{equation}
This factor dilutes the spectrum, simulating that all the observed flux originates from a fraction of the planet's dayside, while the remaining area emits no light. We find $A_\mathrm{HS}$ to be compatible with unity within $1.7\,\sigma$ for WASP-121\,b. We also tested the need for detector offsets between NIRISS, NIRSpec, and MIRI/LRS. These offsets were parametrized with a constant flux contribution to NIRSpec and MIRI/LRS, using the NIRISS eclipse spectrum as an anchor. While detector offsets for NIRSpec ($260\pm120\,$ppm) and MIRI/LRS ($237\pm75$\,ppm) were found to be inconsistent with zero by $2.2$ and $3.2\,\sigma$, our model selection criteria disfavor the inclusion of the extra parameters for the offsets. For the chemistry parametrization, we tested decoupling vanadium enrichment from refractory enrichment, analogous to how we treated titanium. This retrieval returns slightly different volatile and refractory reservoirs: It overall shows smaller refractory and volatile enrichments, while only vanadium is enriched above $10\times$ solar. Since vanadium is strongly affected by the correlated noise of the NIRISS detector (see Sect.~\ref{subsec:detection}), we conclude that these findings are unreliable. Finally, the MIRI/LRS spectrum (see Fig.~\ref{fig:jointspectrum} and Appendix~\ref{app:GP}) shows multiple minor model-data mismatches, apart from the main noise structure around $\SI{9}{\micro\meter}$. Therefore, we tested adding a second local kernel to the Gaussian processes. This did not improve the retrievals and will be discussed further in Sect.~\ref{subsec:GPresult}.

\begin{figure*}
    \centering
    \includegraphics[width=0.32\linewidth]{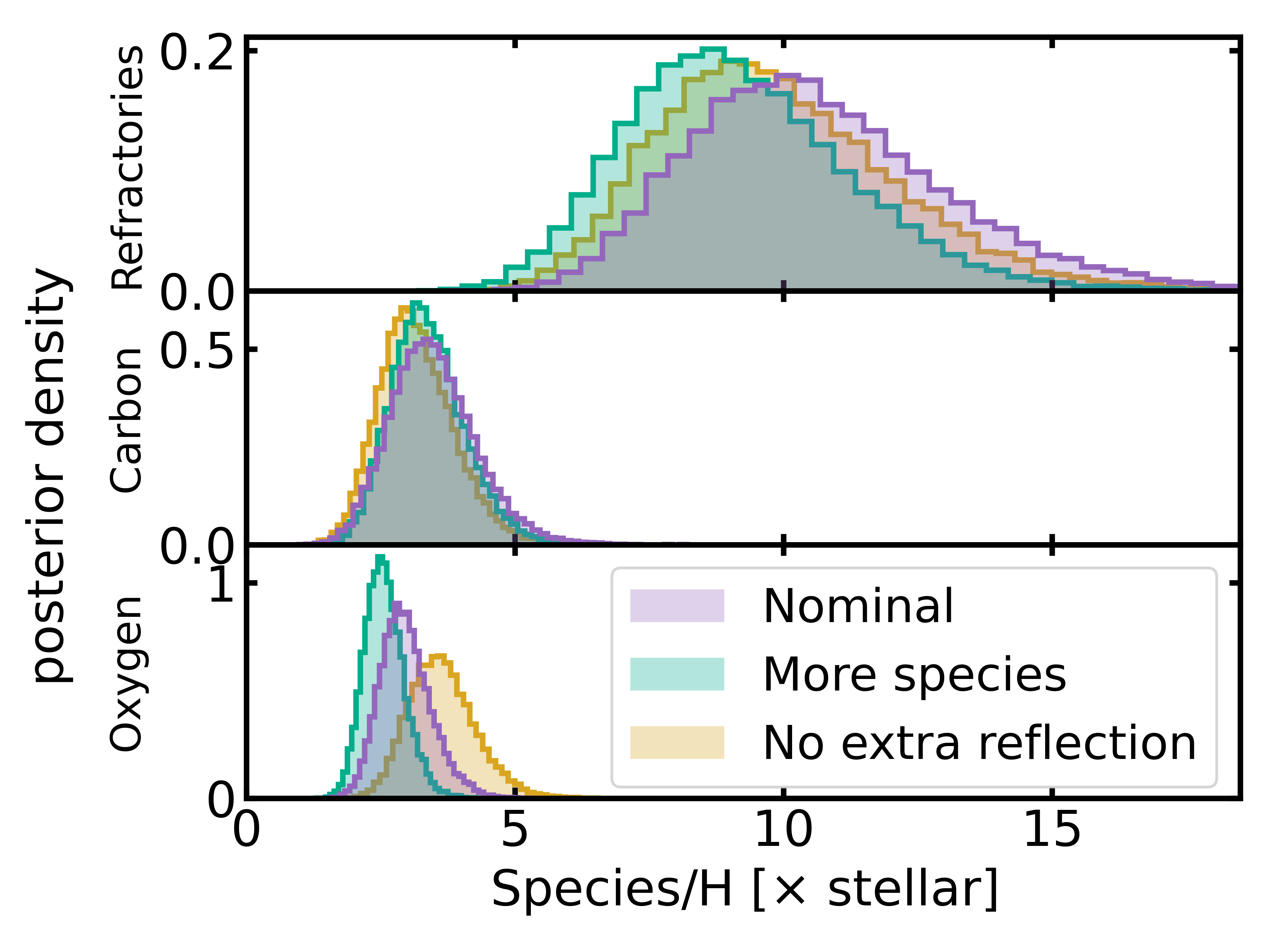}
    \includegraphics[width=0.32\linewidth]{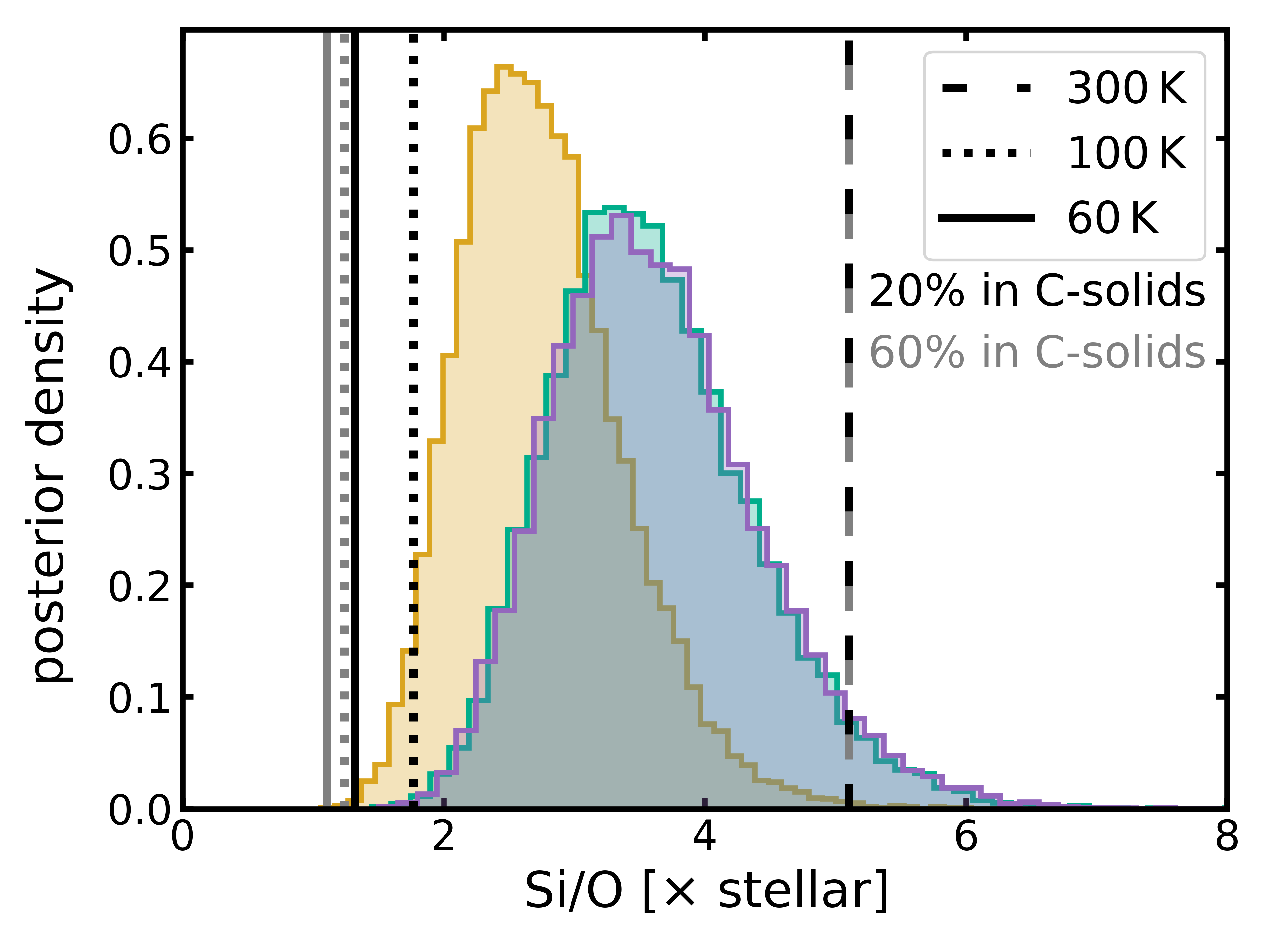}
    \includegraphics[width=0.32\linewidth]{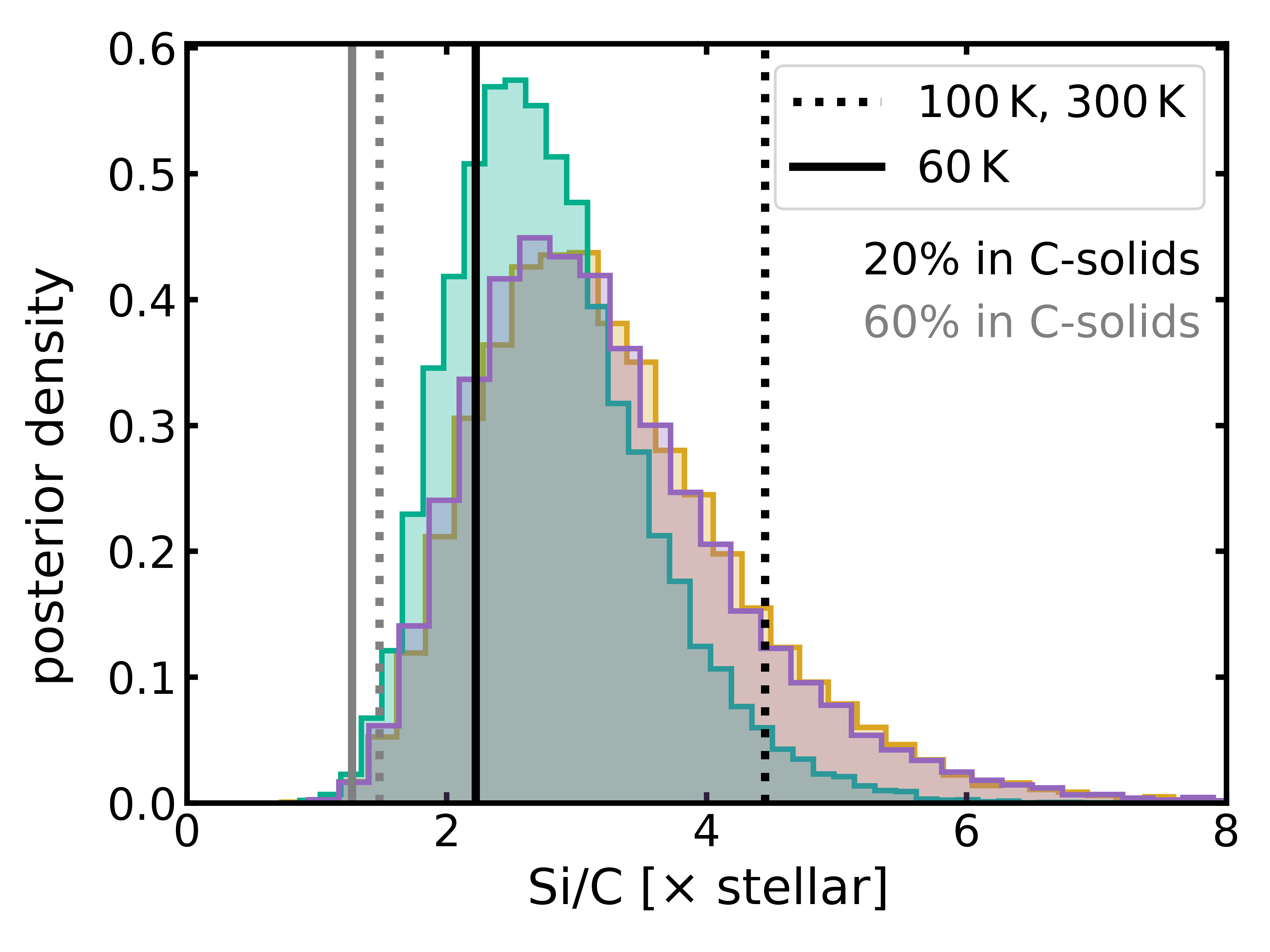}
    \caption{Posterior density of elemental abundances and refractory-to-volatile ratios in WASP-121\,b. Different colors indicate different retrieval setups (see Sect.~\ref{subsec:retrieval:variations}), as indicated in the left panel. Predictions for solid compositions in the protoplanetary disk at regions with different temperatures (compare to Fig.~\ref{fig:interpretation}) are shown as vertical lines and will be discussed in Sect.~\ref{subsec:accretionlocation}. The compositions were computed assuming 20\% (black) and 60\% (gray) of carbon to be bound in carbonaceous solids.}
    \label{fig:formation:moving}
\end{figure*}
\subsection{Retrieval results}\label{subsec:retrievalresults}
An overview of all retrieval results is presented in Table~\ref{tab:retrievalresults}, which also provides an overview of the fit qualities and Bayesian model comparison according to the logarithmic Bayesian evidence~\citep[the Bayes factor $\ln(Z)$,][]{Jeffreys_1935}, reduced $\chi^2$, Bayesian Information Criterion~\citep[BIC,][]{Schwarz1978BIC}, Akaike Information Criterion~\citep[AIC,][]{Akaike1974AIC}, and the simplified Bayesian Predictive Information Criterion~\citep[BPICS][]{Ando2011BPICS,Thorngren2026bayesian}. An overview of these comparison criteria was compiled by~\citet{Thorngren2026bayesian}, and our interpretation is based on their Table 1. In general, the best models show large values for the logarithmic evidence $\ln(Z)$, and small values of the BIC, AIC, and BPICS.

The largest change in chemical abundance and fit quality is driven by the choice of the temperature-pressure parametrization: Retrievals that include a simple~\citet{Guillot2010guillot} parametrization result in a smaller C/O ratio and a smaller enrichment in refractory species (see Table~\ref{tab:retrievalresults}). The reason for this discrepancy might be that the Guillot temperature profile does not allow for the complexity needed to describe the data accurately. The molecular species in WASP-121\,b's atmosphere contribute to the observed panchromatic JWST emission spectrum in pressure layers between $10^{-1}$ and $10^{-4}$ bar (see Fig.~\ref{fig:TPcontribution}), covering a broad pressure range. While the flexible temperature-pressure structure of the nominal model can finely tune the temperature gradients in the relevant layers, the~\citet{Guillot2010guillot} profile is limited to a single temperature gradient across the upper atmosphere. As shown in Fig.~\ref{fig:TPcontribution}, extra flexibility might be needed to account for differences in the temperature gradients between $10^{-2}$ and $10^{-3}$\,bar and between $10^{-3}$ and $10^{-4}$\,bar retrieved by the nominal model. The emission contribution indicates a strong contribution of SiO in the mid-IR for pressures between $10^{-3}$ and $10^{-4}$\,bar, which might contribute to the steepness of the corresponding temperature gradient. As the strength of an observed emission feature depends both on the species abundance and the temperature gradient in the emitting layer, fixing the temperature gradient can lead to a bias in the retrieved abundances. We note that the steep temperature gradient between $10^{-3}$ and $10^{-4}$\,bar is not present in predictions from general circulation models~\citep[e.g.,][]{parmentier2018dissociation}. The reason for this difference might be that these models do not typically include all relevant absorbers.

Our nominal retrieval setup is limited to molecular species with strong vibrational bands. While the inclusion of opacities from additional species changes the retrieved elemental reservoirs only within one sigma, the corresponding model is significantly preferred over our nominal model (see Table~\ref{tab:retrievalresults}). This might be expected, as it provides a more accurate chemical description without the need for extra free parameters. As can be seen in Fig.~\ref{fig:TPcontribution}, the inclusion of the complete chemical inventory further changes the temperature structure in the upper atmosphere, as compared to the nominal model: The temperature decline with increasing altitude above $10^{-4}$\,bar is less steep, and the uncertainty of the derived temperature structure is slightly smaller. This is likely caused by the atomic species, which show major contributions at these lower pressures. In contrast, the temperature structure resembles that of our base retrieval for pressures below $10^{-3.5}$\,bar, where the majority of our molecules emit. This consequently results in a similar atmospheric enrichment, independent of the included atomic absorbers. While the inclusion of all previously detected absorbers represents the more complete approach, the inferred chemical composition can therefore be constrained with the reduced chemical inventory. However, the retrieved temperature from the chemical complete retrieval is likely to be more accurate than the one from the nominal retrieval. 

Models with additional free parameters are generally disfavored compared to our nominal retrieval setup. In particular, the absence of a non-unity hotspot area scaling factor is consistent with the shallow temperature gradients seen in the eclipse mapping of WASP-121\,b, which is presented in Sect.~\ref{sec:mapping}.

While the inclusion of detector offsets is not preferred, the retrieved NIRSpec offset of $237\pm75$\,ppm is inconsistent with zero to 3.2 sigma. This could hint that we are not fully reconstructing the NIRSpec eclipse depths with our correction method applied in Appendix~\ref{app:nirspec}. However, as this NIRSpec offset does not significantly affect the retrieved atmospheric elemental enrichments, refitting the NIRSpec phase curve is out of scope for this work. 

\subsection{Retrieved elemental inventories}\label{subsec:inventories}
The models considered in our retrieval analysis consistently produce an enrichment in both volatile (O, C) and refractory (Si, V) containing species, with $[\mathrm{Vol}/\mathrm{H}]=0.795^{+0.086}_{-0.095}\times$ solar and $[\mathrm{Ref}/\mathrm{H}]=1.220^{+0.068}_{-0.067}\times$ solar for our nominal model. These values are stable within $1\,\sigma$ for most model parameterizations, indicating that these results are likely robust. The only models that produce slightly different values include the model variation without both the analytical reflection of star light and Gaussian processes (``No reflection no GP’’), and the variation in which we freely fit for $[\mathrm{V}/\mathrm{H}]$ (``fit VO’’). We reject the results of these retrievals, as the retrieval without analytical reflection parametrization and without Gaussian processes resulted in unphysically high atmospheric temperatures, while the separate fitting of VO might not be reasonable given the low detection significance of the species (see Sect.~\ref{subsec:detection}) and the correlated noise in the NIRISS detector (see Appendix~\ref{app:GP}). 

The retrieved atmospheric enrichments are relative to the Sun~\citep{Asplund2009}. We converted these values to enrichments relative to WASP-121, according to the measured stellar abundances listed in Table~\ref{tab:star}. We accounted for uncertainties in the stellar abundances by randomly drawing values from a Gaussian distribution centered on the best-fit abundances, with widths corresponding to the uncertainties. The derived enrichment of WASP-121\,b's atmosphere relative to its host-star is shown in Figure~\ref{fig:formation:moving}. For our nominal model, we derive atmospheric elemental enrichments of $\mathrm{C}/\mathrm{H}=3.41^{+0.85}_{-0.71}\times$ stellar, $\mathrm{O}/\mathrm{H}=2.92^{+0.51}_{-0.42}\times$ stellar, and $\mathrm{Ref}/\mathrm{H}=10.4^{+2.6}_{-2.1}\times$ stellar. Assuming the silicon enrichment to be equal to the derived refractory enrichment, this results in refractory-to-volatile abundance ratios of $\mathrm{Si}/\mathrm{O}=3.54^{+0.86}_{-0.69}\times$ stellar and $\mathrm{Si}/\mathrm{C}=3.05^{+1.12}_{-0.80}\times$ stellar.

In contrast to the refractory-to-volatile abundance ratio, the retrieved C/O abundance ratio remains model-dependent and varies between $\mathrm{C/O}=0.36$ and $0.69$ when considering the $1\,\sigma$ intervals of models with the largest discrepancy (see Table~\ref{tab:retrievalresults}). As can be seen in the left panel of Fig.~\ref{fig:formation:moving}, this is caused by model-dependent oxygen enrichments, while the carbon enrichment is mostly consistent across different models. This could be an artifact of our parametrization, as we set the atmospheric C/O abundance ratio by scaling the oxygen mass fraction. For determining the oxygen enrichment, we therefore have to modify the volatile enrichment with the C/O abundance ratio, transmitting the model-dependency.

While the posterior distributions of the atmospheric enrichments vary slightly between our retrieval parametrizations, they overall lead to the same implications for WASP-121\,b's formation. These will be discussed further in Sect.~\ref{subsec:disc:formation}.

\subsection{The impact of Gaussian Processes and scattering}\label{subsec:GPresult}
Our nominal retrieval is aided by Gaussian processes, which treat correlated noise and allow a single Gaussian model-data mismatch in each dataset. When excluding Gaussian processes from our base retrieval, we continued to apply a simple uncertainty inflation by a constant factor for each detector. The retrieved chemical abundances are consistent between the two approaches (see Table~\ref{tab:retrievalresults}), possibly as a consequence of the permitted constant uncertainty inflation.

In our nominal retrieval, the global GP kernel for the NIRISS data converges to a length scale of $\SI{0.03}{\micro\meter}$ with an amplitude of $34\,$ppm. This length scale is characteristic of NIRISS and originates from the shape of the instrument PSF on the detector~\citep{Rotman2025GPs}. The NIRISS and NIRSpec local GP kernels converge to wavelengths of $2.35$ and $\SI{3.57}{\micro\meter}$, respectively, with amplitudes that converge toward the lower prior bound. This indicates minor model-data mismatches, which only weakly affect the retrieved abundances (see Table~\ref{tab:retrievalresults} and Appendix~\ref{app:GP}). 

The local kernel for MIRI/LRS shows a broad posterior distribution: The kernel is located at $8.10^{+1.13}_{-1.30}\,\mu$m, and generally allows for a wide range of length scales within the prior bounds. While the kernel amplitudes are skewed towards smaller values, the impact of the Gaussian process can clearly be seen in the residual panel of Fig.~\ref{fig:jointspectrum}. The local MIRI/LRS kernel does not find a single Gaussian model-data mismatch, but instead places varying kernels for different sample parameter sets. As a result, the two and three sigma model percentiles include noise structure between $8.5$ and $9.5\,\mu$m, as well as the outlier data point at $6.2\,\mu$m. The use of a second local kernel for MIRI/LRS is not preferred over a single local kernel and did not result in more concentrated kernel posteriors. Instead, both local kernels yield a posterior distribution similar to that of the single-kernel retrieval. This is expected, as both kernels were initialized with the same priors, making them interchangeable. 

The capability of Gaussian processes to debias model-data mismatches is demonstrated by the retrieval without the analytical scattering description. While this retrieval resulted in similar atmospheric enrichments and a similar temperature-pressure structure as our nominal retrieval, the best-fit spectrum revealed that the local GP kernel for the NIRISS dataset added extra continuum and replaced the need for scattering.

When excluding both the Gaussian process and scattering, the retrieval instead fits the blue end with an unphysically large temperature gradient, reaching 7000\,K at $10^{-5}$\,bar. This automatically leads to stronger emission from TiO and VO, especially at the shortest wavelengths. In addition, all chemical enrichments are higher. As the retrievals without GP and without extra scattering only yield unphysical solutions, we conclude that the scattering component identified by us and~\citet{Pelletier2026NIRISSWASP121b} is significant and of astrophysical origin. We will discuss possible origins in Section~\ref{disc:reflection}.

\subsection{Detection significance of SiO and VO}~\label{subsec:detection}
As shown in Fig.~\ref{fig:jointspectrum}, the signatures of H$_2$O and CO are visible by eye in the NIRISS and NIRSpec bandpasses, while SiO is seen best in the MIRI/LRS dataset. While all three species have been significantly detected in the NIRSpec spectrum by~\citep{Evans-Soma2025SiO}, we tested if we can verify the detection of the refractory tracers VO and SiO on the joint dayside spectrum of WASP-121\,b. For this, we recomputed the nominal retrieval and the retrieval without Gaussian processes, but individually excluded the opacities of VO or SiO.

To detect a given species, we compared a model that included all species to the same model with the species of interest excluded, using Bayesian model comparison. The resulting model comparisons in the framework of the differences in logarithmic evidence $\Delta\ln(Z)$, $\Delta$BIC, $\Delta$AIC, and $\Delta$BPICS are shown in Table~\ref{tab:retrievalresults}. For species detection,~\citet{Thorngren2026bayesian} suggests that the model including the species should be preferred by $\Delta \ln(Z) > 5.9$, approximately corresponding to a difference $> 11.8$ for the other criteria. However, the authors note that these are not hard cutoffs for species detections.

We find that we robustly detect SiO, regardless of whether our retrievals are allowed to compensate model-data mismatches with Gaussian processes (SiO preferred by $\Delta\ln(Z)=23.2$), or not (SiO preferred by $\Delta\ln(Z)=44.7$). We highlight that this result is further supported by additional tests of significance, including $\Delta$BIC, $\Delta$AIC, and $\Delta$BPICS. The evidence estimates for our retrievals with Gaussian processes might be biased by the sampling with constant efficiency. In contrast, our retrievals without Gaussian processes were conducted in both flexible and constant efficiency modes. They suggest that the results from the model comparison criteria $\Delta$BIC, $\Delta$AIC, and $\Delta$BPICS are more robust against the sampling method, and can therefore be reliably used for our retrievals with Gaussian processes. 

The detection significance of VO is not as clear as that of SiO. When we do not account for correlated uncertainties in the data, the model that includes VO opacities is preferred by $\Delta\ln(Z)=23.1$ over a model without VO opacities. In contrast, when we include Gaussian processes in the tested models, there is no clear evidence for or against including VO under the tested comparison criteria. We notice an increase of $A_\mathrm{GP,NIRISS}$ from $33.9 \pm 4.7\,$ppm in the nominal retrieval to $37.6 \pm 5.0\,$ppm in the retrieval without VO opacity, while $L_\mathrm{GP,NIRISS}$ remains unchanged. The increased amplitude of the global kernel might hint at the Gaussian process absorbing parts of the signal from the somewhat regular VO bands. In contrast, the apparent VO bands might just be part of the correlated noise from the detector. 

We generally expect the presence of VO on WASP-121\,b, as it was detected on both hotter~\citep{Simonnin2025TOI1518dragVO} and colder~\citep{Pelletier2023VO} ultra-hot Jupiters. Therefore, we further tested the use of more recent VO opacities from~\citet{Bowesman2024VO}. Models that include Gaussian processes and the updated VO opacities are favored by $\Delta\mathrm{BIC}=9.2$, $\Delta\mathrm{AIC}=9.2$, $\Delta\mathrm{BPICS}=11.5$, and $\Delta\ln (Z) = -1.6$ as compared to models without VO opacities. However, even with the updated opacities, the inclusion of Gaussian processes strongly alters the significance of the VO detection. This reliance of the VO detection on the uncertainty treatment serves as a cautionary tale for further analysis of spectra with potential wavelength-dependent correlations.

\section{Eclipse mapping}\label{sec:mapping}
To further investigate the dayside of WASP-121\,b, we fitted the broadband MIRI/LRS light curve for an eclipse mapping signal. The philosophy of eclipse mapping is that with the high-cadence and high-flux precision of JWST, the phases of partial eclipse during ingress/egress are well-sampled. As the planet is successively covered/revealed, its emission profile is imprinted onto the eclipse light curve in these regions. By fitting for these signals, we can therefore indirectly recover a spatially-resolved temperature map of the dayside atmosphere.

We used \texttt{ThERESA} \citep{theresa} to fit these signals, which uses the spherical harmonic basis to model planetary brightness patterns \citep{Luger2019starry}. With input system parameters, the spherical harmonics maps can be transformed into basis light curves and fit as a weighted linear sum to the data, with the same weighted sum of the corresponding basis maps giving the eclipse map. However, whilst the spherical harmonic basis is orthogonal, the corresponding light curve basis is not, which leads to signal degeneracy complications since the light curves are fit to the data. To overcome this, \texttt{ThERESA} adopts the eigenmapping approach \citep{rauscher2018}, performing principal component analysis (PCA) on the spherical harmonic basis (sampled up to a user-specified degree, $l_\mathrm{max}$) to transform it to a new basis in which the light curves (``eigencurves’’) are orthogonal. Moreover, the corresponding ``eigenmaps’’ are ranked in order of their variance, with the most observable signals being fit first. \texttt{ThERESA} optimizes the mapping model complexity that best matches the information content of the data by selecting the one that minimizes the BIC.

We tested all models of maximum spherical harmonic degree, $l_\mathrm{max} \leq 5$, and number of eigenmaps, $2 \leq N \leq 5$. We excluded one-component models because the PCA process results in the first eigenmap hotspot being located at the longitude scanned during the central phase of the observation, rather than at the substellar longitude. In this case, that corresponds to 2.5$^{\circ}$ east. A minimum of two eigenmaps, where the second eigenmap allows the longitudinal location of the hotspot to vary, are therefore required to produce an unbiased result \citep{valentine2026}. We adopted the same system parameters as in Sect.~\ref{subsec:EurekaFitting}, and also jointly model the systematics, as it has been shown that they can be correlated with the astrophysical eclipse mapping signals \citep{Valentine2024WASP17bRamp, lally2025}. This is especially pertinent for WASP-121\,b due to the evident phase signal in the light curve. We used the same systematic model as in Equation \ref{eq:systematicsFit}, with the best-fit and 1$\sigma$ uncertainties from the \texttt{batman} analysis as priors.

\begin{figure}
    \centering
\includegraphics[width=1.0\linewidth]{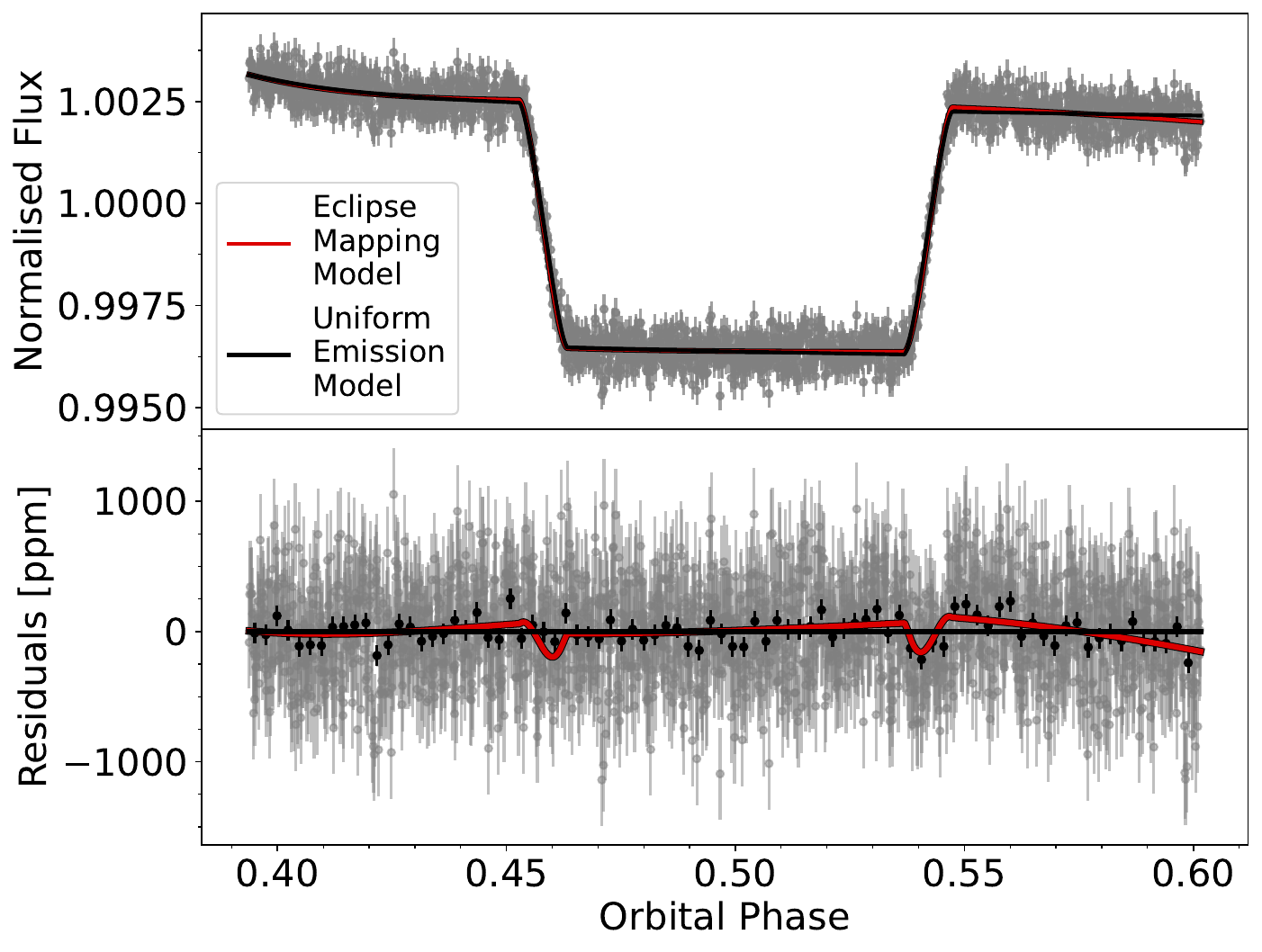}
    \caption{Eclipse mapping versus uniform emission model fits and residuals to the broadband MIRI/LRS light curve.}
\label{fig:eclipse_map_lc_fit}
\end{figure}

We found that all eclipse mapping models were preferred over our null hypothesis of spatially uniform emission by at least $\Delta \mathrm{BIC} \sim 10$ \citep[i.e., at strong statistical significance,][]{bic}. The preference for a mapping model decreases with increasing $N$, with two-component models being most preferred at $\Delta \mathrm{BIC} \sim 28$. Regarding the optimal $l_\mathrm{max}$, we found equal preference across all tested values, which is a common result in \texttt{ThERESA} \citep[e.g., see][]{theresa,Valentine2024WASP17bRamp, valentine2026}. This is because the BIC only formally penalizes the inclusion of additional basis components, $N$, not their complexity, which is set by $l_\mathrm{max}$. Hence, models with the same $N$ but different $l_\mathrm{max}$ are treated as equally complex, which means that the metric cannot discriminate between such models if they fit the data equally well. We therefore manually inspected the maps and found that the $l_\mathrm{max} \geq 2$ models are, in fact, morphologically consistent. As such, we elected for the lowest degree model of this physically consistent set (i.e., the two-component  $l_\mathrm{max}= 2$ model, hereafter referred to as L2N2).
In the top panel of Figure~\ref{fig:eclipse_map_lc_fit}, we show the fit of both this L2N2 model and that of our null hypothesis of a spatially uniform emission model to the light curve. In the bottom panel, we show the residuals of both the data and the L2N2 model when subtracting the null hypothesis model. This isolates the residual mapping signal to which the eclipse mapping model fits, showing that the map is mainly driven by the signal in egress, along with the post-eclipse phase signal as the dayside of the planet begins to rotate out of view. The mapping signal is less evident pre-eclipse and in ingress, potentially due to degeneracies with the systematic instrumental ramp.
\begin{figure}
    \centering
\includegraphics[width=1.0\linewidth]{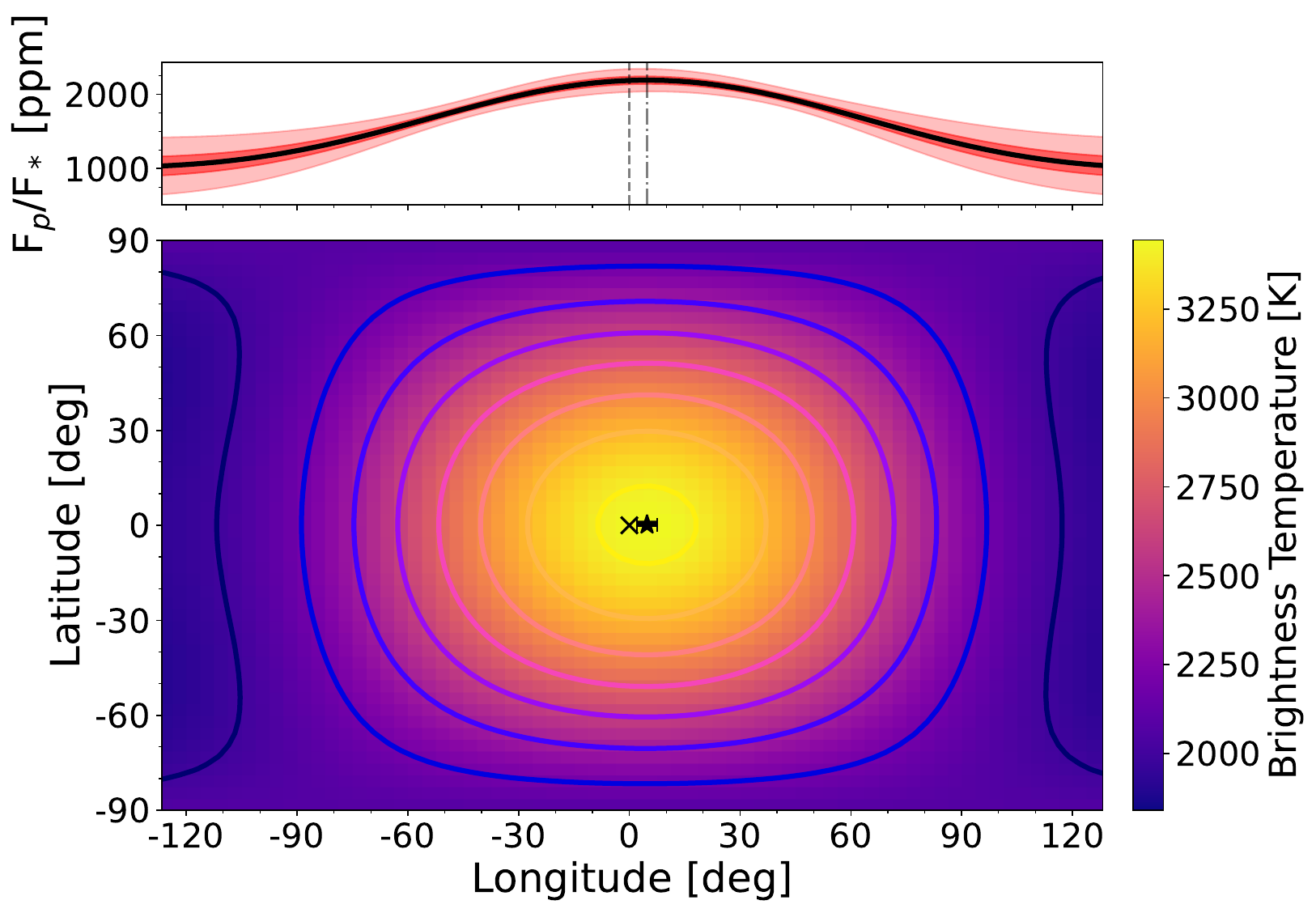}
    \caption{MIRI/LRS broadband eclipse map of WASP-121\,b. The substellar point is marked with a cross and the hotspot location is marked with a star and shows 1$\sigma$ uncertainties. The top panel shows the flux profile as a function of longitude with $1$ and $3\,\sigma$ credibility bounds.}
    \label{fig:eclipse_map}
\end{figure}
In Figure \ref{fig:eclipse_map}, we show the L2N2 eclipse map, centered on the substellar point. The map was converted from normalized flux to brightness temperature using
\begin{equation}
    T_{b} = \frac{hc}{\lambda k} \left[\mathrm{ln}\left(1+ \frac{2 \pi hc^{2} (R_{\mathrm{planet}}/R_\mathrm{star})^{2}} {\lambda^{5} F_\mathrm{star} (F_{\mathrm{planet}}/F_\mathrm{star})} \right)\right]^{-1},
\end{equation}
\noindent where $\lambda$ is the filter mean-wavelength ($\lambda$=8.12\,$\upmu$m), the stellar flux is calculated using a \texttt{PHOENIX} model \citep{Husser2013phoenix}, and the system parameters are provided in the lower part of Table~\ref{tab:star}. We show the regions of the planet that were visible during the observations, calculated using the visibility function of \texttt{ThERESA}, which extend significantly beyond the dayside due to the rotation of this short-period planet during the observation. Atop the map, we plot the flux profile as a function of longitude, which has been weighted by the squared cosine of the latitudes to account for both decreasing emitting area and reduced viewing geometry.

The MIRI-derived eclipse map shows a small longitudinal hotspot offset of $4.8^{+2.7\circ}_{-2.8}$ eastward, which is consistent with the $2.95^{+0.11\circ}_{-0.12}$ and $2.39^{+0.12\circ}_{-0.13}$ eastward values measured from the NIRSpec/G395H NRS1 and NRS2 phase curves, respectively \citep{Evans-Soma2025SiO}. We measure a hotspot temperature of 3444$\pm$79\,K and a (latitudinally-weighted) median dayside temperature of 2889$\pm$58\,K. These agree well with the irradiation temperature of 3400\,K \citep{Splinter2025WASP121bAlbedo}, the 2700--2800\,K dayside temperatures measured from the NIRISS/SOSS and NIRSpec/G395H phase curves \citep{Mikal-Evans2023WASP121bNIRSpecPhaseC, Splinter2025WASP121bAlbedo}, and the $2600-3200$\,K temperature range from the corresponding emission spectra \citep{Evans-Soma2025SiO, Pelletier2026NIRISSWASP121b}. As shown in Fig.~\ref{fig:jointspectrum}, the MIRI/LRS spectrum derived in Sect.~\ref{subsec:EurekaFitting} also shows brightness temperatures around 2900\,K, consistent with the results of the eclipse mapping. We further find limb temperatures of $\sim$2200\,K, with our map reaching a minimum temperature of $\sim$1800\,K over the visible longitudes ($-126.6^{\circ}\leq\phi\leq128.2^{\circ}$). 

The small hotspot offset is not unexpected, as ultra-hot Jupiters can have a strongly reduced wind speed, affecting the global heat circulation~\citep[see e.g.,][]{Perna2010drag,Beltz2022drag,Splinter2025WASP121bAlbedo,Bloecker2026anisotropic}. The reduced winds would also lead to a strong day-night temperature contrast, which is somewhat incompatible with our relatively small measured substellar-to-limb temperature gradient. This provides an interesting test for climate models for ultra-hot Jupiters,  which are highly sensitive to the coupling between the atmosphere and the planet's magnetic field~\citep[e.g.,][]{Hindle2019westHotspot,Christie2025drag,Bloecker2026anisotropic} and to additional factors like H$_2$ dissociation~\citep{bell2018}. 

\section{Discussion}\label{sec:discussion}
\begin{figure*}[t]
	\centering
\includegraphics[width=0.99\textwidth]{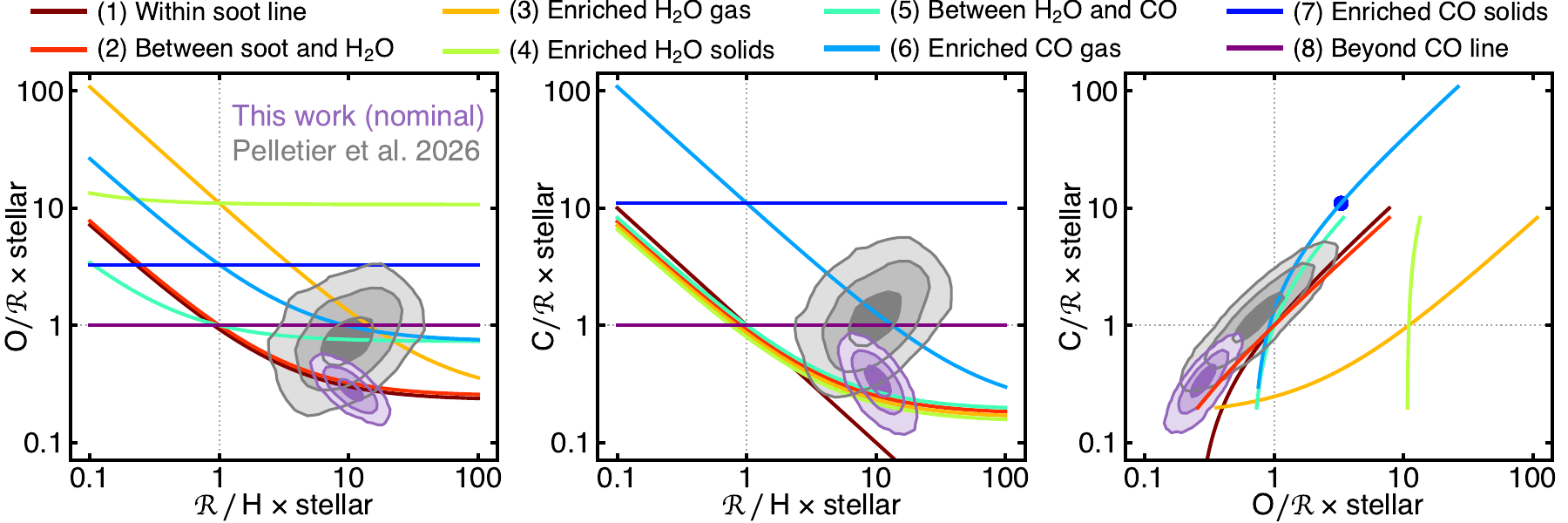}
	\caption{Posterior distribution of the oxygen-to-refractory and carbon-to-refractory abundance ratio from the joint retrieval analysis compared to different planet formation scenarios from~\citet{chachan2023formationmodel}. Colored lines show predictions of atmospheric enrichment as indicated in the top legend. Projections are indicated by colored dots. The posterior distributions of~\citet{Pelletier2026NIRISSWASP121b} are shown in gray. The refractory reservoir is labeled as ``$\mathcal{R}$’’.}
	\label{fig:formation:stationary}
\end{figure*}
Our results present the first panchromatic ($0.6 - 10.5\,\mu$m) estimate of the refractory-to-volatile abundance ratio of WASP-121\,b. Consistent with results from single-instrument transmission and emission spectroscopy~\citep{Gapp2025SiO,Evans-Soma2025SiO,Pelletier2026NIRISSWASP121b}, we find an atmospheric enrichment in both volatile and refractory species on WASP-121\,b, and we confirm the depletion of titanium on the planet's dayside. Furthermore, we also find an inverted temperature structure on the planet's dayside, as well as the need for a reflected light contribution to explain the large eclipse depths observed at short wavelengths. 

We are able to robustly constrain the elemental enrichments and abundance ratios of volatile and refractory elements: Our joint retrieval of all JWST dayside spectra of WASP-121\,b results in elemental enrichments of $\mathrm{C}/\mathrm{H}=3.41^{+0.85}_{-0.71}\times$ stellar, $\mathrm{O}/\mathrm{H}=2.92^{+0.51}_{-0.42}\times$ stellar, and $\mathrm{Ref}/\mathrm{H}=10.4^{+2.6}_{-2.1}\times$ stellar. In addition, we find that refractory species, mostly traced by silicon, are enhanced relative to volatile inventories by $\mathrm{Si}/\mathrm{O}=3.54^{+0.86}_{-0.69}\times$ stellar and $\mathrm{Si}/\mathrm{C}=3.05^{+1.12}_{-0.80}\times$ stellar. We do not recover the previously reported, highly super-stellar C/O abundance ratio, which~\citet{Evans-Soma2025SiO} derived to be $0.92^{+0.02}_{-0.03}$. Instead, our nominal retrieval finds a planetary C/O of $0.571^{+0.067}_{-0.084}$, which is only slightly enhanced as compared to the derived stellar C/O of $0.48\pm 0.05$ (see Table~\ref{tab:star}). Unlike the overall volatile and refractory reservoirs, we further find that the C/O ratio is highly model-dependent, varying between $0.36$ and $0.69$ within the $1\,\sigma$ uncertainties of the most diverging models.

We present the suggested formation and accretion scenarios of WASP-121\,b in Sect.~\ref{subsec:disc:formation}, and discuss the impact of stellar abundance uncertainties and 3D planetary effects in Sects~\ref{disc:star} and~\ref{disc:represent}. Sections~\ref{subsec:obliquity} and~\ref{subsec:hem} discuss how WASP-121\,b might have acquired its present-day high-obliquity orbit. Finally, Sects.~\ref{disc:reflection} and~\ref{subsec:disc:lit} discuss the observed reflected light and compare our findings with previous analyses of WASP-121\,b in the literature.

\subsection{The formation of WASP-121\,b}\label{subsec:disc:formation}
The elemental inventories derived in Sect.~\ref{subsec:inventories} provide insights about the formation history of WASP-121\,b. In particular, the measured super-stellar atmospheric C/H, O/H, and Si/H ratios imply significant accretion of heavy elements during the planet's formation. The ratios between the different heavy element inventories additionally allow us to make predictions about how these volatile and refractory species arrived at WASP-121\,b.

The two dominating models for giant planet formation are the direct gas collapse via gravitational instability~\citep[e.g.,][]{Boss1997gravInst} and core accretion mechanisms~\citep[e.g.,][]{Pollack1996accretion,Mordasini2008overview,Mordasisi2009population,Burn2021model,Drazkowska2023formationoverview}. While it was shown that disk instability can result in significant atmospheric enrichment~\citep{Boley2011DIenrichment}, this mechanism is likely to occur mostly for planets that orbit their star at distances $>100\,$au~\citep{Boley2009twomodes}. We therefore discuss the formation of WASP-121\,b in the framework of core accretion. In this scenario, the planetary core forms first via planetesimal~\citep{Pollack1996accretion} or pebble accretion~\citep{Johansen2010pebbles,Ormel2010pebbles}. In the pebble accretion scenario, the planetary core grows via pebble accretion until the pebble-isolation mass is reached and the incoming pebble flux is halted by a pressure maximum outside the planet's orbit~\citep{Morbidelli2012pebbles,Lambrechts2014pebbleisolationmass,bitsch2018scalinglaw}. Gas giants then accrete a large fraction of their mass from the gas phase of the disk and also from planetesimals~\citep{Lambrechts2014pebbleisolationmass}. Depending on a planet's entropy, the accreted material could be convectively mixed during the early history of the planet~\citep{Knierim2024mixing}.

\subsubsection{Accretion locations}\label{subsec:accretionlocation}
To qualitatively assess possible planet formation and accretion scenarios in WASP-121\,b, we apply the modeling approaches of~\citet{Schneider2021IpebblesCO} and~\citet{chachan2023formationmodel}. In these, the composition and elemental ratios of the present-day planetary atmosphere are assumed to reflect those of the accreted disk material. Both frameworks calculate the solid and gas compositions in the protoplanetary disk as a function of the underlying temperature structure. Consequently, volatile species, such as H$_2$O, CO$_2$, and CO, are condensed into the solid phase outside their corresponding evaporation fronts~\citep[also called ice lines, see e.g.,][]{oeberg2011snowlines, Schneider2021IpebblesCO,chachan2023formationmodel, houge2025disk}. We note that disks are 3D objects and evolve with time. Therefore, these evaporation fronts are not at a fixed distance from the star, but rather vary as a function of distance from the disk mid-plane, and their exact position changes with time \citep[see e.g.,][]{Miotello2023,Bergner2026JEDI}. Both modeling approaches can take this into account, as they rely on the accretion region's temperature rather than the planet's distance from the star. Furthermore,~\citet{Schneider2021IpebblesCO} allow the planet to migrate during its accretion phase, while~\citet{chachan2023formationmodel} instead assume a stationary planet and move the position of the evaporation fronts by evolving the disk temperature with time. As planets can form at different disc locations, planetary migration and/or scattering events could be needed to bring WASP-121\,b to its currently observed short-period high-obliquity orbit. This will be discussed further in Sects.~\ref{subsec:obliquity} and \ref{subsec:hem}. 

The posterior distributions of the planetary elemental enrichments are shown in Figs.~\ref{fig:formation:moving} and~\ref{fig:formation:stationary}, alongside model predictions from~\citet{Schneider2021IpebblesCO} and~\citet{chachan2023formationmodel} for solid compositions and formation locations, respectively. We calculated the solid compositions at disk temperatures of 60\,K, 100\,K, and 300\,K, which loosely correspond to accretion regions outside the CO$_2$ ice line, between H$_2$O and CO$_2$ ice line, and between H$_2$O and soot line~\citep{Kress2010sootline,houge2025disk}. For this, we used a simple stoichiometric model that distributes the different elements (Fe, Mg, S, Si, C, and O, according to the present-day stellar abundances of WASP-121 derived in Sect.~\ref{subsec:star}) into solids (e.g., Fe$_2$O$_3$, CO, CO$_2$, CH$_4$, H$_2$O) that can be accreted by the growing planet. This model first distributes all elements and then uses the remaining oxygen to form water~\citep[e.g.,][]{Bitsch2020solids,Schneider2021IpebblesCO}. The amount of leftover oxygen crucially depends on the stellar C/O ratio, as carbon is very efficient at binding oxygen, e.g., in CO and CO$_2$. It also depends on the number of carbonaceous solids, which contain little to no oxygen. Standard models include either that 60\% of the carbon is in carbonaceous solids~\citep[interstellar medium,][]{Bergin2015ism} or 20\%~\citep[comets in the solar system,][]{Altwegg2020comet}. These assumptions have a direct consequence on the Si/C abundance ratio within the solids in the inner disc, where part of the C-bearing molecules like CO, CH$_4$, and CO$_2$ have already evaporated. This also influences the Si/O abundance ratio, because the larger amount of carbonaceous solids results in less oxygen bound in CO and CO$_2$, consequently increasing the water ice fraction and reducing the Si/O abundance ratio of solids in the outer disc.

\begin{figure}
	\centering
\includegraphics[width=0.499\textwidth]{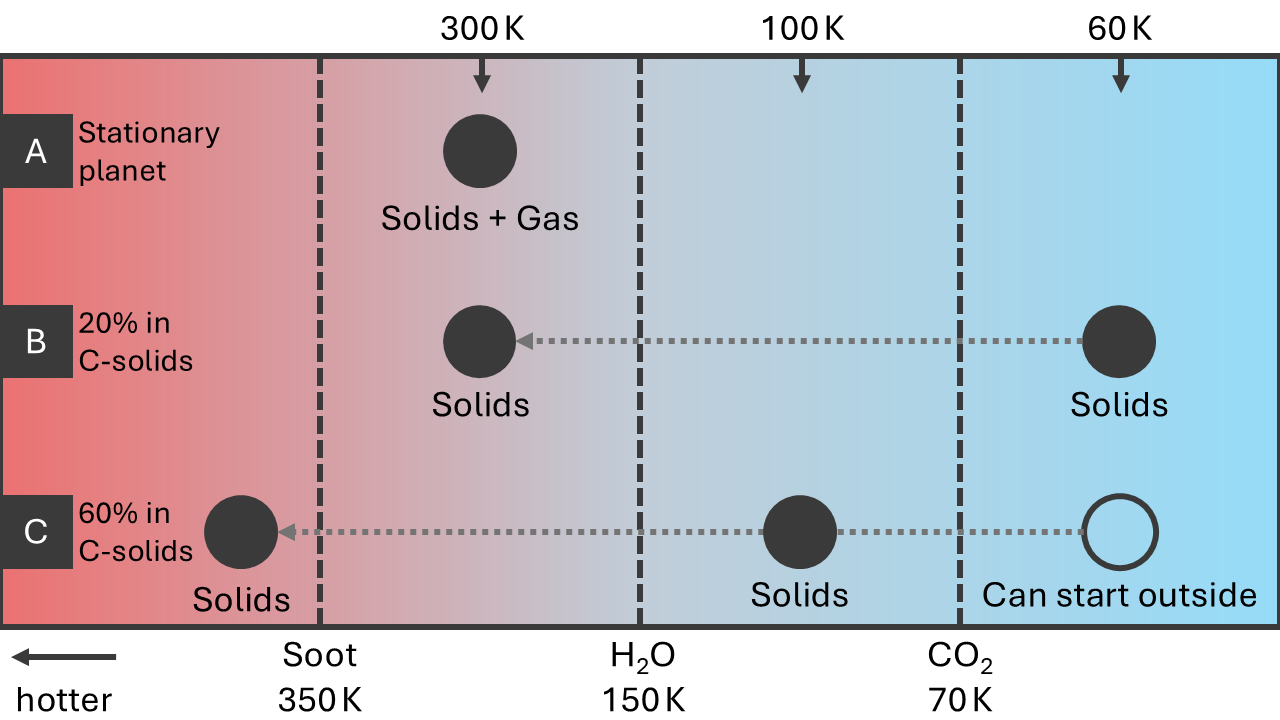}
\caption{Overview of accretion scenarios for WASP-121\,b discussed in Sect.~\ref{subsec:accretionlocation}. We indicate the sources that contribute to the Si/O and Si/C enrichments below the planet positions. Relevant ice lines are indicated with their corresponding temperatures according to Table A.1 of~\citet{houge2025disk}. The top part marks the temperatures at which we computed the solid compositions shown in Fig.~\ref{fig:formation:moving}.}
\label{fig:interpretation}
\end{figure}

Our posterior distributions for the Si/O abundance enrichments, shown in Fig.~\ref{fig:formation:moving}, generally do not align with solid enrichment at one specific location, but instead cluster between the enrichments in 100\,K and 300\,K regions. Since our model assumes discrete changes in solid and gas-phase chemistry at the ice lines, this implies either the need for planet migration to accrete solids with different compositions or an additional volatile enrichment via gas accretion that would reduce high atmospheric Si/O and Si/C abundance ratios to lower values. Figure~\ref{fig:interpretation} schematically shows three possible scenarios that are likely to result in the observed Si/O and Si/C enrichment in WASP-121\,b:\\

\textbf{ Scenario A:} WASP-121\,b formed interior to the water ice line, and accreted the volatile-poor planetesimals with an Si/O abundance ratio of $5.1$ (computed for a temperature of $T=300$\,K). The additional volatile enrichment can be achieved if the planet also accretes the volatile-rich vapor from the gas phase. This scenario agrees well with the predictions of~\citet{chachan2023formationmodel}, who studied how the addition of solids onto a stationary planet can change its abundance. Figure~\ref{fig:formation:stationary} highlights their eight tested formation locations, and shows that our derived posterior distributions for WASP-121\,b's elemental ratios are only compatible with their predictions for a formation between soot and H$_2$O ice line. As will be discussed further in Sect.~\ref{subsec:disc:lit},~\citet{Smith2024IGRINSWASP121b}, derived a similar scenario based on high-resolution observations from the ground.

\textbf{Scenario B:} The planet accreted volatile-rich planetesimals exterior to the water ice line (Si/O ratio of $1.8$ for $T=100$\,K and $1.3$ for $T=60\,$K), and then migrated interior to the water ice line, where it accreted the remaining silicon-rich material from solids. For this, volatile-poor planetesimals could be able to deliver significant amounts of silicates to the planet. Alternatively, a flow of small dust grains originating from the exterior of the planetary gap~\citep{Stammler2023leakydusttraps,Houge2026leaky} could enrich the vapor accreted inside the water ice line. If the viscosity of the disk is sufficiently high, these small dust grains could then also enrich WASP-121\,b's refractory content during gas accretion~\citep{Bitsch2023viscosity,Morbidelli2023insitu}. When we additionally consider the Si/C enrichment, both formation scenarios generally remain valid if we assume that 20\% of carbon is locked up in carbonaceous solids. In this case, the observed Si/C enrichment additionally constrains scenario B, in which the planet needs to start its migration outside the CO$_2$ ice line to accrete volatile-rich solids with $\mathrm{Si/C}=2.2$ (for $T=60\,$K).

\textbf{Scenario C:} This scenario assumes that 60\% of carbon is locked up in carbonaceous solids. As a consequence, the Si/C ratio of the solids does not exceed $1.5$ exterior to the soot line. To obtain the measured Si/O and Si/C enrichments, the planet would have needed to accrete volatile-rich solids beyond the H$_2$O ice line (Si/C ratio of $1.3$ for $T\leq100$\,K) and subsequently migrate to the interior of the soot line. Solids within the soot line do not contain carbon and are therefore able to elevate the atmospheric Si/C abundance ratio. 

Scenarios B and C can both explain the Si/C and Si/O enrichment of WASP-121\,b purely with solid accretion. However, to become a gas giant, WASP-121\,b additionally needed to accrete gas from the disk~\citep{Lambrechts2014pebbleisolationmass}. Therefore, the scenarios only explain the accretion of WASP-121\,b's atmosphere if the accreted gas does not strongly alter the planetary refractory-to-volatile abundance ratio. As a consequence, we think scenario A provides the most likely explanation for WASP-121\,b's formation, as it links the complete accretion history to the present-day atmosphere.

Alternatively to our main scenarios, the planet might have accreted gas in the very inner region of the disc, within the silicate rock-vapor lines~\citep{Danti2023pebbles}. This gas is enriched in refractory species during the very early formation stages, when large amounts of pebbles drift inwards and evaporate. As more silicon-rich vapor is accreted onto the star and more volatile-rich gas migrates inwards, the planet starts to accrete more volatiles, reaching its present-day enrichment. While we cannot fully reject this scenario, it requires the planet to accrete gas at the earliest stages of disk formation, very close to its host star. N-body simulations show that such a formation of giant planets in the very inner disk is unlikely~\citep{Poon2021insituFormation}, which is why we do not favor this pathway. Finally, we can also not exclude the possibility that the planetary core dissolved itself into the atmosphere~\citep{Helled2017fuzzy,Helled2022fuzzy}. As the core contains large amounts of silicon, this could lead to the enhancement of the atmosphere with refractories. The carbon and oxygen then mostly originate from the vapor accretion. 

\subsection{The impact of stellar abundance uncertainties}\label{disc:star}
The stellar abundances derived in Sect.~\ref{subsec:star} have non-negligible uncertainties. We account for these when computing our posteriors of the elemental enrichments relative to the stellar value (shown in Fig.~\ref{fig:formation:moving}). This was done by randomly drawing stellar abundance values from a Gaussian distribution centered around the best-fit values from Sect.~\ref{subsec:star}, with widths according to the corresponding uncertainties.

Our derived stellar abundances for Fe, Mg, S, and Si agree well with the findings of~\citet{Evans-Soma2025SiO}, while we find a higher stellar enrichment of carbon and oxygen. These differences might be explained by differences in the underlying line lists and atomic data. For oxygen, the larger difference might be due to the fitting approach: While~\citet{Evans-Soma2025SiO} applied NLTE corrections to posteriors derived from the equivalent-width method, we directly performed NLTE synthesis during the fitting. While applying the stellar abundances of~\citet{Evans-Soma2025SiO} would result in lower Si/O and Si/C enrichments of WASP-121\,b relative to its star, the conclusions for the planet's formation in Sect.~\ref{subsec:disc:formation} are not affected. Going forward, we recommend conducting a more thorough 3D NLTE modeling of the host star WASP-121, which is beyond the scope of this work. 

\subsection{Are dayside elemental enrichments representative for the entire planet?}\label{disc:represent}
The interpretation of our results requires the assumption that the dayside gas-phase abundances on WASP-121\,b are representative of the entire planet. Given the hot dayside temperatures of WASP-121\,b, we do not expect any of the main tracers, Si, O, and C, to be condensed into clouds. The saturation vapor pressures of silicon-bearing species, such as MgSiO$_3$, Mg$_2$SiO$_4$, and SiO$_2$, are above the expected partial pressures for the relevant photospheric pressures, rendering the retrieved dayside temperatures of WASP-121\,b too hot for these clouds to condense~\citep[see Fig.~\ref{fig:TPcontribution} and also Fig.~2a of][]{Evans-Soma2025SiO}. However, it is unclear what fraction of silicon could be cold-trapped on the planet's cold night side. A cold-trapping scenario on WASP-121\,b possibly takes place for titanium~\citep[e.g.,][]{Hoeijmakers2024tio,Prinoth2025TiO,Pelletier2026NIRISSWASP121b}, which is clearly depleted on WASP-121\,b's dayside with respect to the stellar enrichment. In contrast, the enrichment of dayside Si indicates the absence of a silion cold-trap. However, we can not decisively rule out that a fraction of silicon is frozen out on the planet's nightside. If this were the case, our ratio between refractory and volatile species abundances would mark a lower limit.

In general, our one-dimensional retrievals are oblivious to the effects of the three-dimensional physics on the planet. By averaging over the dayside hemisphere, we cannot account for temperature and abundance gradients between the terminators and the substellar point or for possible asymmetries. The joint dayside spectrum might not be sensitive to these effects, since the retrievals do not favor the inclusion of a hotspot-area scaling factor (see Table~\ref{tab:retrievalresults}). Furthermore, the eclipse mappings of Sect.~\ref{sec:mapping}, \citet{Mikal-Evans2023WASP121bNIRSpecPhaseC}, and~\citet{Frazier2026WASP121bNIRISSphaseCModel} disfavor large ($>10^\circ$) hotspot offsets, making strong asymmetries on WASP-121\,b's dayside unlikely. A more careful 3D treatment of the planet should consider the full phase curves of the planet, and is out of the scope of this work. 

\subsection{The high obliquity of WASP-121\,b}\label{subsec:obliquity}
WASP-121\,b orbits its host star in a near-polar orbit \citep{delrezWASP121HotJupiter2016, bourrierHotExoplanetAtmospheres2020}. Standard disk migration (Type I or II) produces planets on prograde, equatorially aligned orbits because the host star and disk share angular momentum from birth \citep[see e.g.,][]{dawsonOriginsHotJupiters2018}. WASP-121\,b's extreme orbital tilt therefore requires a disruption to this standard picture. This could occur early on, if a stellar flyby torques and warps the protoplanetary disk out of alignment, allowing standard migration to deliver the planet on an oblique orbit~\citep{Batygin2012warp}. Alternatively, if the disk remained aligned, the tilt points to a post-formation dynamical event. Focusing on the latter, two main high-eccentricity migration (HEM) mechanisms are known to generate such obliquities:
\begin{enumerate}
    \item \textbf{Von Zeipel--Lidov--Kozai (ZLK) Cycles:} A far-off, massive companion pulls on the inner planet, causing its orbit to become tilted and eccentric \citep{vonZeipel1910, Lidov1962, Kozai1962, Naoz2016}. Over time, tidal interactions can reorient the planet's orbit to a polar or even retrograde configuration \citep{wuPlanetMigrationBinary2003, Fabrycky2007shrinking}.
    \item \textbf{Planet-Planet Scattering and Secular Chaos:} Gravitational interactions among multiple planets can destabilize the system, potentially resulting in the ejection of one planet and leaving the remaining planet on a highly eccentric and misaligned orbit \citep{rasioDynamicalInstabilitiesFormation1996, chatterjeeDynamicalOutcomesPlanetPlanet2008}. Even without an ejection, long-term secular chaos can drive orbital misalignments as planets slowly exchange angular momentum, leaving the inner planet on a highly eccentric trajectory \citep{naozHotJupitersSecular2011,Lithwick2014secular}. A close stellar flyby could act as an external trigger for this kind of evolution, either directly exciting the planet's eccentricity and inclination in a single passage or destabilizing a multi-planet inner system.
\end{enumerate}
Both of these dynamical pathways initially drive the planet into a highly eccentric orbit. Subsequent tidal dissipation during close periastron passages removes orbital energy, gradually shrinking and circularizing the orbit \citep{jacksonTidalEvolutionClosein2008}. Currently, WASP-121\,b follows a nearly circular path \citep{delrezWASP121HotJupiter2016, Bourrier2020parameters}, as expected for an old, close-in giant planet. Because tidal circularization is a slow process, typically requiring hundreds of millions to billions of years \citep{wuPlanetMigrationBinary2003, Fabrycky2007shrinking}, a very young system would not have had time to erase its initial eccentricity. To confirm WASP-121\,b had sufficient time to circularize, we derived the host star's age using the Bayesian STellar Algorithm \citep[BASTA;][]{silvaaguirreAgesFundamentalProperties2015, aguirreborsen-kochBAyesianSTellarAlgorithm2022} combined with Gaia parallax and photometry. We find a stellar age of $\tau_\star = 1.53^{+0.30}_{-0.24}$\,Gyr, similar to previous age estimates of $1.5\pm1.0\,$Gyr~\citep{delrezWASP121HotJupiter2016} and $1.11\pm 0.14\,$Gyr~\citep{Sing2024wasp121bagemass}, and likely sufficient for tidal effects to erase post-migration eccentricity.

Among the post-formation mechanisms that could drive WASP-121\,b's near-polar orbit, a stellar flyby or planet-planet scattering by an ejected body leave no present-day signature and cannot be falsified with current data. In contrast, a ZLK pathway with a presently bound outer companion is, in principle, detectable today. Therefore, we searched for evidence of an outer companion capable of driving ZLK cycles. Such a companion would induce a periodic photocenter motion detectable in \textit{Gaia}~\citep{Gaia2016spacecraft} data as an elevated Renormalized Unit Weight Error (RUWE). The observed \textit{Gaia} DR3~\citep{Gaia2021DR3} value of $\mathrm{RUWE} = 0.884$ is fully consistent with a single star, showing no significant excess noise. We used this non-detection to set quantitative exclusion limits on companion mass and separation via an injection-recovery analysis following \citet{penoyreAstrometricIdentificationNearby2022}.

\begin{figure}
    \centering
    \includegraphics[width=\columnwidth]{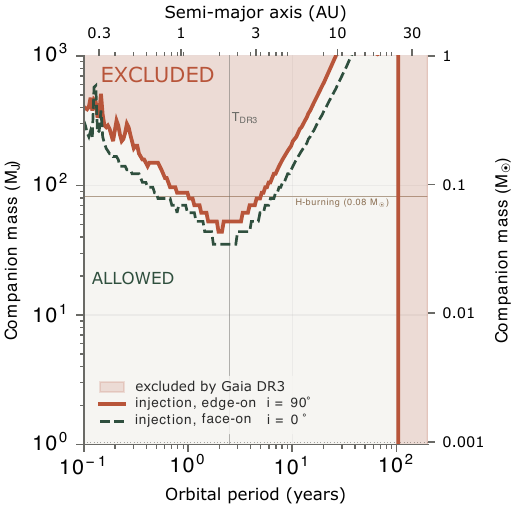}
    \caption{Outer-companion exclusion for WASP-121 from
        \textit{Gaia} DR3 astrometry. The rust-shaded region above the
        solid red edge-on curve ($i = 90^\circ$) is ruled out at
        $\mathrm{RUWE} > 1.25$ from a 30-phase median injection--recovery
        analysis; the dashed dark green curve shows the more optimistic
        face-on bound ($i = 0^\circ$). The vertical
        line is the DR3 baseline ($t_\mathrm{DR3} = 2.5$\,yr). WASP-121's measured $\mathrm{RUWE} = 0.884$ places any real companion in the
        non-shaded ALLOWED region. Stellar companions with orbital periods $>140\,$years were excluded by~\citet{bohnYoungSunsExoplanet2020}.}
    \label{fig:wasp121_gaia_exclusion}
\end{figure}

We simulated synthetic astrometric signals for circular orbits and injected them into the actual \textit{Gaia} DR3 scanning epochs (49 field-of-view scans of WASP-121 across Gaia's focal plane between 2014.8 and 2017.3). For each scan, we added Gaussian noise according to the star's brightness (an along-scan uncertainty of $\sigma_\mathrm{AL} = 0.15$\,mas at a \textit{Gaia} $G$-band magnitude of $10.4$, using the \texttt{astromet} package; \citealt{everallCompletenessGaiaverseIII2021}) and refit the standard 5-parameter astrometric model ($\alpha, \delta, \mu_\alpha, \mu_\delta, \varpi$). We set $\mathrm{RUWE} > 1.25$ as our detection threshold, a standard limit for identifying unresolved companions in \textit{Gaia} DR3 \citep{penoyreAstrometricIdentificationNearby2022}. For each grid point in companion mass and orbital period, we averaged results over 30 independent realizations, each with random orbital phase and independent noise, and used the median RUWE as the detection statistic. We assumed circular orbits ($e = 0$), a non-luminous companion, and a host mass of $M_\star = 1.353\,M_\odot$ \citep{delrezWASP121HotJupiter2016}. While a stellar-mass companion would contribute its own light to the system, this extra flux shifts the observed center of light toward the companion, reducing the amplitude of the astrometric signal measured by \textit{Gaia}. Therefore, assuming a non-luminous companion ensures we are simulating the maximum possible signal, making our derived mass limits strictly conservative. We considered both edge-on ($i = 90^\circ$) and face-on ($i = 0^\circ$) orbits to cover typical geometry. The longitude of the ascending node $\Omega$ and the argument of periastron $\omega$ are degenerate with the orbital phase under the circular-orbit assumption and were held fixed.

Figure~\ref{fig:wasp121_gaia_exclusion} shows the resulting exclusion limits. Since WASP-121's observed $\mathrm{RUWE}$ is significantly below the 1.25 threshold, we detect no evidence for an outer stellar companion. Any stellar-mass companion ($M > 0.08\,M_\odot$) with orbital periods between approximately 1 and 6 years (corresponding to separations of $1$--$4$\,AU), depending on orbital orientation, can be excluded. However, we cannot exclude any stellar companions outside this range. \citet{bohnYoungSunsExoplanet2020} excludes any stellar companions with periods longer than about 140 years ($\gtrsim 30$\,AU). Thus, a stellar companion could remain undetected in the intermediate orbital period gap between roughly 6 and 140 years. In this regime, the companion is too close to the host star to be spatially resolved by direct imaging, while its orbital period far exceeds the \textit{Gaia} DR3 baseline, rendering it difficult to detect with current astrometric data. 

\subsection{High-eccentricity migration assessment}\label{subsec:hem}
Having established that a hidden outer perturber is dynamically allowed and that the system is old enough for tidal circularization to complete, we tested whether ZLK-driven HEM can actually reproduce WASP-121\,b's present-day state (a circular, near-polar orbit at $a=0.026$\,AU) without stripping its atmosphere. We stress that our goal is a qualitative assessment of the viability of this mechanism rather than a full exploration of the parameter space or a formal constraint on the companion: we therefore seek a single self-consistent evolutionary track that lands on the observed architecture, and use it to understand the importance of the underlying physical assumptions.

We used the \texttt{JADE} code \citep{Attia2021, Attia2025}\footnote{\url{https://github.com/JADE-Exoplanets/JADE}}, which integrates the secular, hierarchical three-body Hamiltonian expanded to hexadecapole (fourth) order. \texttt{JADE} couples the resulting orbital evolution to the short-range forces that bind the inner orbit (general relativity, equilibrium tide, and spin- and tidally-induced distortion of the planet) as well as to a self-consistent treatment of the planet's atmosphere and interior. Atmospheric mass loss was computed from the XUV-driven \citep[stellar parameters constrained by][]{Bourrier2020parameters, Sing2024wasp121bagemass}, energy-limited hydrodynamic escape prescription introduced in \citet{Attia2021}, while the interior was resolved by the structure integrator of \citet{Attia2025} as an iron core and silicate mantle beneath a cosmic-abundance H/He envelope. The planetary radius, and hence both the strength of the tidal response and the evaporative cross-section, evolved self-consistently with the orbit and the cooling history. We started the planet on a post-disk orbit ($a_\mathrm{init}=0.5$\,AU, well within the regime of strong star--planet coupling) under the influence of an outer companion, adopting the present-day masses and an Earth-like 2:1 rock-to-iron core.

\begin{figure}
    \centering
    \includegraphics[width=\columnwidth]{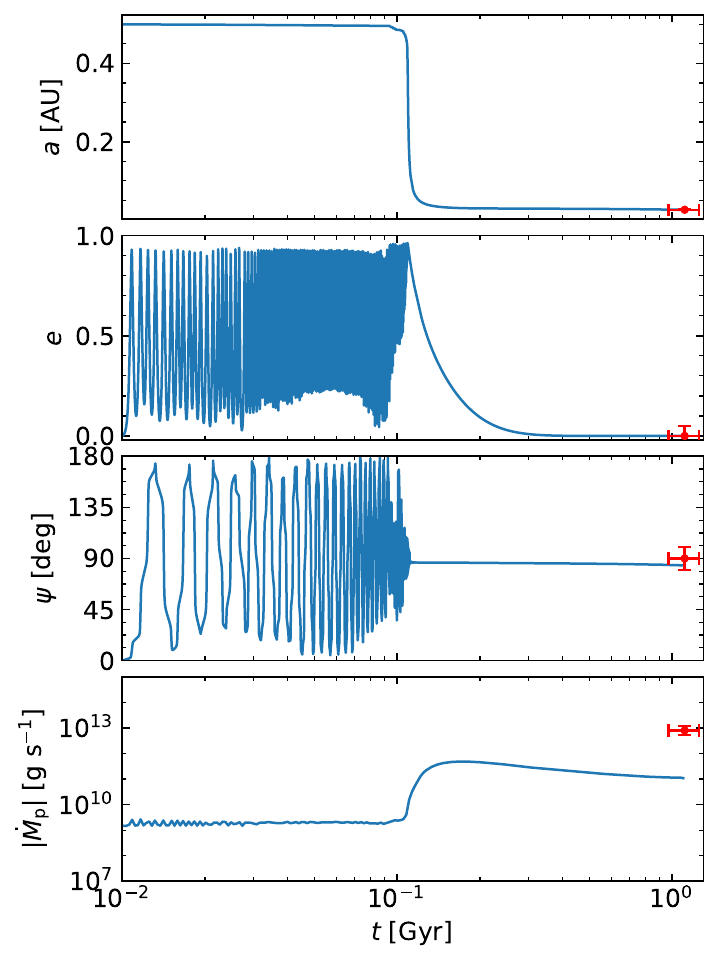}
    \caption{A representative HEM track for WASP-121\,b,
        computed with \texttt{JADE}. From top to bottom: semimajor axis $a$,
        eccentricity $e$, true obliquity $\psi$, and atmospheric mass-loss rate
        $|\dot{M}_\mathrm{p}|$, as a function of time. A $5\,M_\mathrm{Jup}$
        companion at $22$\,AU ($e=0.8$, mutual inclination $89^\circ$,
        $\epsilon_\mathrm{oct}\approx0.05$) drives octupole-level ZLK cycles
        that, combined with tidal dissipation, circularize the orbit and shrink
        it from $0.5$\,AU to $0.026$\,AU while tilting it to a near-polar
        configuration by $1.1$\,Gyr. Red error bars mark the observed
        constraints at the stellar age. The predicted mass-loss rate falls below the
        observed range because \texttt{JADE} does not reproduce the planet's
        inflated radius (see text).}
    \label{fig:hem}
\end{figure}

A heuristic scan of the relevant parameters, each varied within its observational or theoretical uncertainty, readily yielded configurations that match the observed end-state: we fixed the planetary Love number to $k_2=0.2$, compatible with the tentative value derived for WASP-121\,b by \citet{Hellard2020}, and treated the tidal quality factor $Q_\mathrm{p}$ as a free lever within the range $10^4$--$3\times10^5$ spanned by post-ZLK giants \citep{Fabrycky2007shrinking,Hallatt2026}; the initial semimajor axis and the companion's mass, separation, eccentricity, and mutual inclination were varied likewise. The representative example in Fig.~\ref{fig:hem} is driven by a $5\,M_\mathrm{Jup}$ companion at a $a_\mathrm{out} = 22$\,AU separation (eccentricity $e_\mathrm{out}=0.8$, mutual inclination $89^\circ$). Being sub-stellar, it lies comfortably within the \textit{Gaia}-allowed region of Fig.~\ref{fig:wasp121_gaia_exclusion}. The companion excites large-amplitude ZLK cycles that pump the planet's eccentricity to near-unity values, and tidal dissipation at successive periastron passages then removes orbital energy, circularizing and shrinking the orbit. By $1.1$\,Gyr, the planet has migrated from $0.5$\,AU to $a=0.026$\,AU -- matching the observed value to within $0.3\sigma$ -- onto a circular ($e\lesssim10^{-6}$), near-polar ($\psi\simeq84^\circ$) orbit, while its envelope loses less than $0.1\%$ of its mass. HEM is therefore a viable origin for WASP-121\,b's architecture. The few-degree offset from the measured $88.1^\circ$~\citep{bourrierHotExoplanetAtmospheres2020} is not significant for a qualitative assessment, and we find the exact final obliquity to be sensitive to the assumed core mass. An essentially identical scenario was recently invoked for the polar warm Neptune GJ\,436\,b \citep{Attia2025, Im2026}, underscoring that ZLK migration from a hidden sub-stellar perturber is a generic route to polar, close-in planets. Such a misalignment, once imprinted, is not subsequently erased: WASP-121\,b's hot host lies above the Kraft break with a nearly absent convective envelope, giving a very low tidal efficiency factor and a corresponding misalignment probability of $\theta\approx50\%$ from the empirical relation of \citet{Attia2023}, so that tides cannot realign the orbit over the system's lifetime.

The decisive ingredient behind this near-polar outcome is the companion's eccentricity: in the quadrupole limit of the secular interaction (a circular companion), the obliquity stalls below $\sim70^\circ$, and a nonzero $e_\mathrm{out}$ is required to drive the spin--orbit angle into the polar regime. The responsible physics is the octupole term of the hierarchical potential, whose strength relative to the quadrupole is measured by $\epsilon_\mathrm{oct}=(a/a_\mathrm{out})\,e_\mathrm{out}/(1-e_\mathrm{out}^2)$ and which vanishes for a circular companion. Our representative example gives $\epsilon_\mathrm{oct}\approx0.05$, within the narrow band ($\epsilon_\mathrm{oct}\sim0.04$--$0.06$) where the octupole drives large-amplitude orbit flips that carry the mutual inclination through $90^\circ$ \citep{naozHotJupitersSecular2011,Liu2015}; it is these flips that pump the obliquity into the polar regime. The depth of the final migration is instead set by the short-range forces, as anticipated analytically by \citet{Liu2015}: a lower $k_2$ (weaker spin--orbit precession) and a higher $Q_\mathrm{p}$ (slower eccentricity damping) both delay the point at which these forces detach the planet from the resonance, allowing the orbit to reach $0.026$\,AU rather than stalling at wider plateaus. Such a late HEM phase is moreover fully consistent with the formation pathways discussed in Sect.~\ref{subsec:accretionlocation}: it supplies exactly the post-disk migration that those models invoke to bring the planet to its present orbit, and does not require WASP-121\,b to have formed at $0.026$\,AU. Crucially, because the envelope loses less than $0.1\%$ of its mass along this track, the late HEM phase preserves the elemental inventory engraved during accretion: the refractory-to-volatile abundance ratios interpreted in Sect.~\ref{subsec:accretionlocation} as a formation tracer are not subsequently sculpted by evaporation, but remain a faithful record of where the planet acquired its material. Consistently, our adopted starting orbit ($a_\mathrm{init}=0.5$\,AU) already lies interior to the water ice line, in line with the favored formation scenarios. The composition is therefore set during disk accretion, while HEM merely transports the planet inward to its present orbit.

We close with one caveat. \texttt{JADE}'s interior model does not include the deep-heating mechanisms responsible for the inflated radii of ultrahot Jupiters, and caps the planet's radius at $\approx1.05\,R_\mathrm{Jup}$, well below the observed $1.77\,R_\mathrm{Jup}$. Because the photoevaporation rate scales roughly as $\dot{M}_\mathrm{p}\propto R_\mathrm{p}^3$, the predicted present-day mass-loss rate ($\sim10^{11} - 10^{12}$\,g\,s$^{-1}$, see Fig.~\ref{fig:hem}) falls one to two orders of magnitude below the observed and subsequently modeled range of $10^{12} - 10^{14}$\,g\,s$^{-1}$ \citep{Yan2021masslossHalpha, Huang2023, Czesla2024masslossHe, Wang2026}. This reflects the missing radius inflation rather than the migration physics. The same limitation was encountered by \citet{Im2026}, who bracket it by externally rescaling the planetary radius, and it does not affect our dynamical conclusion that HEM can reproduce WASP-121\,b's circular, polar orbit at the age of the system, given the massive gravitational reservoir binding the atmosphere and the ensuing negligible cumulative mass loss.

\subsection{The origin of the extra reflection}\label{disc:reflection}
The need for an extra source of reflection on WASP-121\,b agrees with the retrieval analysis of~\citet{Pelletier2026NIRISSWASP121b}. They can explain the NIRISS dataset either with an extra scattering term or alternatively with a stellar [Ti/H] enrichment and a stronger temperature inversion. In contrast, our joint retrieval of the panchromatic spectrum necessitates the depletion of titanium to achieve a physically plausible solution. This leaves scattered light as the only available opacity source for the short wavelength slope. The retrieved geometric albedo of $0.22\pm0.03$ is in line with the findings of~\citet{Pelletier2026NIRISSWASP121b} and~\citet{Splinter2025WASP121bAlbedo}, who derive a geometric albedo of $0.16\pm 0.02$ and a Bond albedo of $0.277\pm 0.016$, respectively. Differences in the geometric albedo might be attributed to the power law index introduced in Eq.~\ref{eq:scatter}, which the nominal retrieval constrains to be $\delta=-0.3\pm 0.3$. Similar to the nominal retrieval, the power law index is found to be consistent with zero across all retrieval variations, indicating a similar reflectivity of WASP-121\,b across the observed bands. 

The exact source of the scattering opacity that causes the high reflectivity is unclear. One possibility~\citep[also discussed by][]{Pelletier2026NIRISSWASP121b} is a partial dayside coverage with reflective clouds. As WASP-121\,b's dayside is too hot for clouds to form, these would need to be transported from the nightside over the morning terminator~\citep{Kempton2017terminatorclouds,Helling2021clouds} to the dayside. A possible mechanism for this could be a super-rotational jet, as suggested by the findings of \citet{Seidel2025WASP121bJet} for high altitudes of WASP-121\,b. Alternatively, anisotropic magnetic drag could result in localized winds, which could enable a redistribution of cloud material over the morning limb~\citep{Bloecker2026anisotropic}. 

Inspecting our posterior distribution for the geometric albedo, we find that our analytic reflection prescription (see Eq.~\ref{eq:scatter}) contributes between 100 and 300\,ppm to the shortest observed wavelengths. Assuming the clouds to be perfect Lambertian surfaces with a geometric albedo of 1, this would correspond to a cloud coverage fraction between approximately 10\% and 30\%. While general circulation models show that a large fraction of WASP-121\,b's morning limb could potentially be covered by cloud material~\citep{Roman2021hot3Dclouds}, it is still not certain how long the clouds could actually survive at the extreme temperatures in this region. Dedicated microphysical kinetic cloud models could help to address this question. Alternatively, the observed reflection might originate from the outflowing structure around WASP-121\,b, whose corresponding helium absorption was found to span roughly 60\% of WASP-121\,b's orbit~\citep{Allart2025Heoutflow}. While the outflowing gas on its own is mostly optically thin, the large extent of the outflow might lead to a measurable contribution of extra scattered light.

\subsection{Comparison with literature findings}\label{subsec:disc:lit}
To identify if our chemical abundance constraints are data-driven or due to our parametrization, we repeated our nominal retrieval on the isolated NIRISS and NIRSpec datasets. For NIRISS alone, the retrieval results in volatile and refractory enrichments of [Vol/H]=$0.99\pm0.24$ and [Ref/H]=$1.30\pm0.30$, which are within $1\,\sigma$ consistent with the findings of~\citep{Pelletier2026NIRISSWASP121b}. This result shows that the differences in the derived refractory-to-volatile ratio between~\citet{Pelletier2026NIRISSWASP121b} and this work are data-driven and rely on our broader wavelength coverage. In contrast, our NIRISS retrieval finds a low C/O ratio of $0.33\pm0.10$, which is smaller than the value derived by the retrievals with parametrized reflection of~\citet{Pelletier2026NIRISSWASP121b}. We can reproduce a high C/O ratio of $0.83\pm 0.05$, consistent with~\citet{Pelletier2026NIRISSWASP121b}, by excluding the Gaussian processes from the uncertainty treatment. Therefore, the retrieved C/O abundance ratio is sensitive to the applied uncertainty treatment.

For considering NIRSpec alone, we applied our nominal retrieval setup to the phase-corrected data (see Appendix~\ref{app:nirspec}). The results of the NIRSpec retrieval ([Vol/H]=$0.70 \pm 0.16$, [Ref/H]=$1.13 \pm 0.12$, C/O=$0.58 \pm 0.12$) are comparable to those from our panchromatic spectrum (see Table~\ref{tab:retrievalresults}). We tested whether our analysis can replicate the results of~\citet{Evans-Soma2025SiO}, and found that the exact parametrization of the temperature structure is important. Similar to \citet{Evans-Soma2025SiO}, we retrieve a high C/O abundance ratio of $0.84 \pm0.05$ and equally enriched volatile and refractory reservoirs ([Vol/H]=$0.84 \pm 0.23$, [Ref/H]=$0.79 \pm 0.18$), when we use a~\citet{Guillot2010guillot} parametrization. However, we needed to modify the priors such that they did not allow for temperatures above $\sim3600\,$K. It is not completely clear if the temperature priors alone drive these differences, as~\citet{Evans-Soma2025SiO} used the five-parameter extension of~\citet{Parmentier2014guillot} to the~\citet{Guillot2010guillot} temperature structure, while we implemented the three-parameter version. \citet{Evans-Soma2025SiO} also tested the six-parameter temperature implementation of~\citet{Madhusudhan2009seager}, which also resulted in high C/O ratios, consistent with their nominal model. Another possible difference between \citet{Evans-Soma2025SiO} and our modeling approach might be the parametrization of the carbon and oxygen enrichments. While we retrieve [Vol/H] and C/O, and alter the atomic reservoirs in~\texttt{easyCHEM} accordingly, \citet{Evans-Soma2025SiO} directly retrieve for [C/H] and [O/H]. As already indicated by our retrieval results in Sect.~\ref{subsec:retrievalresults}, the retrieved C/O ratio therefore seems to be model-dependent. While both the retrieval setups of \citet{Evans-Soma2025SiO} and us are in agreement with super-stellar C/O (compared to the derived stellar C/O ratio of $0.48\pm 0.05$, see Table~\ref{tab:star}), the exact value is not yet robustly determined. 

We also compare our results to the joint NIRISS and NIRSpec retrievals of~\citet{Saha2025Titanate}, who found a super-stellar C/O ratio of $0.96\pm0.024$, a sub-stellar Si/O ratio
of $0.034\pm0.024$, and an overall metallicity of $4.7^{+1.99}_{-1.38}\times$ solar. These results, particularly their finding of a sub-stellar Si/O ratio, are inconsistent with ours. There are multiple differences in our analysis that may be driving this discrepancy. First, for deriving the bulk metallicity of the planet,~\citet{Saha2025Titanate} do not decouple the abundance estimates for refractory, volatile, and titanium elemental inventories. As seen in Table~\ref{tab:retrievalresults}, these differ significantly and can therefore not be reliably retrieved as a single parameter. Further, their reported Si/O ratio is based on the outputs of their free abundance retrieval. We found this free abundance parametrization to be unreliable (see Appendix~\ref{app:freeAbund}), due to the effect of photodissociation in the upper atmosphere of WASP-121\,b. Finally, \citet{Saha2025Titanate} include CaTiO$_3$ clouds in their nominal model. We do not expect these clouds to be stable at the hot dayside temperatures of WASP-121\,b, as their corresponding saturation vapor pressures are above the expected partial pressures for the probed atmospheric pressure layers~\citep[see Fig.~\ref{fig:TPcontribution} and also Fig.~2a of][]{Evans-Soma2025SiO}. Models that include clouds are likely the preferred parametrization, because they provide the necessary broadband reflected light contribution (see e.g., Sect.~\ref{disc:reflection}) to explain the spectrum at the shortest wavelengths. However, CaTiO$_3$ clouds are not expected at these temperatures, and the specific choice of one cloud species could bias the retrieved abundances.

Finally, we compare our results to measurements from facilities other than JWST. From HST optical transmission spectroscopy, \citep{Lothringer2021VolRef} obtained a super-solar refractory-to-volatile ratio of $5.0^{+6.0}_{-2.7}\times$ solar. Similarly, ground-based high-resolution observations of~\citet{Smith2024IGRINSWASP121b} with IGRINS at Gemini South II also derive a super-stellar refractory-to-volatile ratio of $3.83^{+3.62}_{-1.67}\times$ stellar. Furthermore, they retrieve a C/O ratio of $0.70^{+0.07}_{-0.10}$ that agrees well with our value derived from the ``More species’’ model. Analogous to our discussion in Sect.~\ref{subsec:accretionlocation}, they also conclude that the planet likely formed between the water ice line and the soot line. In contrast, ~\citet{Pelletier2025CRIRESWASP121b} observed WASP-121\,b with the VLT and found it to be volatile-rich, with a volatile-to-refractory enrichment of $1.75^{+0.57}_{-0.41}\times$ stellar.~\citet{Pelletier2025CRIRESWASP121b} derive the value using detections of H$_2$O, CO, and atomic Fe and Ni, while~\citet{Smith2024IGRINSWASP121b} use detections of H$_2$O, CO, and OH, and argue that their refractory inventory is mostly driven by Fe, Mg, and Ca. The silicon reservoir observed by our JWST observations remains undetected from the ground for WASP-121\,b. Therefore, species selection might affect the measured refractory-to-volatile abundance ratio and, consequently, the discrepancy between ground-based studies. Compared with high-resolution spectra of WASP-121\,b, we also probe slightly lower atmospheric altitudes~\citep[see Fig.~\ref{fig:TPcontribution}, and Fig. 1 of][]{Seidel2025WASP121bJet}. Chemical differences along atmospheric altitudes could therefore also cause the differences in the derived values. Finally, abundances of high-resolution spectra might be affected by non-corrected reflected light, which could skew the retrieved abundance estimates~\citep{Vaughan2026wasp121b}.

\section{Summary}\label{sec:summary}
This analysis provides the mid-infrared dayside spectrum of the ultra-hot Jupiter WASP-121\,b, observed with MIRI/LRS on JWST. We used this data to perform an eclipse mapping of the planet, and combined the dayside spectrum with data from NIRISS and NIRSpec for a joint retrieval analysis. We further measured the stellar abundances of WASP-121\,b's host star and used the retrieved elemental reservoirs together with the stellar abundances to explore possible formation and accretion scenarios. Lastly, we investigated the possibility of planetary migration via ZLK cycles. 

The MIRI/LRS data were reduced with \texttt{Eureka!}, and show the typical instrument systematics for this detector. We recover an eclipse mapping signal using the \texttt{TheRESA} code and find that a non-uniform eclipse mapping model is preferred by at least $\Delta \mathrm{BIC} = 10$ over a uniform model. The mid-infrared maps of WASP-121\,b suggest the presence of an eastward longitudinal hotspot offset of $4.8^{+2.7\circ}_{-2.8}$, consistent with previous findings at shorter wavelengths~\citep{Mikal-Evans2023WASP121bNIRSpecPhaseC,Frazier2026WASP121bNIRISSphaseCModel}. 

We recalculated the stellar abundances and the stellar age. Our 1D LTE and direct NLTE synthesis fits to the ESPRESSO stellar spectrum yield similar abundances as previousy reported for silicon, iron, magnesium, and sulfur. We find slightly higher abundances of carbon and oxygen, possibly due to differences in the fitting routine. We used data from Gaia DR3 and \cite{Sing2024wasp121bagemass} to confirm the stellar age of WASP-121 with \texttt{BASTA}. We find a stellar age of $1.5^{+0.29}_{-0.24}\,$Gyr, similar to previously reported age estimates. 

Our joint retrievals were performed using the atmospheric radiative transfer code \texttt{petitRADTRANS}. Using a 1D model for the dayside, we retrieved WASP-121\,b's temperature structure, elemental inventories, and reflection properties. We assumed equilibrium chemistry and allowed the atmospheric enrichment in volatiles, refractories, and titanium to freely adjust. We implemented a dedicated treatment for correlated and uncorrelated uncertainties in the retrievals and tested the impacts of a variety of different model assumptions.

Our joint retrievals of WASP-121\,b's dayside resulted in an enrichment of the planet's atmosphere in both volatile and refractory inventories. We identified enhancements of the refractory-to-volatile abundance ratios $\mathrm{Si}/\mathrm{O}=3.54^{+0.86}_{-0.69}\times$ stellar and $\mathrm{Si}/\mathrm{C}=3.05^{+1.12}_{-0.80}\times$ stellar. In contrast, we find that the C/O abundance ratio of the planet is strongly model-dependent, and varies between $0.36$ and $0.69$ within the $1\,\sigma$ uncertainties of the most diverging models. Similar to previous analyses of individual datasets, we find a strong temperature inversion in the dayside atmosphere of WASP-121\,b. The retrieved temperature structure shows a change in temperature gradients between $10^{-1}$ and $10^{-4}\,$bar, which can not be satisfactorily modeled with a parametrized~\citet{Guillot2010guillot} profile. We further confirm the titanium depletion on the dayside of WASP-121\,b, and find the inclusion of additional reflected starlight crucial to model the joint spectrum. Lastly, we find that the detection significance of VO depends on the uncertainty treatment and the applied opacities.

The enrichment of the WASP-121\,b's atmosphere with refractory and volatile species implies significant accretion of heavy elements during the planet's formation. We compared our resulting elemental inventories and abundance ratios of WASP-121\,b to predictions from core-accretion formation models. In this scenario, we conclude that the planet either accreted solids both inside and outside the water ice line, or that it accreted both solids and gas between the water ice line and the soot line. 

To reach its present-day high-obliquity orbit, WASP-121\,b could have migrated via ZLK cycles or a planet--planet scattering event. Using the \texttt{JADE} code, we verified the feasibility of HEM and find that a $5\,$M$_\mathrm{Jup}$ companion on an eccentric orbit at $22\,$AU is not ruled out by \textit{Gaia} and could have driven the planet to its present-day orbit. The atmosphere survives this migration: WASP-121\,b's large mass binds the envelope against escape, and because the inward migration sets in late, the planet stays far from the star throughout the first $\sim100$\,Myr of saturated stellar XUV emission, limiting the cumulative envelope loss.

\section{Data availability}
The JWST data used in this paper are associated with the programs GO 2961 (P.I., P.~Molli\`ere), GO 1729 (P.I.:~T.~Evans-Soma, co-PI: T. Kataria), and GTO 1201 (P.I.:~D.~Lafrenière). They are publicly available via
the Mikulski Archive for Space Telescope at \url{https://mast.stsci.edu}. The JWST MIRI/LRS light curves, spectra used in this work, and the posterior distributions of the retrieval variations are available via Zenodo at~\url{https://doi.org/10.5281/zenodo.20767846}. Additional data products can be made available upon request.

\begin{acknowledgements}
K.A.K thanks Jeroen Bouwman and Mark Hammond for a preliminary analysis of the MIRI/LRS data. K.A.K also wishes to thank Lorena Acu\~na, Channon Visscher, Antonia von Stauffenberg, Ji Wang, Vivien Parmentier, and Katy Chubb for providing helpful inputs and insightful discussions, which improved the quality of the paper. This work is based on observations made with the NASA/ESA/CSA James Webb Space Telescope. The data were obtained from the Mikulski Archive for Space Telescopes at the Space Telescope Science Institute, which is operated by the Association of Universities for Research in Astronomy, Inc., under NASA contract NAS 5-03127 for JWST. These observations are associated with programs GO 2961, GO 1729, and GTO 1201. Furthermore, this work is based on observations made with ESO Telescopes at the La Silla Paranal Observatory under program ID 106.21QM. K.A.K.~gratefully acknowledges support from the DLR via project P.S.ASTR1508. H.J.H. acknowledges support from eSSENCE (grant number eSSENCE@LU 9:3), the Swedish National Research Council (project number 2023-05307), The Crafoord foundation and the Royal Physiographic Society of Lund, through The Fund of the Walter Gyllenberg Foundation. S.P.\ acknowledges support from the Swiss National Science Foundation under grant 51NF40\_205606 within the framework of the National Centre of Competence in Research PlanetS.  L.-P.C. acknowledges financial support from Mitacs through the Mitacs Accelerate program, in partnership with the Montreal Planetarium.
\end{acknowledgements}
\bibliographystyle{aa} 
\bibliography{WASP121b} 
\begin{appendix}
\section{Extracting the NIRSpec eclipse spectrum}\label{app:nirspec}
To cover the wavelength range between $2.7$ and $\SI{5.2}{\micro\meter}$, we use data from the NIRSpec phase curve reduced by~\citet{Evans-Soma2025SiO}. In their analysis,~\citet{Evans-Soma2025SiO} produced emission spectra of WASP-121\,b in 36 phase bins of 46\,min. These represent the planet emission at different out-of-eclipse and out-of-transit phases. Consequently, none of the provided bins cover the phase probed by the NIRISS/SOSS and MIRI/LRS eclipse spectra. 

\begin{figure}
    \centering
    \includegraphics[width=1.0\linewidth]{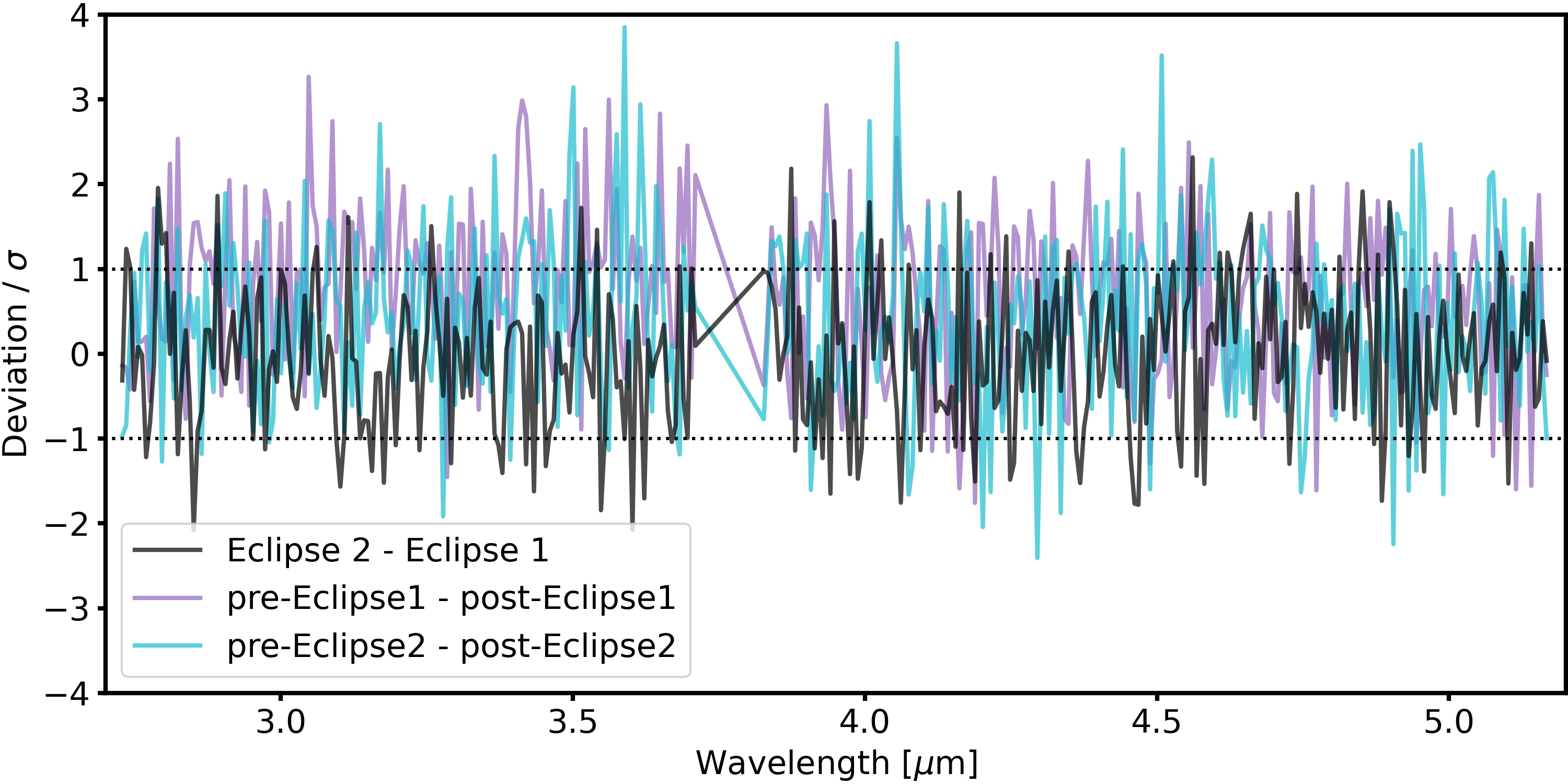}
    \caption{Deviation between emission spectra of different bins from the NIRSpec phase curve of~\citet{Evans-Soma2025SiO}.}
    \label{fig:app:eclipsecompare}
\end{figure}

To obtain an equivalent NIRSpec spectrum at the phase of eclipse mid-point, we start by obtaining a weighted average of pre-eclipse and post-eclipse phase bins for both eclipses. As shown in Fig.~\ref{fig:app:eclipsecompare}, NIRSpec pre-eclipse spectra have a systematically higher flux than the post-eclipse specta. This offset is approximately $1\,\sigma$ for the first eclipse, and $0.3\,\sigma$ for the second eclipse. This offset might be caused by the small hotspot offset observed in the NIRSpec phase curve~\citep{Mikal-Evans2023WASP121bNIRSpecPhaseC}. The larger offset for the first eclipse could be due to instrument systematics, which are stronger during the start of the observations. In contrast, we do not see a strong systematic offset between the averaged spectra of the first and secondary eclipse, which is why we proceed with the average spectrum of both eclipses. 

\begin{figure}
    \centering
    \includegraphics[width=1.0\linewidth]{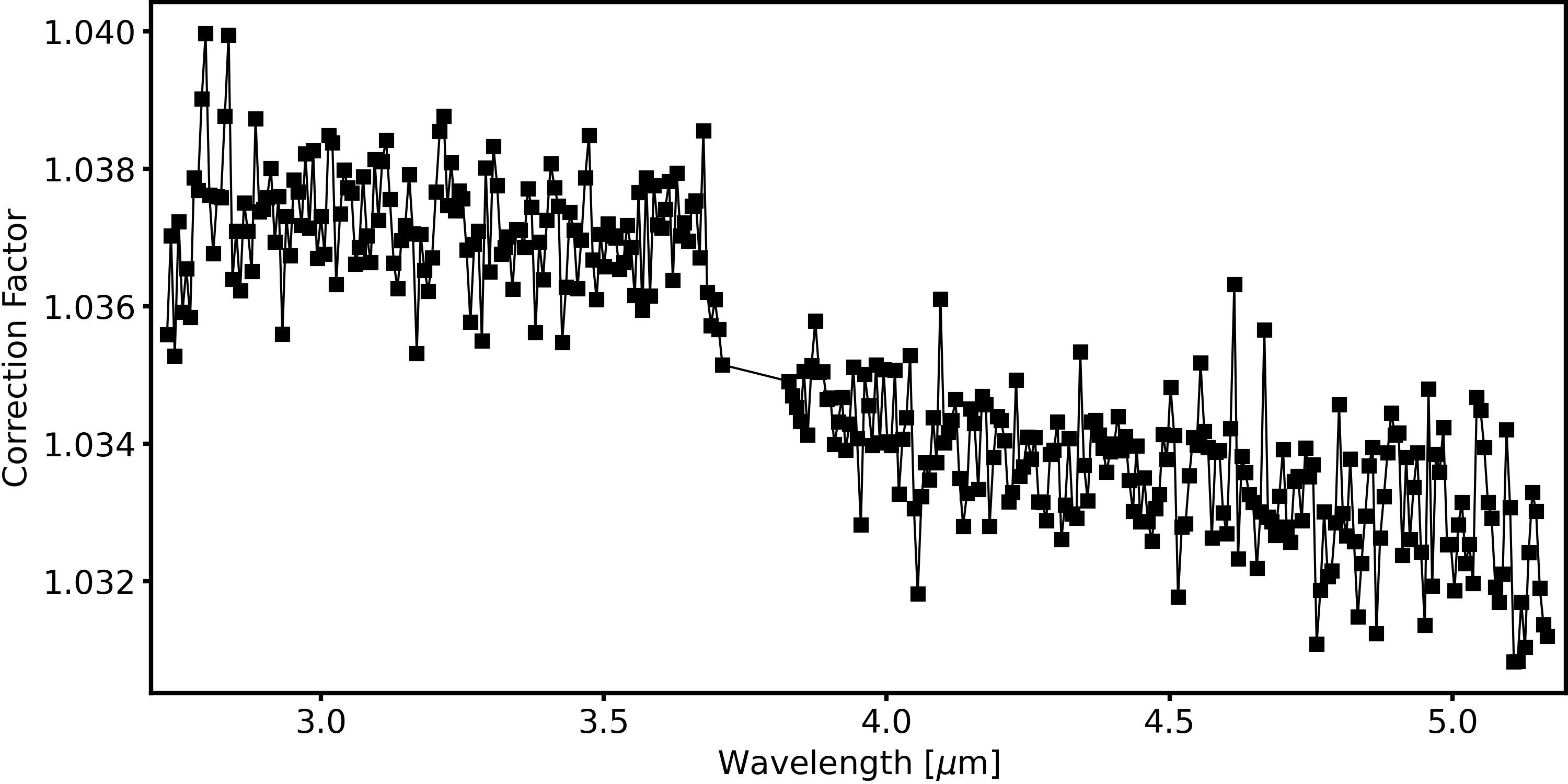}
    \includegraphics[width=1.0\linewidth]{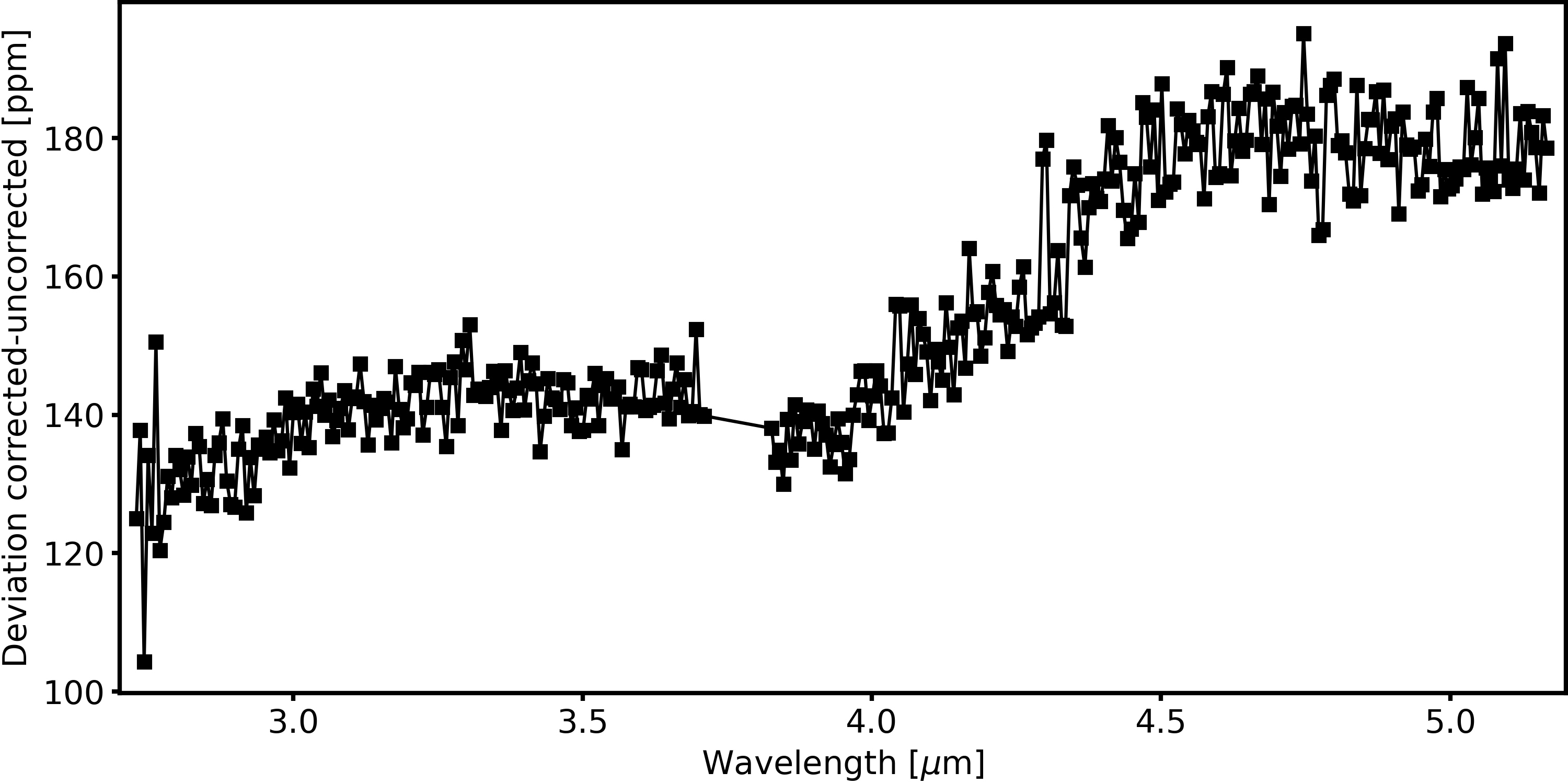}
    \caption{Impact of the interpolation correction factor on the eclipse spectrum. The top panel shows the wavelength dependence of the derived correction factor. The bottom panel shows the difference between corrected and uncorrected spectrum in ppm.}
    \label{fig:app:correction}
\end{figure}

The average of pre and post-eclipse spectra is equivalent to a linear interpolation between these phase bins. This interpolation underestimates the dayside flux of WASP-121\,b during mid-eclipse, around which the sub-stellar point of the planet emits the most flux. Therefore, we corrected the average eclipse spectra according to the sinusoidal shape of the full NIRSpec phase curve. For this, we interpolated the best-fit spectroscopic models of~\citet{Evans-Soma2025SiO} to the in-eclipse intervals. This was done by fitting the best-fit models with the light curve generation tool \texttt{starry}~\citep{Luger2019starry} using the non-linear least squares fitter implemented in the \texttt{curve\_fit} function of \texttt{scipy.optimize}~\citep{Virtanen2020SciPy-NMeth}. Analogous to~\citet{Evans-Soma2025SiO}, we model the light curves with a dipole map of degree $l=1$, where only $Y_{1,0}$ deviates from zero. By fitting for $Y_{1,0}$, the map amplitude, a map rotation offset, and the rotation period, it is possible to replicate all best-fit spectroscopic light curves of~\citet{Evans-Soma2025SiO} up to the numerical precision. These replicated light curves include the in-eclipse time, allowing us to derive a correction factor for the linear interpolated spectra of WASP-121\,b. For this, we divide the flux of the replicated model light curve into the same bins as the~\citet{Evans-Soma2025SiO} emission spectra, and also apply the same phase-bin averaging. The inverse of the ratio between the resulting value and the model flux at eclipse mid-time was then used as a correction factor for each spectral bin, respectively. As shown in Fig.~\ref{fig:app:correction}, this wavelength-dependent factor varied between $1.03$ and $1.04$, resulting in extra emission between $120\,$ppm and $200\,$ppm. The correction factor is mostly constant for the NRS1 detector, while it shows a linear slope for the NRS2 detector. The absolute flux difference between the corrected and uncorrected spectrum mimics the original spectral shape, since the mostly constant correction factor is applied multiplicatively. The corrected eclipse spectrum shows a similar flux level as the adjacent data sets (see Sect.~\ref{sec:retrieval}), indicating that we successfully replicated WASP-121\,b's spectrum during eclipse mid-time.

During this analysis, we propagated uncertainties using Gaussian error propagation of the original phase-binned emission spectra uncertainties. As all of the measurements originate from the same phase curve observation, these spectra might be correlated, violating the assumption of uncorrelated noise needed for this strategy. This could lead to an underestimation of uncertainties for the corrected NIRSpec spectrum used in Sect.~\ref{sec:retrieval}. We account for this possibility during the retrievals by fitting an uncertainty inflation factor.

\section{Free abundance retrievals}\label{app:freeAbund}
\begin{figure*}
    \includegraphics[width=0.999\textwidth]{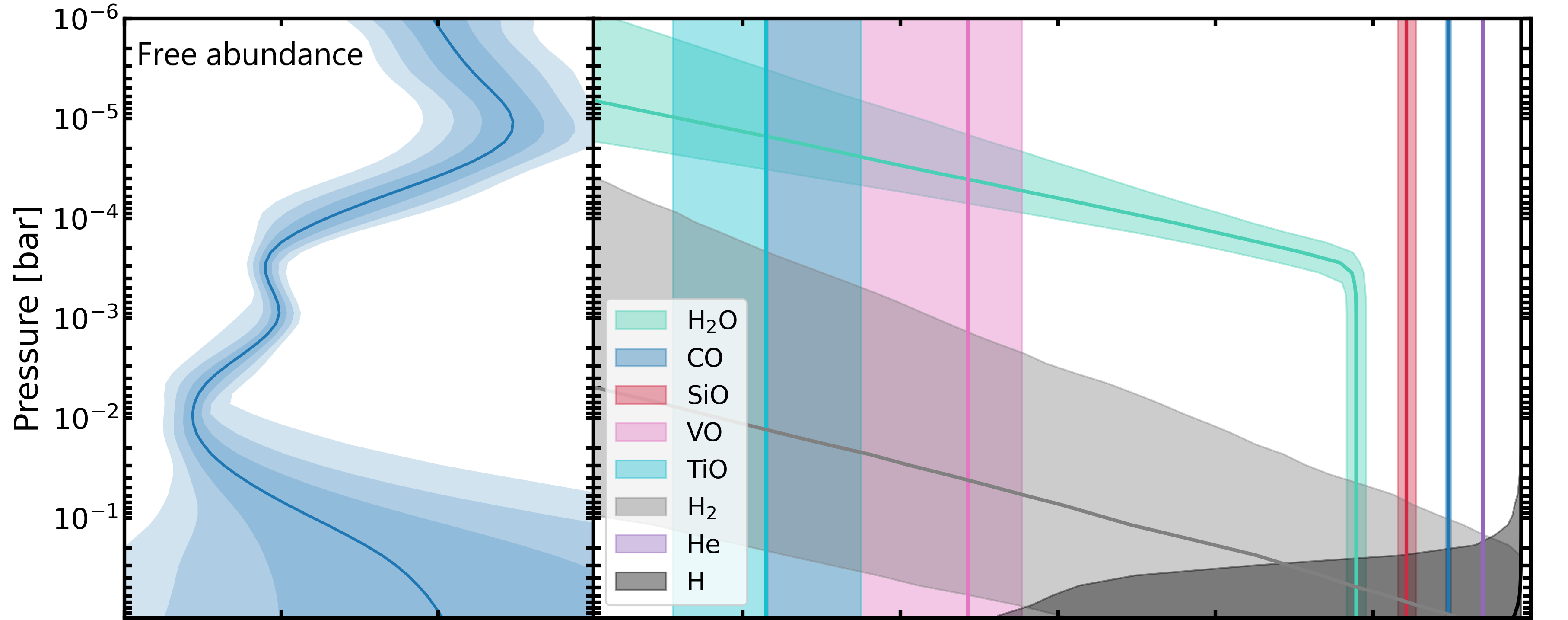}\\
    \includegraphics[width=0.999\textwidth]{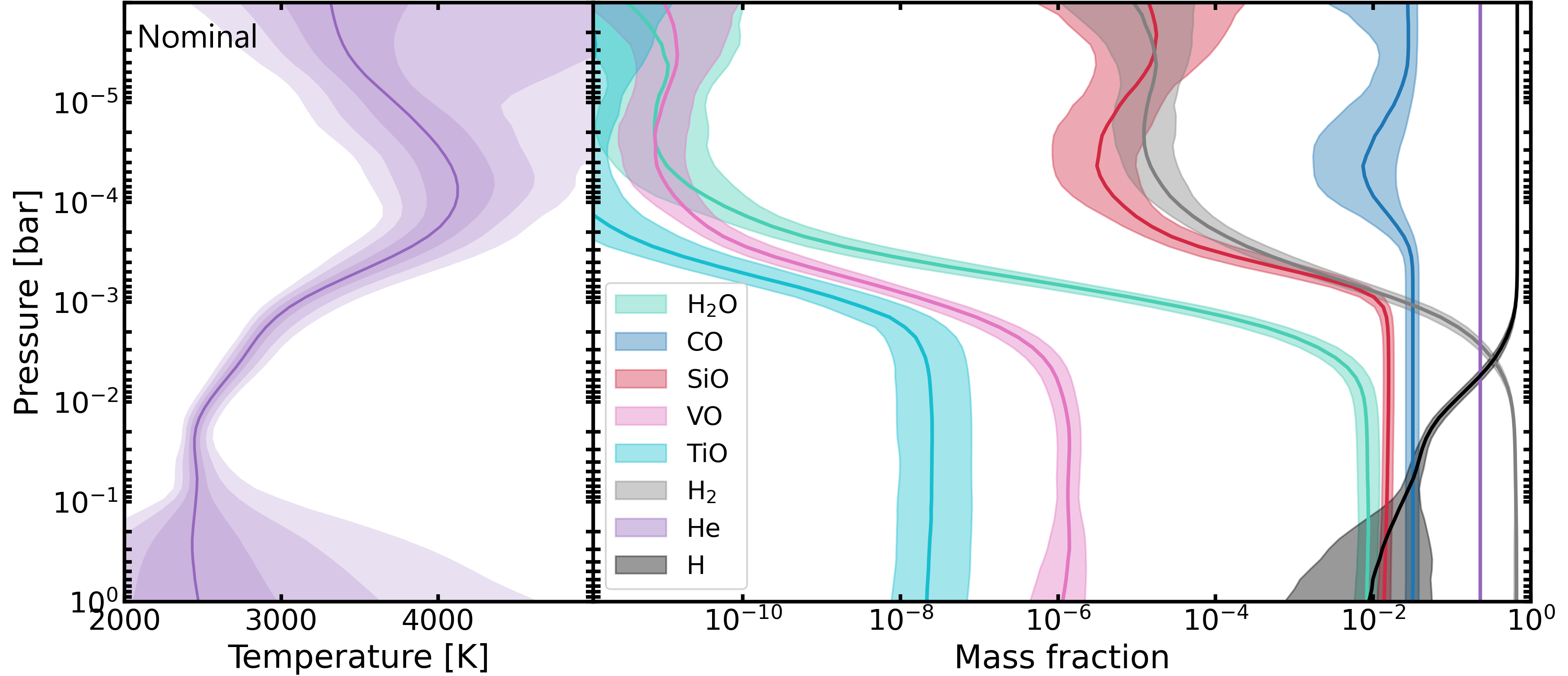}
    \caption{Retrieved temperature structures (left panels) and abundance profiles (right panels). The top panel shows the retrieved profiles for the free abundance parametrization, while the bottom panel shows the results for the nominal retrieval. Solid lines show the median retrieved profiles, while shaded regions show $1\,\sigma$ percentiles for the retrieved mass fractions, and 1, 2, and $3\,\sigma$ percentiles for the temperature profiles.}
    \label{app:fig:freeAbund}
\end{figure*}
In addition to our nominal chemistry description with \texttt{easyCHEM}, we also tested freely retrieving the abundances of the major contributing species (including H$_2$O, CO, SiO, VO, TiO, OH, CO$_2$, CH$_4$). Abundances of these species were allowed to vary freely between $10^{-12}$ and $10^{0}\,$dex, and are vertically constant for all molecules but H$_2$O and H$_2$. After adding mass fractions of all species, the remaining fraction was filled with a mixture of $75.4$\% H$_2$ and $24.6$\% He~\citep{Asplund2009}. Solutions with a total mass fraction above 1 were rejected.

Since the hot temperatures of WASP-121\,b's atmosphere are expected to dissociate dissociate many of the molecular species, we implemented the dissociation of H$_2$O and H$_2$. Similar to~\citep{Gapp2025SiO}, we retrieved the pressure at which the dissociation starts, $p_\mathrm{dissociate}$, and a pressure power-law coefficient $\gamma$ that controls the abundance decline of the species:
\begin{align}
    X_{\mathrm{H}_2\mathrm{O}} [p < p_{\mathrm{dissociate},\,\mathrm{H}_2\mathrm{O}}] &= X_{\mathrm{H}_2\mathrm{O, \,base}} \cdot \left(\frac{p}{p_{\mathrm{dissociate},\,\mathrm{H}_2\mathrm{O}}}\right)^{\gamma_{\mathrm{H}_2\mathrm{O}}} \\
    X_{\mathrm{H}_2} [p < p_{\mathrm{dissociate},\,\mathrm{H}_2}] &= X_{\mathrm{H}_2\mathrm{,\, base}} \cdot \left(\frac{p}{p_{\mathrm{dissociate},\,\mathrm{H}_2}}\right)^{\gamma_{\mathrm{H}_2}}
\end{align}
Where $p$ is the pressure of a respective atmospheric layer. The vertically constant mass fraction $X_{\mathrm{H}_2\mathrm{O, \,base}}$ of water below the dissociation pressure was allowed to vary freely between $10^{-12}$ and $10^{0}\,$dex. For H$_2$ dissociation, we assumed that the hydrogen mass fraction is converted into molecular hydrogen. The free electrons and H$^-$ abundances were decoupled from this parametrization and are retrieved with a vertically constant abundance. Apart from the chemistry description, the remaining parametrization follows our nominal retrieval described in Sect.~\ref{subsec:ret:nominal}. For computational reasons, we did not retrieve the dissociation of all molecular species. We furthermore did not include species that only weakly affect the spectrum, such as FeH or HCN. We acknowledge that this biases the retrieved abundances, which is why we prefer to work with the results from Gibbs free energy minimization in the main analysis. 

To ensure convergence despite the increase of free parameters to 44, we used 4000 live points for the sampling with \texttt{PyMultiNest}. Compared with our nominal retrieval, the free-abundance retrieval is strongly disfavored, with $\Delta\ln(Z)=-3260$. This is partially linked to the large number of free parameters, as excluding Gaussian processes for the retrieval decreased the rejection significance to $\Delta\ln(Z)=-53$. While the retrievals still manage to reproduce the data well with $\chi^2_\mathrm{red}=0.93$, the drawbacks of the free-abundance parametrization can be seen in Fig.~\ref{app:fig:freeAbund}: The constant abundance profiles are compensated by a more pronounced sub-structure of the temperature profile. Consequently, this can result in biased mass fraction estimates. An example for this is CO, which the free-abundance retrieval finds to make up 10\% of the planet's mass fraction with only a small uncertainty. 

While the exact abundances of individual species may be biased, the free-abundance parametrization qualitatively produces an atmosphere similar to that of the chemically consistent retrieval. It measures a temperature inversion and identifies CO, SiO, and H$_2$O as the dominant species in WASP-121\,b's atmosphere. The measured mass fractions of TiO and VO are consistent with the mass fractions predicted from the chemically consistent retrievals at probed pressures around $10^{-3}$\,bar. The dissociation of H$_2$O is reproduced, although it is measured to occur at pressures that are approximately one dex lower than found by the chemically consistent retrieval. In contrast, the H$_2$ dissociation is largely unconstrained by the free-abundance retrievals and found to start already in the lowest altitudes. Since there is no characteristic feature of H$_2$ in the spectrum and H$^-$ was retrieved separately, the observed H$_2$ dissociation might be driven by an attempt of the retrieval to lower the mean molecular weight of the atmosphere. 

\section{The impact of Gaussian progresses}\label{app:GP}
\begin{figure}
    \includegraphics[width=0.499\textwidth]{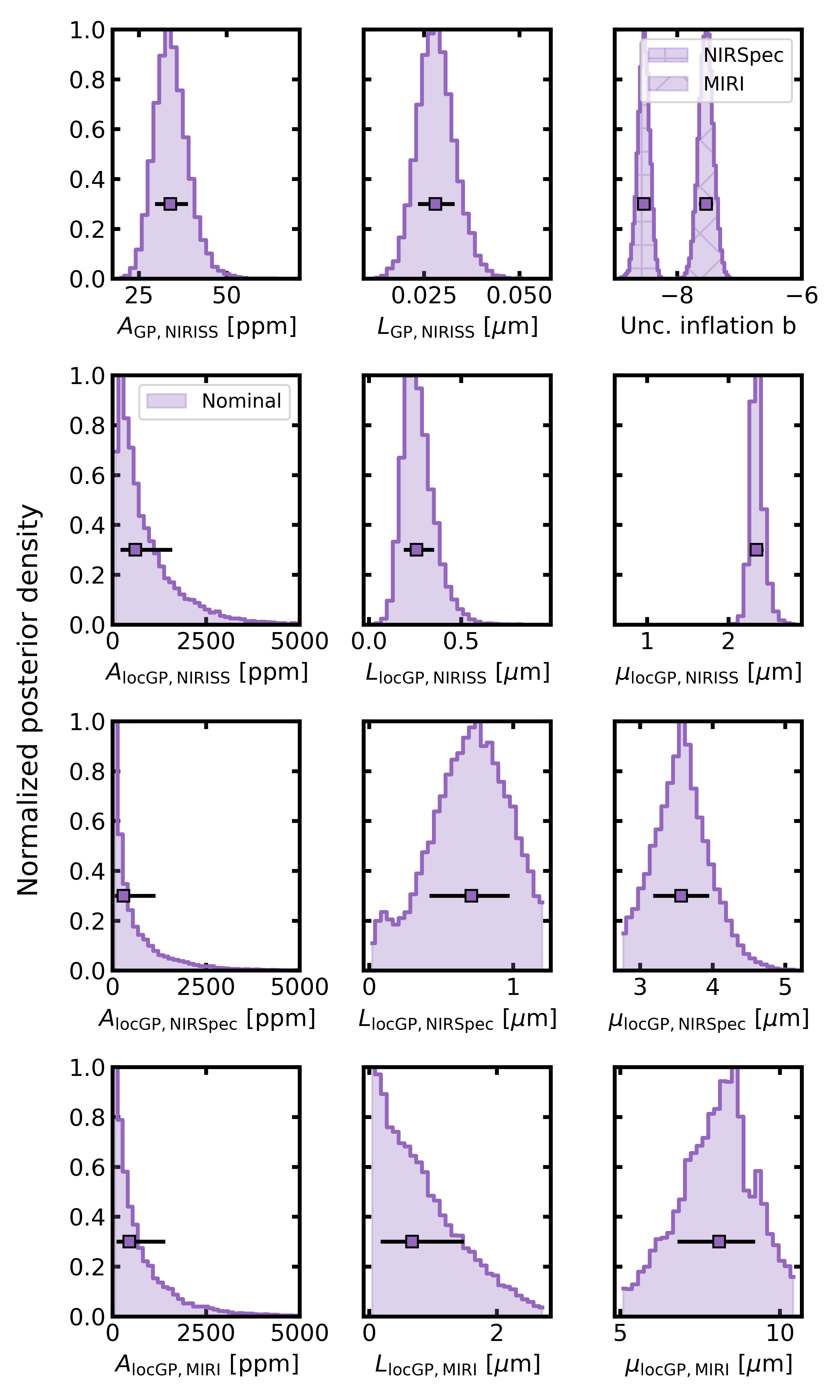}\\
    \caption{Normalized posterior distributions for parameters of the Gaussian processes and uncertainty inflation applied in the nominal model (see Table~\ref{tab:retrievalprior}).}
    \label{app:fig:GPposteriors}
\end{figure}

The posteriors for the parameters of the Gaussian processes and the uncertainty inflation explained in Sect.~\ref{sec:retrieval} are shown in Fig.~\ref{app:fig:GPposteriors}. The global NIRISS GP kernel converges to length scales of $\SI{0.03}{\micro\meter}$, consistent with correlated noise from the instrument PSF~\citep{Rotman2025GPs}. The amplitudes for the local kernels of MIRI and NIRSpec converge to the lower prior bounds and show broad posterior distributions in position and length scale. Therefore, they do not identify a single narrow Gaussian feature, but instead contribute at various wavelengths. For NIRIRSS, the position posterior of the local kernel converges to $\mu_\mathrm{locGP, NIRISS} = 2.35 \pm 0.08\,\mu$m, indicating a more defined model-data mismatch at the corresponding wavelength.

\begin{figure*}
    \includegraphics[width=0.999\textwidth]{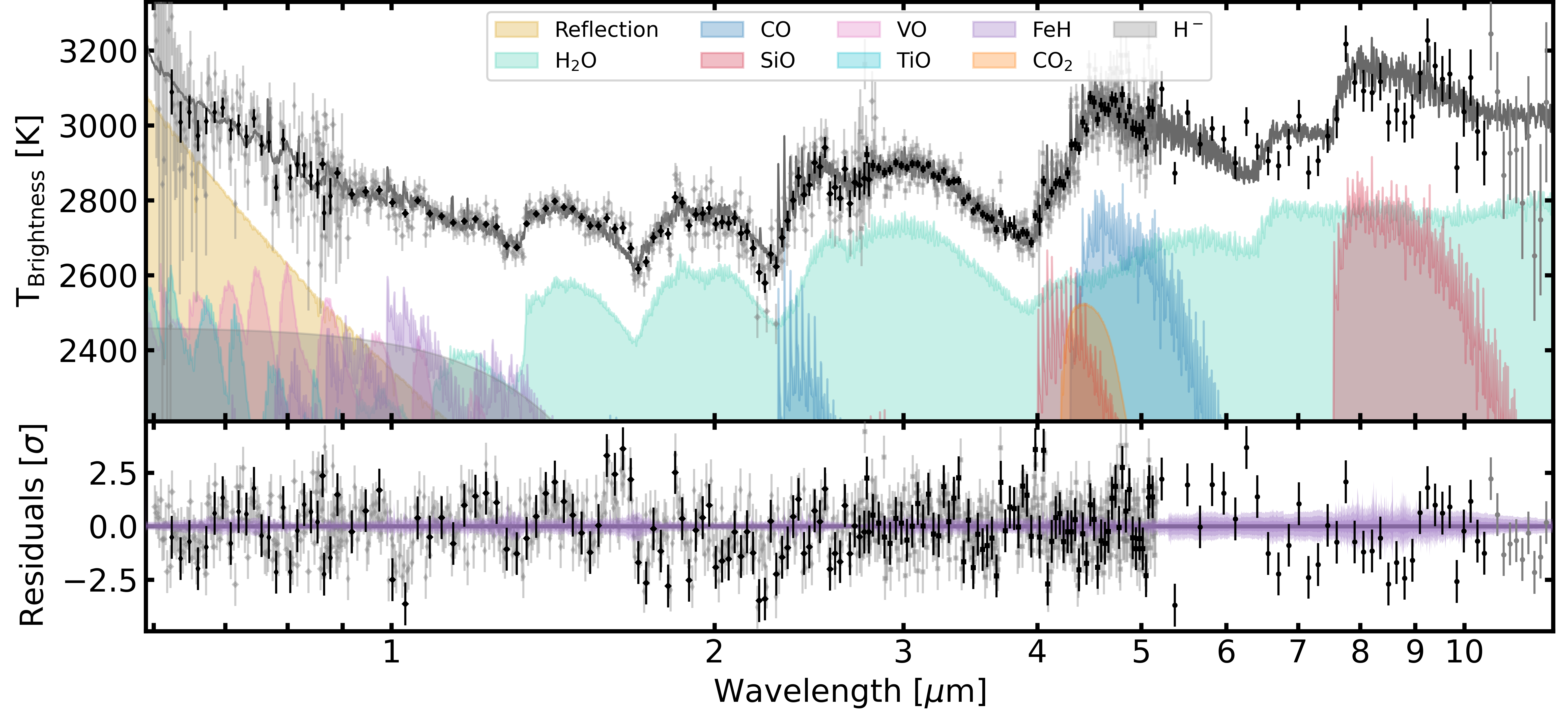}\\
    \caption{Panchromatic spectrum of WASP-121\,b and opacity profiles from the most relevant species. The shown models and residuals correspond to the retrieval without Gaussian processes. All markers in the figure are analogous to Fig.~\ref{fig:jointspectrum}.}
    \label{app:fig:noGPspec}
\end{figure*}

We show the models and residuals for the retrieval without Gaussian processes in Fig.~\ref{app:fig:noGPspec}. The wavelength region between $0.9$ and $2.4\,\mu$m shows a possibly correlated noise structure, which is accounted for by the Gaussian processes in our nominal model (see Fig.~\ref{fig:jointspectrum}). We also tested to account for this noise structure by freely fitting for VO and TiO abundances, whose opacities have a similar structure. We further tested the use of updated opacities for VO (see Sect.~\ref{subsec:detection}). Neither test removed the observed residual structure in this wavelength region. 

\end{appendix}
\end{document}